\documentclass{aa}  

\newcommand{\citeg}[1]{\citep[e.g.,][]{#1}}

\usepackage{graphicx}
\usepackage{xcolor}
\usepackage{amsmath}
\usepackage[english]{babel} 
\usepackage{multirow}

\usepackage{adjustbox}

\usepackage{amsmath}
\usepackage{amsfonts}
\usepackage{amssymb}
\usepackage{wasysym}

\usepackage{natbib}  

\begin{document}

\title{Morphology-density Relation, Quenching, and Mergers in CARLA Clusters and Proto-Clusters at $1.4<z<2.8$}
\titlerunning{CARLA Clusters and Proto-Clusters at $1.4<z<2.8$}
\authorrunning{Mei et al.}

 \author{Simona Mei  \inst{1,2}, 
             Nina A. Hatch \inst{3},
             Stefania Amodeo \inst{4},
             Anton V. Afanasiev \inst{1},
             Carlos De Breuck\inst{5}, 
             Daniel Stern\inst{2},
             Elizabeth A. Cooke\inst{6},  
              Anthony H. Gonzalez\inst{7},
              Ga\"el Noirot\inst{8}, 
              Alessandro Rettura\inst{2},
              Nick Seymour\inst{9}, 
              Spencer A. Stanford\inst{10}, 
              Jo\"el Vernet\inst{5},
              Dominika Wylezalek\inst{11}
                   }
       
 \institute{Universit\'e Paris Cit\'e, CNRS(/IN2P3), Astroparticule et Cosmologie, F-75013 Paris, France \email{mei@apc.in2p3.fr}
\and Jet Propulsion Laboratory and Cahill Center for Astronomy \& Astrophysics, California Institute of Technology, 4800 Oak Grove Drive, Pasadena, California 91011, USA
  \and School of Physics and Astronomy, University of Nottingham, University Park, Nottingham NG7 2RD, UK
\and Universit\'e de Strasbourg, CNRS, Observatoire astronomique de Strasbourg, UMR 7550, F-67000 Strasbourg, France
 \and European Southern Observatory, Karl-Schwarzschildstrasse 2, 85748 Garching, Germany
  \and  National Physical Laboratory, Hampton Road, Teddington, Middlesex, TW11 0LW, UK      
   \and       Department of Astronomy, University of Florida, Gainesville, FL 32611-2055, USA
     \and  Department of Astronomy \& Physics, Saint Mary's University, 923 Robie Street, Halifax, NS B3H 3C3, Canada
\and International Center for Radio Astronomy Research, Curtin University, GPO Box U1987, 6102 Perth, Australia
\and       Department of Physics, University of California, One Shields Avenue, Davis, CA 95616, USA
\and Zentrum f\"ur Astronomie der Universit\"at Heidelberg, Astronomisches Rechen-Institut, M\"onchhofstr 12-14, D-69120 Heidelberg, Germany
   }




\date{Received 15 March, 2022; accepted 5 October, 2022}
\authorrunning{Mei et al.}

\abstract{

At redshift $z\lesssim1.3$, early-type (ETG) and passive galaxies are mainly found in dense environments, such as galaxy clusters. What is less clear is whether these well-known morphology-density and passive-density relations are already established at higher redshifts. To address this question, we perform an in-depth study of galaxies in 16 spectroscopically confirmed clusters at $1.3<z<2.8$ from the Clusters Around Radio-Loud AGN survey. Our clusters span a total stellar mass in the range $11.3<\log(\frac{M^c_*}{M_{\odot}})<12.6$ (approximate halo mass in the range $13.6<\log(\frac{M^c_h}{M_{\odot}})<14.6$). Our main finding is that the morphology-density and passive-density relations are already in place at $z\sim2$. The cluster at z = 2.8 shows a similar fraction of ETG as in the other clusters in its densest region, however one cluster does not provide enough statistics to confirm that the morphology-density relation is already in place.  The cluster ETG and passive fractions depend on local environment and mildly on galaxy mass. They do not depend on global environment.  At lower local densities, where $\Sigma_N < 700$ gal/Mpc$^2$, the CARLA clusters exhibit a similar ETG fraction as the field, in contradiction to clusters at z = 1 that already exhibit higher ETG fractions. This implies that the densest regions influence the morphology of galaxies first, with lower density local environments either taking longer or only influencing galaxy morphology at later cosmological times. Interestingly, we find evidence of high merger fractions in our clusters with respect to the CANDELS fields, but the merger fractions do not significantly depend on local environment. This suggests that merger remnants in the lowest density regions can reform disks fuelled by cold gas flows, but those in the highest density regions are cut-off from the gas supply and will become passive ETG. The percentages of active ETG, with respect to the total ETG population, are $21 \pm 6$\% and $59 \pm 14\% $ at 1.35 < z <1.65 and  1.65 < z < 2.05, respectively, and about half of them are mergers or asymmetric in both redshift bins. All the spectroscopically confirmed CARLA clusters have properties consistent with clusters and proto-clusters, confirming that RLAGN are lighthouses for dense environments.
The differences between our results and those that find enhanced star formation and star-bursts in cluster cores at similar redshifts are probably due to the different sample selection criteria, which choose different environments that host galaxies with different accretion and pre-processing histories.
}

  \maketitle


\section{INTRODUCTION} \label{sec:intro}

Clusters of galaxies form by mergers of galaxy groups and filament accretion, and typically cross the typical halo mass of virialization of $M^c_h=10^{14} M_{\odot}$ \citep{08ev} in the redshift range $ 1  \lesssim z  \lesssim 3$ \citep{13chiang,18mul}.
In this first phase of assembly, cluster progenitors, also called proto-clusters, are distributed over large areas (e.g., at $z \sim 2 $ over $\approx 35~h^{-1} Mpc^2$ comoving; \citealp{18mul}), and a typical local cluster with mass $M=2-5  \times 10^{14} \ M_{\odot}$ is predicted to attain the typical halo mass of virialization at $z \sim1.5-2$ \citep{13chiang}.
While few clusters and proto-clusters at $1.5<z<3$ were detected by the early 2010's, tens of them were observed in the following five years (Castellano et al. 2007, 2011; Eisenhardt et al. 2008; Ashby et al. 2009; Kurk et al. 2009; Andreon et al. 2010, 2011; Chiaberge et al. 2010; Papovich et al. 2010; Tanaka et al. 2010; Gobat et al. 2011, 2013; Santos et al. 2011; Hayashi et al. 2012, 2016; Muzzin et al. 2013; Galametz et al. 2012; Stanford et al. 2012; Zeinman et al. 2012; Ashby et al. 2013;  Wylezalek et al. 2013, 2014; Mei et al. 2015), driven by the advent of deep, large surveys in the infrared and mid-infrared. Hundreds more candidates have been detected at present \citep{14rettura, 16baro,18green, 18marti}. 

However, while cluster candidate detection has been very successful, only tens of the confirmed clusters have been spectroscopically confirmed and observed with high resolution space imaging with the {\it Hubble Space Telescope} ({\it HST}) to study in detail their galaxy morphology and structural properties \citep{12stan, 12zei,13newman,14dela,15mei,16stra}. A unique sample of this kind is the {\it HST} follow-up of the Clusters Around Radio-Loud AGN (CARLA; \citealp{13wyle,14wyle}).
The CARLA survey includes 420 fields observed with the  {\it Spitzer Space Telescope} (hereafter {\it Spitzer}) IRAC 3.6$\mu m$ (hereafter IRAC1) and 4.5$\mu m$ (hereafter IRAC2) channels around Radio-Loud Active Galactive Nuclei (RLAGN). The RLAGN  include 211 RLQs (Radio Loud Quasars) and 209 HzRGs (High Redshift Radio Sources, and were selected from the Miley \& De Breuck (2008) compendium, built by using both flux-limited radio surveys (e.g., MRC, 3C, 6C, 7C) and ultra-steep spectrum surveys (e.g., Rottgering et al. 1997, De Breuck et al. 2001). Since high-redshift clusters and proto-clusters are often found around radio sources (e.g. Castignani et al. 2014, Hatch et al. 2014, Daddi et al. 2017, Paterno-Mahler et al. 2017), the main aim of CARLA was the discovery of galaxy clusters and proto-clusters at $z>1.3$ by selecting galaxy overdensities with  {\it Spitzer}/IRAC color $(IRAC1-IRAC2)>-0.1$ (Wylezalek et al. 2013, 2014).  \citet{13wyle} found that 46\% and 11\% of the CARLA fields are overdense at a 2$\sigma$ and a 5$\sigma$ level, respectively, with respect to the field surface density of sources in the {\it {\it Spitzer}} UKIDSS Ultra Deep Survey (SpUDS; \citealp{13gal}).  

Twenty of the CARLA fields from  \citet{14wyle} with the highest {\it Spitzer} density were followed up with a 40-orbit {\it HST} WFC3 {\it G141} spectroscopy and F140W (hereafter $H_{140}$) imaging. These observations led to the spectroscopic confirmation of sixteen CARLA overdensities at redshift $1.34 \le z \le 2.8$, which are classified as {\it probable} or {\it highly probable} clusters based on their spectroscopic member overdensities \citep{18noirot}. Hereafter, we will call all these structures {\it CARLA confirmed clusters} with the caveat that they are classified as {\it probable} or {\it highly probable} clusters, and have not yet been confirmed by the presence of hot gas in their potential well. \citet{15cooke} have followed up twenty-three of the densest CARLA overdensities at $1.3 \le z \le 3.2$ (overdense at a 4-8$\sigma$ level) with the {\it William Herschel Telescope} ({\it WHT}) auxiliary-port camera  ({\it ACAM}) in La Palma, and fourteen with  the {\it Gemini Multi-Object Spectrograph South instrument} ({\it GMOS-S}; Hook et al. 2004) on {\it Gemini-South} in Chile. They have studied their galaxy stellar population formation histories using a statistical background subtraction, and found that galaxies in their CARLA sample have an approximately constant red observed-frame ({\it i}-band  - IRAC1) color across the examined redshift range.  This indicates that star formation should have been fast in their sample to produce these red colors. \citet{16noirot} have shown that the stellar populations of two of the CARLA confirmed clusters at $z\sim2$ are very different, with one being an already evolved cluster with a clearly defined red sequence and the other being dominated by a star-forming galaxy population. 

In this paper, we study the stellar populations and morphology of the entire sample of the CARLA confirmed clusters at $1.34<z<2.8$ from \citet{18noirot}. Our main result is that the morphology-density and passive density relations are already in place at $z\sim3 \ {\rm and } \ 2$, respectively. The cluster ETG and passive fractions depend on local environment and mildly on galaxy mass. They do not depend on global environment quantified as the cluster core total stellar mass, or density contrast. Our sample is described in Sec.~\ref{sec:obs}, the sample photometry in  Sec.~\ref{sec:phot}, the galaxy sample selection in  Sec.~\ref{sec:sample},  the mass estimation in  Sec.~\ref{sec:mass}, the local and global environment measurements in Sec.~\ref{sec:over}, the morphological classification in Sec.~\ref{sec:morpho} and the passive galaxy selection in Sec.~\ref{sec:color}. Our results are presented in Sec.~\ref{sec:results} and  discussed in Sec.~\ref{sec:discussion}. The summary is in Sec.~\ref{sec:summary}.

We adopt a $\Lambda CDM$ cosmology, with $\Omega_M  =0.3$,
 $\Omega_{\Lambda} =0.7$, $h=0.72$. All magnitudes are given in the AB system (Oke \& Gunn
1983; Sirianni et al. 2005). The photometry and structural parameters in this paper were measured adopting the empirical PSF model released for $H_{140}$\footnote{\url{https://www.stsci.edu/hst/instrumentation/wfc3/data-analysis/psf}}. 
Stellar masses are estimated with a Chabrier initial mass function (IMF; Chabrier 2003). Logarithms are with base ten.
The uncertainty on the fractions in this paper are calculated following Gehrels (1986; see Section 3 for binomial statistics). These approximations apply even when the ratios of different events are calculated from small numbers,  and yield the lower and upper limits of a binomial distribution  within the 84\% confidence limit, corresponding to 1$\sigma$. Using the Gehrels's conservative approach our uncertainties are slightly overestimated (Cameron 2011).

\begin{table*}
\caption{CARLA confirmed cluster sample. From \citet{18noirot} (Table~2 and 4) and \cite{14wyle}:  the CARLA name of the cluster, the AGN redshift $z_{AGN}$, the average cluster spectroscopic redshift $z_{clus}$, other names for the cluster (when this applies), the number of spectroscopically confirmed members $N_{spec}$, the significance of the detection from the Spitzer color selection  $\sigma_{det}$ from \cite{14wyle}, the significance of the spectroscopic selection  {\it HST} (1+$\delta_{spec}$), the AGN type and the cluster classification. HzRG labels high redshift radio galaxies, and RLQ labels radio loud quasars.  HPC and PC indicate {\it highly probable confirmed cluster} and {\it probable confirmed cluster}, respectively, as defined in \cite{18noirot}.  \label{tab:sample}}
\center
\resizebox{!}{4cm}{
\begin{tabular}{lcclcccccccccccccc}
\hline \hline\\
Name & $z_{AGN}$ &$z_{clus}$& Other Name& $N_{spec}$ & $\sigma_{det}$ & 1+$\delta_{spec}$&AGN Type&Classification\\
 \hline \hline\\
 CARLAJ1358+5752 & 1.370 & 1.368&J1358+5752 & 14 &19.4 & 6.8   &   RLQ &PC\\
 CARLAJ0958-2904 & 1.411 & 1.392&MRC0955-288 &8& 19  &3.5  &  HzRG&PC\\
CARLAJ0116-2052 &1.417 &1.425&MRC0114-211 &12&18  &6.3   & HzRG&PC\\
CARLAJ1103+3449 & 1.444 & 1.442&6CE1100+3505 &8 & 21.0 & 3.9   & HzRG &PC\\
CARLAJ1131-2705 & 1.444 & 1.446&MRC1128-268 & 9 &17.8  &5.1   & HzRG&PC\\
CARLAJ2355-0002 & 1.487 & 1.490&TXS2353-003 & 12  &19.4 & 6.3   & HzRG&PC\\
CARLAJ1129+0951 & 1.520 &1.528& J1129+0951 & 12 &21.0 & 6.3  &  RLQ &PC\\
CARLAJ1753+6310 & 1.576 &1.582& 7C1753+6311 & 5  &17.5 & 2.8 &  HzRG &HPC\\
CARLAJ1052+0806 & 1.641 &1.646& J1052+0806 &6&19& 6.2 &  RLQ&PC \\
CARLAJ2227-2705 & 1.684 & 1.692&MRC2224-273 & 7 & 18.5  &6.2  &  HzRG&PC\\
CARLAJ1300+4009 & 1.669 &1.675& J1300+4009 & 8 &18.1 & 7.2   & RLQ &PC\\
CARLAJ1510+5958 & 1.719 & 1.725&J1510+5958 & 6 &19.4 & 6.2 &   RLQ &PC\\
CARLAJ1018+0530 & 1.949 &1.952&J1018+0530&8 & 18  &8.6&   RLQ& PC\\
CARLAJ2039-2514 & 1.997 & 1.999&MRC2036-254 &9 & 22.6 & 9.8 &  HzRG &HPC \\
CARLAJ0800+4029 & 2.004 &1.986 &J0800+4029 &10&21 & 11.0  & RLQ& PC\\
CARLAJ1017+6116 & 2.80 & 2.801& J1017+6116 &7&21 & 47.9 &  RLQ& HPC\\
 \hline \hline\\
 \end{tabular}}
\end{table*}

\section{Method}
\subsection{OBSERVATIONS} \label{sec:obs}
In this work we use 16 CARLA clusters from the \citet{18noirot} sample, which we list in  
Table~\ref{tab:sample}. Each cluster has a {\it HST}/WFC3 $H_{140}$ image, {\it G141} spectroscopy, and {\it Spitzer} IRAC1 and IRAC2 images. Nine clusters also have {\it i-} or  {\it z}-band images, which  correspond to the  rest-frame{\it NUV}/{\it U} band. We provide details of these observations below.

\subsubsection{{\it Spitzer and HST Observations}}

{\it Spitzer} IRAC1 and IRAC2 images were obtained over a common $5.2 \times 5.2$~arcmin$^2$ field of view during 
{\it Spitzer} Cycle 7 (P.I. D. Stern). The total exposure times of 800~s/1000~s in IRAC1 and 2000~s/2100~s in IRAC2, for clusters at $z<2$/$z>2$, provide a similar depth in both channels. The 95\% completeness limit is obtained at IRAC1=22.6~mag and IRAC2=22.9~mag. Regions with limited coverage ($< 85\%$) were rejected from our analysis. The IRAC point spread function (PSF) is 1.95~\arcsec \  and 2.02~\arcsec \ in IRAC1 and IRAC2, respectively (as described in the IRAC Instrument Handbook). Full details of the image processing can be found in \citet{13wyle, 14wyle}.

{\it HST} observations of the clusters were obtained between October 2014 and April  2016 (Program ID: 13740; P.I.: D. Stern). These observations consisted of both $H_{140}$ images and {\it G141} grism spectroscopy, both performed with the {\it WFC3} instrument. The total exposure time was 1000~s for the $H_{140}$ images, and 4000~s for the {\it G141} grism spectroscopy.

The details of the {\it HST} data reduction are provided in \cite{16noirot}, whilst  \citet{18noirot} describes how the cluster candidates were confirmed using emission line cluster members detected with the grism spectroscopy. Our morphological measurements of the  cluster galaxies were obtained from the {\it WFC3} $H_{140}$ images, so we repeat important details of these data here. Each {\it WFC3} image has a field of view of 2 x 2.3 \arcmin$^2$ with a resolution of 0.13 \arcsec pix$^{-1}$. The images were dithered and processed using the software aXe (v2.2.4; K${\rm \ddot{u}}$mmel et al. 2009, 2011) to obtained images with a final resolution of 0.06 \arcsec pix$^{-1}$. Our HST image  5$\sigma$ magnitude limit within an aperture of radius  $0.17$\arcsec is $H_{140}=27.1$~mag.

\subsubsection{Optical Ground-Based Observations} \label{gbobs}

Ground-based {\it i}-band images of 8 CARLA clusters were obtained with either {\it ACAM} or {\it GMOS-S} \citep{15cooke} and a {\it z}-band image of 1 cluster was obtained with {\it VLT/ISAAC} (run ID 69.A-0234)  (see Noirot et al. 2016 for details of this image). The entire field of view of the {\it HST} images was covered by these ground based images and the pixel scale ranged from 0.25~\arcsec pixel$^{-1}$ for {\it ACAM} to 0.146~\arcsec pixel$^{-1}$ for {\it GMOS-S}. The exposure times were adapted to obtain a consistent depth across all fields to deal with variable seeing and sky conditions. Photometric calibration was performed either using available Sloan Digital Sky Survey photometry or standard stars observed before and after the cluster observations. Full details on these observations and the image processing can be found in \citet{15cooke}, and we provide a summary of the optical imaging used in this work in Table~\ref{tab:ground}.

\begin{table*}
\begin{center}
\caption{CARLA confirmed cluster sample ground-based observations from \citet{15cooke}. For more details on these observations, please refer to \citet{15cooke}. Most of the imaging was taken in the {\it i}-band. One cluster was observed in the {\it z}-band. The 5~$\sigma$ magnitude limit is measured in circular regions with radius equal to 2.5 the full width half maximum of the composed images. \label{tab:ground}}
\vspace{0.3cm}
\resizebox{!}{3cm}{
\begin{tabular}{lcccccccc}
\hline \hline\\
Name & $z_{clus}$ &Other name&Instrument& Bandpass&Exp. Time& Seeing&5$\sigma$ mag. limit\\
&&&&&sec.&\arcsec&mag\\
 \hline \hline\\
CARLAJ1358+5752 & 1.368 & J1358+5752 & {\it ACAM}&{\it i}&8400&0.9&24.95\\
CARLAJ1103+3449 & 1.442 & 6CE1100+3505 &{\it ACAM}&{\it i}&9600&1.1&25.14\\
CARLAJ2355-0002 & 1.490 & TXS2353-003 &{\it ACAM}&{\it i}&6000& 0.8&24.99\\
CARLAJ1129+0951 & 1.528 & J1129+0951 & {\it GMOS-S}&{\it i}&2645&0.4&24.78\\
CARLAJ1753+6310 & 1.582 & 7C1753+6311 & {\it ACAM}&{\it i}&6000&0.7&25.08\\
CARLAJ1052+0806 & 1.646 & J1052+0806 & {\it GMOS-S}&{\it i}&2645&0.6&25.04\\
CARLAJ1018+0530 & 1.952 & J1018+0530&{\it ACAM}&{\it i}&7200&0.8&25.19 \\
CARLAJ2039-2514 & 1.997 & MRC2036-254 & {\it ISAAC} & {\it z}&4800&0.7&23.20&\\
CARLAJ0800+4029 & 2.004 & J0800+4029 &{\it ACAM}&{\it i}&6600&0.9&25.16\\
 \hline \hline\\
 \end{tabular}}
\end{center}
\end{table*}

\subsection{Photometry} \label{sec:phot}

\subsubsection{CARLA Cluster T-PHOT Photometry}
We created PSF-matched multi-wavelength photometric catalogues of the {\it Spitzer}, {\it HST} and ground-based images using the software T-PHOT \citep{15merlin,16merlin_a,16merlin_b}. 
T-PHOT performs source deblending in low resolution images by using the positions and surface brightness profiles of the sources measured on a high resolution image as priors. In our case, the $H_{140}$ image is used as the high resolution image and we derive PSF-matched fluxes in the lower resolution {\it Spitzer} and ground-based images. 

We first derived photometric catalogues and segmentation maps for the $H_{140}$ images using the software SExtractor \citep{96bertin}. We used the same input parameters for SExtractor as used by \citet{13gal,13guo} to obtain the CANDELS photometric catalogues. We use both the cold and the hot mode to detect both faint and bright sources (e.g., \citealp{12barden}).  

We then applied the T-PHOT pipeline using the {\it cell-on-object} fitting method to obtain the multi-wavelength photometric catalog. The statistical uncertainty on the photometry was estimated with both T-PHOT and with Monte-Carlo simulations, and we found the two estimates are consistent. The final photometric errors are the sum in quadrature of the statistical error, shot noise, and the error on the photometric zero point. The SExtractor and TPHOT parameters used in this analysis are shown in the appendix.

The final catalogues for the 9 clusters listed in Table \ref{tab:ground} include 4 bands ($H_{140}$, IRAC1 and IRAC2, and either an {\it i} or {\it z}-band). The catalogues for the remaining clusters include only 3 band photometry ($H_{140}$, IRAC1 and IRAC2). All catalogues are $\sim 95\%$ complete at $H_{140}$=24.5~mag. 

\subsubsection{Control sample photometry from CANDELS} \label{sec:candels}
We wish to compare our cluster findings from the CARLA data to a field galaxy baseline. For this purpose, we use the Cosmic Assembly Near-infrared Deep Extragalactic Legacy Survey (CANDELS; PI: S. Faber, H. Ferguson; Koekemoer et al. 2011; Grogin et al. 2011) wide survey, in the GOODS-S region, since the depth and resolution of the CARLA $H_{140}$ images  correspond to the depth and resolution of the {\it HST}/F160W (hereafter $H_{160}$) CANDELS images. In fact, our HST image  5$\sigma$ magnitude limit is $H_{140}=27.1$~mag compared to CANDELS magnitude limit of $H_{160}=27.4$~mag (both were calculated within an aperture of radius  $0.17$\arcsec).

To assemble all the relevant information we combined the following public catalogues: photometry and photometric redshifts from \citet{13guo}, stellar masses and other galaxy parameters from \citet{15san}, and galaxy morphologies from \citet{15karta}. We combined the catalogues by matching data by position and we refer to this catalogue as the \lq combined CANDELS catalogue\rq, hereafter. In this catalogue the photometry is presented as  total fluxes estimated by PSF-matching. For low resolution imaging, such as IRAC, the PSF-matching was performed using the software TFIT \citep{07lai}.

The PSF-match photometry from CANDELS was derived using a different software to our CARLA catalogues. To ensure this does not bias our results, we re-extract photometry from a $\approx$47 \arcmin$^2$ subregion of the  GOODS-S CANDELS Wide field using exactly the same T-PHOT method we performed on the CARLA fields. 

We re-extract photometry using the public  ACS/F775W (hereafter $i_{775}$), ACS/F850W  (hereafter $z_{850}$), IRAC1, IRAC2, $H_{140}$ and $H_{160}$ images from the CANDELS and 3D-{\it HST} archives. Since the CANDELS images are deeper than the CARLA images in some bands, we degrade the CANDELS photometric uncertainty to the CARLA photometric uncertainty as a function of galaxy magnitude and in steps of 0.25 mag. Furthermore, we applied corrections to the $i_{775}$ and $z_{850}$ photometry to account for the different bandpass responses compared to our CARLA ground-based images.

We combine the new T-PHOT photometry with stellar masses and other galaxy parameters from \citet{15san}, and galaxy morphologies from \citet{15karta}. We will refer to this catalogue as the {\it TPHOT-CANDELS catalogue}, photometry in this catalogue will be referred to as CANDELS TPHOT photometry and labeled with subscript {\it TP}: $i^{TP}_{775}$, $z_{850}^{TP}$,  $H^{TP}_{140}$, $H^{TP}_{160}$, IRAC1$^{TP}$, IRAC2$^{TP}$.

\subsection{Cluster member selection}\label{sec:sample}

\begin{figure}
\center
\includegraphics[width=1\columnwidth]{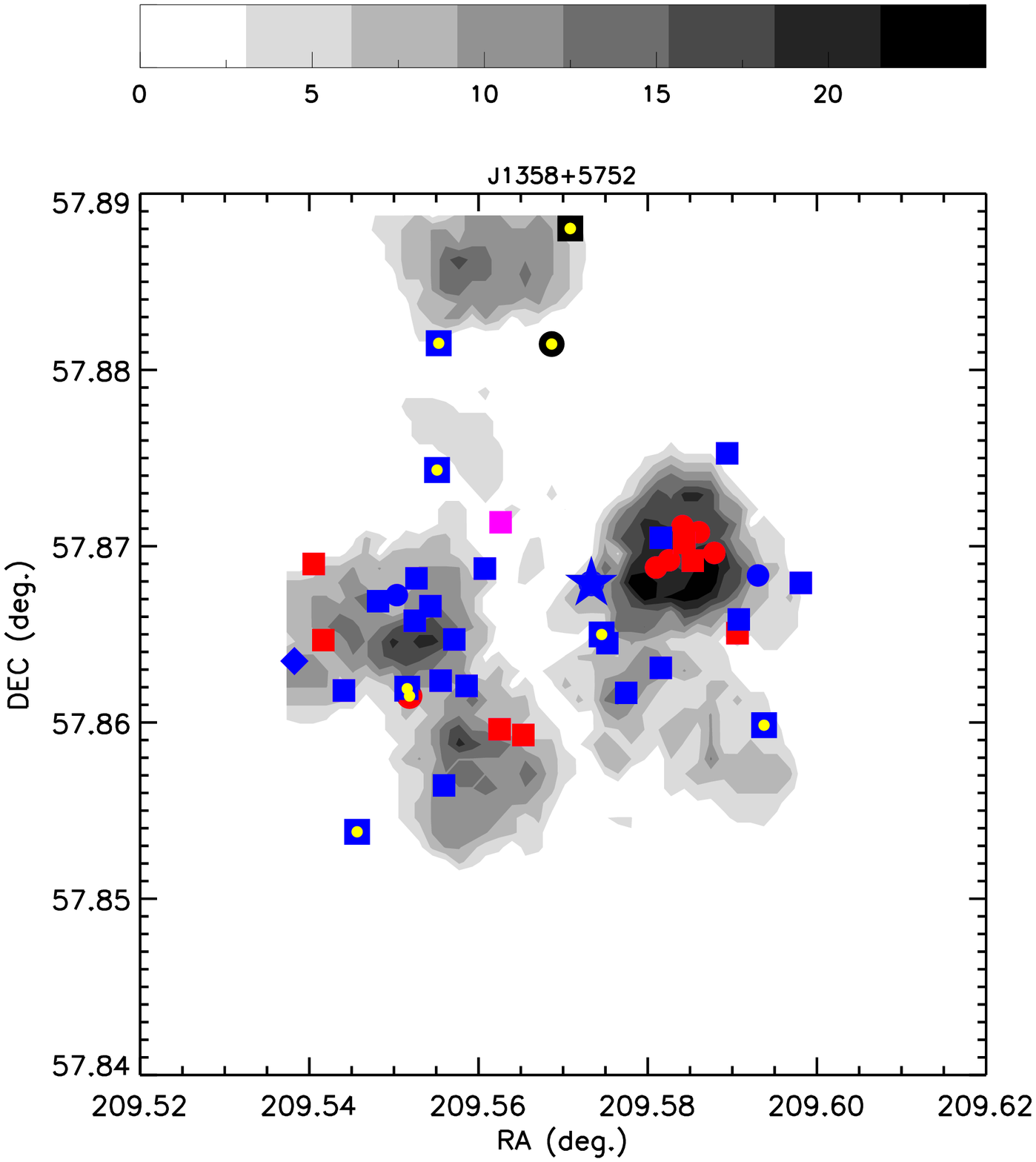}

\caption{ Selected high density regions for CARLAJ1358+5752 at z=1.368. Passive and active galaxies are shown in red and blue, respectively. ETG are shown as circles. LTG are shown as squares (disks) and diamonds (irregulars). The cluster spectroscopic members are marked by a small yellow circle, when they are black they are not included in our magnitude, colour and spatial selection but are shown to better explain our overdensity selection.  The RLQ is shown as stars and triangles, respectively. The bar shows SNR levels. }\label{fig:map}
\end{figure}

 We have between three and five bandpasses for each of the CARLA clusters and proto-clusters, and consequently we cannot perform a precise photometric redshift analysis nor derive galaxy properties from their spectral energy distribution.  We therefore follow Wylezalek et al. (2013) in performing a simple colour-cut of (IRAC1- IRAC2)$>-0.1$ to remove galaxies at $z<1.3$. We furthermore limit our cluster sample to galaxies with $IRAC1 <22.6$~mag, where our sample is $\sim$95\% complete \citep{13wyle,14wyle}. When we perform this magnitude and colour cut with the {\it TPHOT-CANDELS catalogue}, we find that we have selected $\sim~90\%$ of $z>1.3$ galaxies and include only $\sim~10\%$ of contaminants. 
 
\begin{table*}[h]
\center
\caption{CARLA selected overdensities. We show the CARLA name of the cluster, the average cluster spectroscopic redshift $<z_{clus}>$ from \cite{18noirot}, overdensity label (this is 0 is we did not select overdensities, 1 if we select only one overdensity, and 1 or 2 if we select two overdensities), the number of selected galaxies, the SNR in the selected circular region, the SNR scaled to one square arcminute, the cluster total stellar mass, the approximate halo mass, the expected approximate percentage contamination of field galaxies in the selected circular region ($\frac{N_{bkg}}{N_{gal}}$), the position of the AGN: C indicates that the AGN is within $\sim 0.25'$ from the overdensity peak; if it is not, we give the distance from the peak.   \label{tab:densities}}
\resizebox{!}{4.2cm}{
\begin{tabular}{lccccccccccc}
\hline \hline\\
Selected Cluster &$z_{clus}$&Overdensity& $N_{gal}$ &$SNR_{c}$ &log$_{10} (\frac{M^c_*}{M_{\odot}}$) &log$_{10} (\frac{M^c_h}{M_{\odot}}$) &Est. Field Cont.&AGN \\
 \hline \hline\\
CARLAJ1358+5752 & 1.368&1&19&17&12.6&14.3&10\%&C\\
CARLAJ1358+5752 & 1.368&2&18&15&11.2&13.8&10\%&-\\
CARLAJ0958-2904 & 1.392&1&8&6&11.4&13.7&25\%&C\\
CARLAJ0116-2052 &1.425 &1&18&16&11.7&14.0&10\%&$\sim 0.5'$\\
CARLAJ1103+3449  & 1.442&1&12&10&11.5&13.6&15\%&$C$\\
CARLAJ1131-2705 & 1.446&1&9&7&11.3&13.6& 20\%&C\\
CARLAJ2355-0002 & 1.490&1&17&15&11.7&14.0& 10\%&C\\
CARLAJ1129+0951  &1.528&1&16&14&12.5&14.3& 10\%&$\sim 0.5'$\\
CARLAJ1753+6310  &1.582&1&25&23&11.8&13.9& 10\%&C\\
CARLAJ1052+0806  &1.646&1&17&15&12.0&14.1&10\%&C\\
CARLAJ1300+4009  &1.675&1&17&15&12.1&14.1& 10\%&C\\
CARLAJ2227-2705 & 1.692&0&-&-&-&-&-&-\\
CARLAJ1510+5958  & 1.725&1&13&11&12&14.2& 15\%&C\\
CARLAJ1018+0530  &1.952&1&13&11&11.9&14.0&15\%&-\\
CARLAJ1018+0530  &1.952&2&12&10&12.1&14.1& 15\%&C\\
CARLAJ0800+4029  &1.986 &1&19&17&12.3&14.3& 10\%&C\\
CARLAJ2039-2514  & 1.999&1&12&10&11.8&14.0&15\% &C\\
CARLAJ2039-2514  & 1.999&2&10&8&11.6&13.6& 20\%&$\sim 0.7'$\\
CARLAJ1017+6116 & 2.801&1&16&14&12.6&14.6& 10\%&C\\
 \hline \hline\\
 \end{tabular}}
\end{table*}

This sample is still contaminated by galaxies that have redshift $z>1.3$ but are not cluster members. We therefore  restrict our analysis to the densest cluster regions in which the contamination by interlopers is low. To locate these dense regions of the clusters we create galaxy projected overdensity maps across the entire field of view of the {\it $H_{140}$} images.  We count all galaxies, $N_{gal}$, with $IRAC1 <22.6$~mag, and ($IRAC1-IRAC2)>-0.1$~mag within a radius of 0.5~arcmin ($\sim 0.25~{\rm Mpc}$ at our redshifts), which corresponds to the scale of the dense cluster cores at $z>1$ \citeg{05postman}. We obtain an average background of N$_{bkg}=3.0\pm0.6$ galaxies in circles of radius 0.5 square arcminute. 

We define galaxy projected overdensity as $\Sigma$~=~$\frac{N_{gal}-N_{bkg}}{N_{bkg}}$, and  calculate the overdensity signal-to-noise ratio ${\rm SNR}$~$\equiv \frac{N_{gal}-N_{bkg}}{\sigma_{N_{bkg}}}$, which quantifies the density contrast. 
As an example, we display the SNR map of the cluster CARLAJ1358+5752 at z=1.368 in Fig.~\ref{fig:map}. The AGN is close to the strongest density peak, but three other peaks are visible. One of these is only partially covered by the HST image and  another visually seems to be an extension of the second peak. We therefore limit our analysis to the two strongest peaks in this cluster. We provide the SNR maps and details about our overdensity spatial selection for each cluster in Appendix~\ref{overden} and Fig.~\ref{fig:c1}~-~\ref{fig:c3}. Table~\ref{tab:densities} shows the number of overdensities selected within each cluster. 

We now apply a further magnitude limit of $H_{140}=24.5$~mag, which is the depth to which we can reliably classify galaxy morphology (see Sec~\ref{morpho}). We select cluster members as those galaxies with $H_{140}<24.5$, $IRAC1 <22.6$~mag, (IRAC1- IRAC2)$>-0.1$, and lie within 0.5 arcmin of each overdensity peak in the cluster.

For each overdensity we calculated SNR$_c$ as the SNR in circles of 0.5\,arcmin radius when applying our complete cluster sample criterion (i.e., $H_{140}<24.5$, $IRAC1 <22.6$~mag, $IRAC1 - IRAC2 >-0.1$). With this more stringent criterion, we find N$_{bkg}=2\pm1$. SNR$_c$ corresponds to our effective signal-to-noise in each selected overdensity. Table~\ref{tab:densities} shows the number of cluster galaxies selected in each overdensity, and the SNR$_c$ of the overdensity. 
We expect that overdensities with SNR$_c \sim10$, SNR$_c \sim15$, and SNR$_c \sim20$ are contaminated with $\sim 20\%$, $\sim 15\%$, and $\sim 10\%$ of contaminant galaxies, respectively. 

We finally eliminate spectroscopically confirmed outliers from \citet{18noirot}, the photo-spectral analysis of CARLAJ1018+0530 by Werner et al. (in preparation), and of CARLA J1753+6310 by Rettura et al. (in preparation). Our final sample includes a total of 271 galaxies in 19 overdensities in the 16 CARLA confirmed clusters.  In fact, three of our clusters are double structures (see Appendix~B for more details).

\begin{table*}
\center
\caption{CANDELS Mass Calibration. We show the linear coefficients found for the relations : Log$_{10} (M^*)$ = a + b \ $\times (IRAC1$) for the  {\it CANDELS combined} catalog (CC) and the {\it TPHOT-CANDELS catalog} (TCC), and the scatter around these relations $\sigma_{MC}$.\label{tab:masscalib} IRAC1 range is the IRAC1 magnitude range for the fits. For both catalogs the mass variable is the Santini et al. median mass, for CC the IRAC1 is the Guo et al. IRAC1 and for TCC is the TPHOT {\it IRAC1$^{TP}$}.}
\begin{tabular}{c c c c c c c c}
\hline \hline \\
Redshift bin &Catalog &a &b &$\sigma_{MC}$& IRAC1 range \\ 
\hline \hline \\
1.35-1.45 & CC&20.22$\pm$0.08 & -0.467 $\pm$0.003 & 0.2& 20-25\\
1.35-1.45 & TCC & 20.10$\pm$0.02 & -0.467 $\pm$0.003 &  0.4&20-22.6\\
1.45-1.55& CC & 19.97$\pm$0.09& -0.455$\pm$0.004 & 0.2&20-25\\
1.45-1.55 & TCC & 19.87$\pm$0.02 & -0.455$\pm$0.004 &  0.4&20-22.6\\
1.55-1.65& CC & 20.20$\pm$0.07 & -0.462$\pm$0.003 & 0.2&20-25 \\
1.55-1.65 & TCC & 20.03$\pm$0.01 & -0.462$\pm$0.003&  0.5&20-22.6 \\
 1.65-1.75& CC & 20.35$\pm$0.08 & -0.467$\pm$0.003 & 0.2&20-25 \\
1.65-1.75 & TCC & 20.17$\pm$0.01& -0.467$\pm$0.003 &  0.4&20-22.6 \\
 1.75-1.85& CC & 20.72$\pm$0.10 & -0.480$\pm$0.004& 0.2&20-25 \\
 1.75-1.85 & TCC & 20.56$\pm$0.02 & -0.480$\pm$0.004 &  0.4&20-22.6 \\
1.8-2.2& CC & 21.37$\pm$0.06 & -0.496$\pm$0.003 & 0.2&20-25 \\
1.8-2.2 & TCC & 21.01$\pm$0.01& -0.496$\pm$0.003 &  0.4&20-22.6 \\
2.6-3& CC & 21.63$\pm$0.08 & -0.507$\pm$0.003 & 0.2&20-25 \\
2.6-3 & TCC & 21.48$\pm$0.01 & -0.507$\pm$0.003 &  0.5&20-22.6 \\
\hline \hline \\
\end{tabular}
\end{table*}

\subsection{Galaxy stellar mass} \label{sec:mass}

 Different stellar mass estimation methods and the use of different stellar population templates, priors or rest-frame magnitudes, can result in mass differences of a factor of $\sim 1.5-6$ (in the range $\sim 0.1-0.8$~dex) \citep{06vanw,09lee,10mar,11raic,12pforr,18sor,18noirot}, resulting in both underestimating or overestimating masses at $z>1.4$, depending on which methods, templates, priors and rest-frame magnitudes are used.  In our previous analysis of the CARLA sample \citep{16noirot,18noirot}, galaxy stellar masses were derived from {\it Spitzer}/IRAC photometry. 
 
 In this work, we estimate galaxy stellar masses using an empirical method that links our mass estimates to those derived by the CANDELS collaboration \citep{15san} to be able to directly compare our results to theirs. In fact, we cannot derive masses by fitting the galaxy spectral energy distribution.

We estimate the masses of the cluster galaxies by calibrating our CARLA photometry against that of the CANDELS \citet{15san} median galaxy stellar mass.
This catalog has the advantage of having been built using results from ten different teams. This enabled the authors to estimate how different parameter choices of the different teams affected their final results. Their analysis showed that the largest scatter around the median mass was produced by different choices of stellar population synthesis templates (resulting in scatter of $\sim 0.1-0.2$~dex), and the inclusion of nebular emission was crucial for young galaxies (age $<100$~Myr) at $z>2.2$  \citep{15san}. 
The errors on the median stellar masses include the uncertainty due to the different method assumptions and degeneracies, combined with errors on photometric and spectroscopic redshifts, which are the largest contributions to the total uncertainties. When compared to the 3D-{\it HST} published masses \citep{14ske} for the same galaxies in the GOODS-S field, they agree within $\sim 0.1$~dex with a negative offset of $\sim -0.1$~dex for the CANDELS masses.

We use IRAC1 photometry to perform our mass calibration, but we have verified that our results do not change when using IRAC2. IRAC1 corresponds to the rest-frame near-infrared in the redshift range of the CARLA clusters and it is therefore less biased by extinction than  $H_{140}$, which corresponds to the rest-frame $U$ or $V$-band for the CARLA clusters. 

We first use the {\it CANDELS combined catalogue} to obtain a linear relation between the Guo et al. CANDELS IRAC1 photometry and the Santini et al. median galaxy mass  in seven bins chosen in the redshift range covered by our CARLA sample, i.e., $1.35<z<2.8$, and in the magnitude range $20<IRAC1<25$~mag. We account for errors in both parameters in our fit using the code {\it mpfityx} from \citet{09mpfit}. This magnitude range is well within the CANDELS IRAC1 photometric depth. The redshift bin width was chosen to have enough galaxies for stable linear fits.

We then fix the slopes of the relations and re-fit the IRAC1-mass relation using the {\it TPHOT-CANDELS catalog}, constraining the magnitudes to the range corresponding to the depth of the CARLA images: IRAC1=20-22.6~mag. The difference in the offset between the relationships we measure is typically $\sim 0.1-0.3$~mag.
These relations are shown in Table~\ref{tab:masscalib}. 

The average difference between Santini et al.'s median masses and the masses derived from this calibration is $-0.03 \pm0.05$~dex and $0.3 \pm 0.1$~dex, in the redshift range $z=1.35-2$ and $z\sim3$, respectively. 
The scatter increases with magnitude and redshift, from $\sim 0.12$ at redshift $z=1.35-1.45$ to $\sim 0.2$ at redshift $z\sim3$. 

With these new mass estimates, the mass range for the CARLA  spectroscopically confirmed members is $ 2 \times 10^{8} - 3  \times 10^{12} M_{\odot}$, including the AGN, for which the estimated mass is an upper limit. 
These mass estimates are approximately four times smaller ($\approx 0.5$~dex) than those derived in from stellar populations models by Noirot et al. (2018). 
This is consistent with the mass uncertainty estimated in Noirot et al. (2018), and it is well documented that different stellar mass estimation methods, including the use of different stellar population templates, priors or rest-frame magnitudes, can result in mass differences of a factor of $\sim 1.5-6$ (in the range $\sim 0.1-0.8$~dex) \citep{06vanw,09lee,10mar,11raic,12pforr,18sor,18noirot}. 
This difference in mass estimates does not change results from \citet{18noirot}. The conclusions from the SFR vs stellar mass analysis (Fig. 7 in \citealp{18noirot}), for example were also confirmed by \citet{20markov} for one of the clusters, and we verified that they hold with our new mass estimates.

Since our calibration is based on the CANDELS stellar masses, our mass estimates are affected by the same systematic uncertainties as the CANDELS \citet{15san} mass estimates. We estimated our total mass uncertanties by adding in quadrature the scatter around the linear relation fit $\sigma_{MC}=0.4-0.5$~dex for log$_{10} (\frac{M}{M_{\odot}}) < 10.5$, and  $\approx 0.2$~dex for log$_{10} (\frac{M}{M_{\odot}}) > 10.5$, and the average \citet{15san} systematic mass uncertainties, which are in the range $\sim 0.1-0.2$~dex. To obtain the total uncertainty on our mass measurements, we add in quadrature $\sigma_{MC}$ and the average mass uncertainty from \citet{15san} as a function of the redshift and the IRAC1 magnitude, and derive mass uncertainties in the range $\sim 0.4-0.5$~dex  for log$_{10} (\frac{M}{M_{\odot}}) < 10.5$, and  $\approx 0.2-0.3$~dex for log$_{10} (\frac{M}{M_{\odot}}) > 10.5$. 
We consider that differences due to age, dust content and metallicity are taken into account by these mass uncertainties. 

 In particular, most of \citet{15san}'s mass estimate methods assume a Calzetti dust exinction law \citep{00cal}. Given that quiescent ETG at redshifts similar to ours have shown  at least two orders of magnitude more dust at fixed stellar mass than local ETG \citep{18gobat} and that the dust attenuation increases with mass in star-forming galaxies \citeg{15pan}, this prescription introduce uncertainties. More sophisticated radiative transfer models predict a shallower dust attenuation curve than the Calzetti law for larger attenuation optical depths, and this impacts dust attenuation estimations in the rest-frame near-infrared at the redshifts that we are probing \citeg{00charlot, 13chevallard, 18buat,19buat, 20trayford}, and, as a consequence, galaxy stellar mass estimation (e.g. Reddy et al. 2015, 2018). In fact, galaxy stellar masses derived assuming the Calzetti law are predicted to be $\sim$30\% ($\sim$0.2 dex) larger than those derived with new models \citeg{15reddy,18reddy}. This bias impacts both CANDELS mass estimates and ours, and it has been included by \cite{15san} in their estimation of systematic uncertainties. 

\begin{figure}
\center
\includegraphics[width=1\hsize]{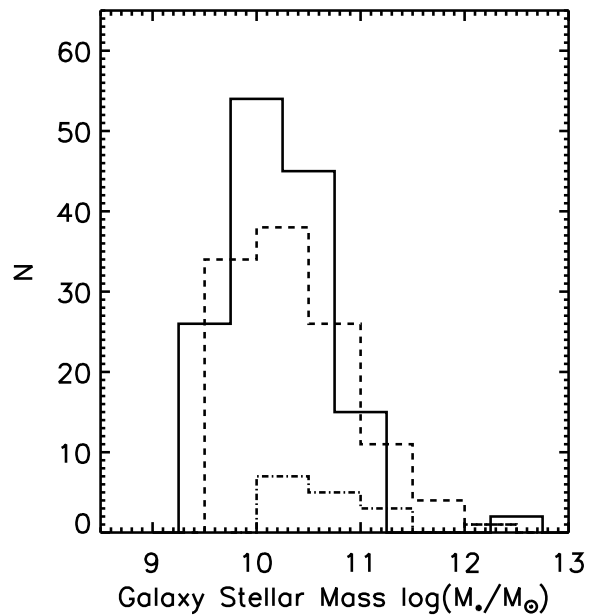}
\caption{The galaxy stellar mass distribution within the cluster overdensities in redshift range $1.3<z<1.6$ (continuous line), $1.6<z<2.1$ (dashed line), and $z=2.8$ (dot-dashed line). The three distributions are consistent.}\label{fig:galmass}
\end{figure}

Fig.~\ref{fig:galmass} shows that the galaxy stellar mass distribution for cluster overdensities in three redshift bins are consistent. The three galaxy mass distributions are similar up to $z\sim2$ and span the range $9.6 \lesssim$log$_{10} (\frac{M_*}{M_{\odot}}) \lesssim 12.6$. The cluster at $z=2.8$ does not show galaxies with log$_{10} (\frac{M_*}{M_{\odot}}) \lesssim 10$.

\subsection{Definition of local and global environment} \label{sec:over}

\subsubsection{Local environment}
We quantify the local environment of the cluster galaxies by deriving galaxy projected surface densities. We select all galaxies in the whole {\it HST} field with $IRAC1 <22.6$~mag and (IRAC1- IRAC2)$>-0.1$. We then apply the method of Nth-nearest neighbor distances: $\Sigma_N=\frac{N}{\pi D^2_N}$. $N$ is the number of galaxy neighbors, $D_N$ is defined as the distance in Mpc to the Nth nearest neighbor. Our results are stable in the range $N=5-10$, and hereafter we use N=7 to be consistent with previous galaxy projected surface density estimates using {\it HST} cluster observations \citeg{05postman}.

\subsubsection{Global environment: Cluster Total Stellar Mass} \label{sec:tot}

\begin{figure}
\center
\includegraphics[width=0.95\hsize]{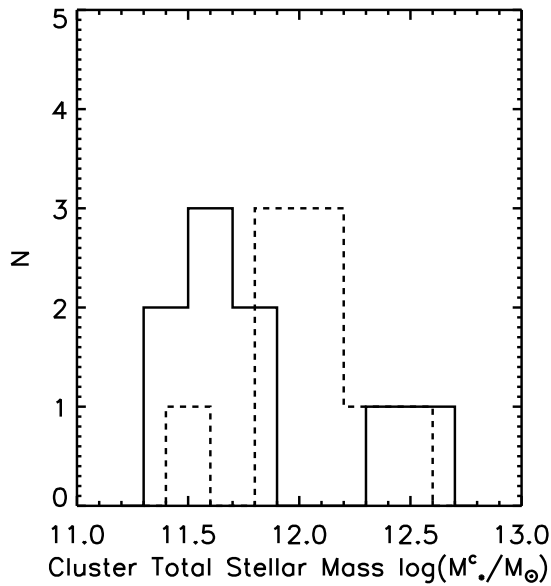}

\caption{The total stellar mass within each overdensity of the CARLA clusters in redshift range $1.3<z<1.6$ (continuous line), and $1.6<z<3$ (dashed line). These stellar masses roughly correspond to a total halo mass of $13.6 \lesssim$log$_{10} (\frac{M^c_h}{M_{\odot}}) \lesssim 14.3$ (see text for details).  }\label{fig:overdenmass}
\end{figure}

We quantify the global environment of each cluster galaxy in terms of the halo mass of the overdensity the galaxy resides in. We first measure the total stellar mass of each overdensity, then use \citet{19beh} stellar to halo mass relations to derive approximate halo masses. 

The total stellar mass within each overdensity was obtained by summing the galaxy stellar masses within 0.5\arcmin\ of each overdensity peak. The total stellar mass of the overdensities are given in Table~\ref{tab:densities} and displayed in
Fig.~\ref{fig:overdenmass}.

We then use the predictions of the average stellar mass over halo mass ratio as a function of redshift from \citet{19beh} (their Fig.~18), assuming that our detections are main progenitors of local clusters,  to derive approximate halo masses. The halo masses of the overdensities are in the range $13.6 \lesssim$log$_{10} (\frac{M^c_h}{M_{\odot}}) \lesssim 14.6$, and are tabulated in Table~\ref{tab:densities}.

\subsection{Morphology} \label{sec:morpho}

\subsubsection{Morphology of galaxies in CARLA images} \label{morpho}

We classify the morphology of all galaxies brighter than $H_{140}=24.5$~mag in the overdensity regions described in Sec.~\ref{sec:over} in each CARLA {\it HST} image. Our classification was performed by a single person based on postage stamps of $128 \times 128$ pixel ($16.6 \times 16.6$ arcsecond). From previous visual classification statistics \citep{05postman,15karta}, a morphological classification is considered stable when $\sim70\%$ of classifiers agree, and we consider this as our uncertainty.
 
In this paper we will describe a galaxy morphologically as an Early Type Galaxy (ETG) or a Late Type Galaxy (LTG). However, we wish our morphological classifications to be directly comparable to the field baseline provided by CANDELS morphologies classified by \citet{15karta}. We therefore have followed the more complex visual morphology classification scheme of  \citet{15karta}, and later convert these classifications into ETG and LTG classes. We classify each galaxy as a:
\begin{itemize}
\item (1) {\it Disk}. These galaxies have a disk even if they don't show clear spiral arms;
\item (2) {\it Spheroid}. There galaxies are resolved spheroids and do not show a disk; 
\item (3) {\it Irregular}. All galaxies that cannot be classified either as Disk or Speroids; 
\item (4) {\it Compact/Unresolved}. These are compact or unresolved galaxies;
\item (5) {\it Unclassifiable}. 
\end{itemize}

The only difference between this scheme and that of \citet{15karta} is that our {\it Disk} class include galaxies classified as either {\it Disk} or {\it Disk Spheroid} by \citet{15karta} (see Sec~\ref{sec:morpho_cc} for more details). 
We furthermore assign a {\it morphology quality flag} to each galaxy classification: 1. Certain; 2. Uncertain. 

In addition to the above morphological classes, we also follow \citet{15karta} in identifying mergers. These are defined as galaxies that show visual tidal features or other structures that may indicate a recent merger has occurred. This class corresponds to galaxies that have Kartaltepe et al. $f_M >2/3$, which are $\sim 1.1 \pm0.1\%$ of the galaxies in our {\it combined CANDELS catalog};

We also add  {\it structure flags} similar to \citet{15karta}:
1. {\it tidal arms}; 2. {\it double nuclei}; 3. {\it asymmetric}; 4. {\it spiral arms/ring}; 5. {\it bar}; 6. {\it point source contamination }; 7. {\it edge-on disk}; 8. {\it face-on disk}; 9. {\it tadpole galaxy}; 10. {\it chain galaxy}; 11. {\it disk- dominated}; 12. {\it bulge-dominated} . Most of these flags do not occur enough in our galaxy magnitude, color and mass range to be statistically significant for this work, except the {\it asymmetric} flag. Therefore, we will not consider them in the statistics that we will present below but these classifications are available from the lead author upon request. 

For this paper, given the small numbers of galaxies in each category for both CANDELS and our sample, we combine {\it Spheroids} and  {\it Compact/Unresolved/Point Sources} in a Early Type Galaxy (ETG) single class, and {\it Disks}, {\it Disk Spheroids}, and {\it Irregulars} in a Late Type Galaxy (LTG) class. The sample used in this paper does not include any {\it Unclassifiable} galaxy.  

\subsubsection{Morphology of galaxies within CANDELS} \label{sec:morpho_cc}

Kartaltepe et al. (2015) visually classified the morphologies of galaxies in the CANDELS GOODS-S field to a depth of $H_{160} = 24.5$~mag. This matches well our magnitude limit of $H_{140}=24.5$~mag in the CARLA images. Both $H_{140}$ and $H_{160}$  are in the optical rest-frame range for galaxies at redshift $1.5<z<2.8$ and, as  \citet{15karta} pointed out, morphology classifications performed in infrared bandpasses at these redshifts do not show significant differences. 

Each CANDELS galaxy was classified by at least three people into the same morphological classes listed above;  {\it Spheroid}, {\it Disk}, {\it Irregular}, {\it Unresolved/Point Source}, {\it Unclassifiable}. They also considered mixed classes, and labelled other features such as face-on disk or tadpole galaxy. Full details of these additional classifications can be found in \citet{15karta}

Kartaltepe et al.'s catalogue also provides the fraction of classifiers that identify each galaxy as belonging to a given class. $f_{sph/disk/irr/tp}$ refers to the fraction of classifiers that identify each galaxy as a spheroid/disk/irregular/tadpole galaxy. $f_{ps}$ refers to the fraction of classifiers that identify the object as a point source, and  $f_{unc}$ refers to the fraction of classifiers that could not classify the object. We use these fractions as proxies of the probability that a galaxy belongs to a given morphological class. 

In the \citet{15karta} classification, each galaxy can be classified into more than one class with high probability. For this reason, we are using the following criteria to place galaxies into each of the five dominant classes:

\begin{itemize}
\item {\it Spheroid}: $f_{sph} > 2/3$ and $f_{disk} < 2/3$, or $0.5 <f_{sph} < 2/3$ and $f_{disk} < 0.5$ and $f_{irr}<2/3$. Spheroids are $\sim 20\%$ of the galaxies in our {\it combined CANDELS catalog}.
\item {\it Disk}: $f_{disk} > 2/3$ and $f_{sph} < 2/3$, or $0.5 <f_{disk} < 2/3$ and $f_{sph} < 0.5$ and $f_{irr}<2/3$. Galaxies with this morphology comprise $\sim 60\%$ of our {\it combined CANDELS catalog}. We add to this class galaxies with $f_{disk} > 2/3$ and $f_{sph} > 2/3$. These disk galaxies have a prominent bulge and comprise $\sim 6\%$ of the galaxies in our {\it combined CANDELS catalog}. Hence, $\sim 66\%$ of the galaxies in our {\it combined CANDELS catalogue} are disks.
\item {\it Irregular}: $f_{irr}>0.5$ and $f_{disk} < 2/3$ and $f_{sph} < 2/3$. Or $f_{tp}>2/3$. Irregulars comprise  $\sim 13\%$ of the galaxies in our {\it combined CANDELS catalog}.

\item {\it Compact/Unresolved/Point Source}:  $f_{ps}>2/3$. These sources comprise $\sim 1\%$ of the galaxies in our {\it combined CANDELS catalog}.

\item {\it Unclassifiable}: $f_{unc}>2/3$, or  galaxies that do not belong to the classes above. This comprises $\sim 1\%$ of the galaxies in our {\it combined CANDELS catalog}.

\end{itemize}

The limit of $2/3\sim 70\%$ corresponds to the typical agreement on galaxy visual morphology from multiple classifiers \citeg{05postman}. The sum of the fractions in each class do not add to 100 because they are rounded numbers. 

\begin{figure*}[ht]
\center
\includegraphics[width=0.329\hsize]{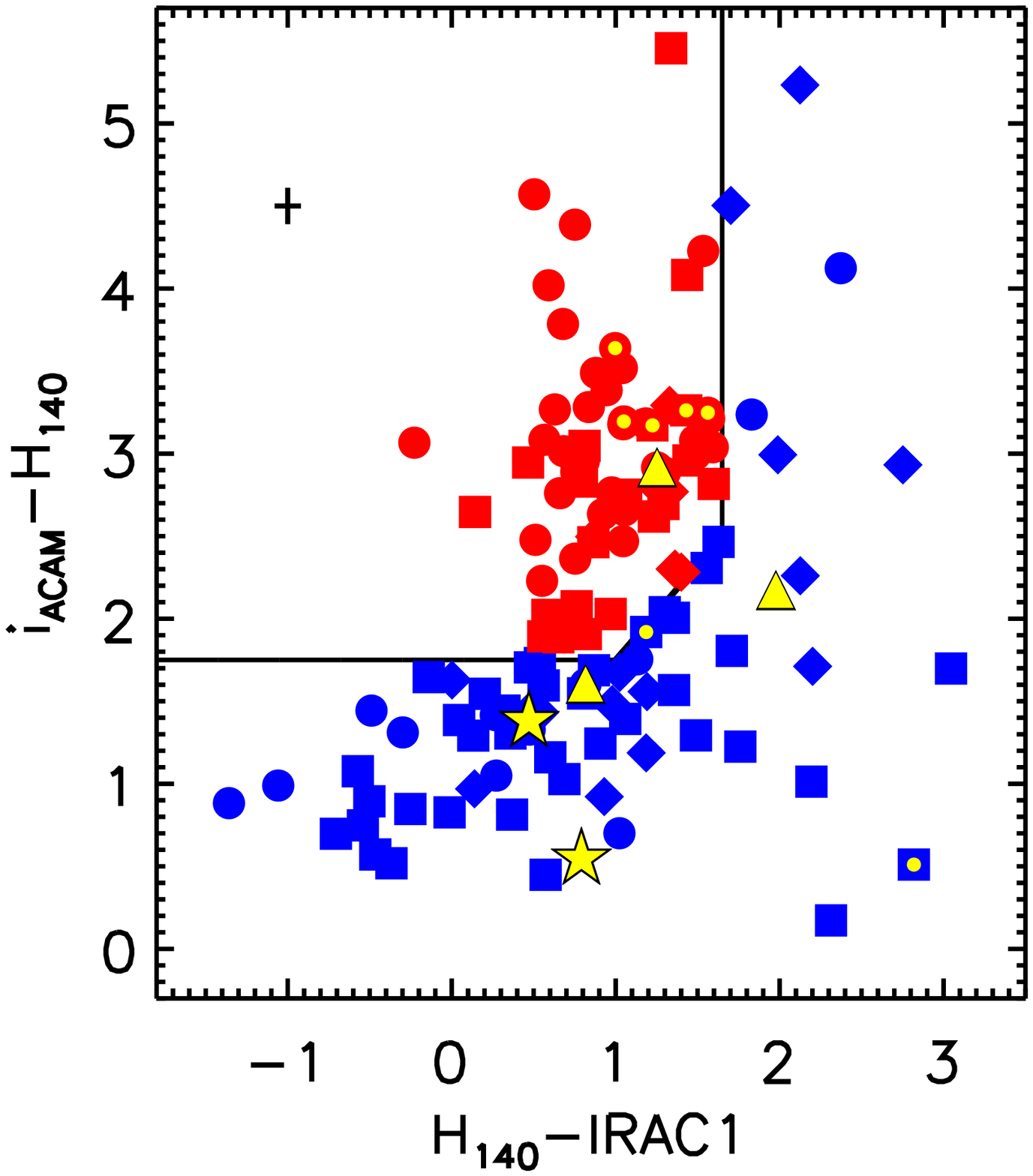}
\includegraphics[width=0.329\hsize]{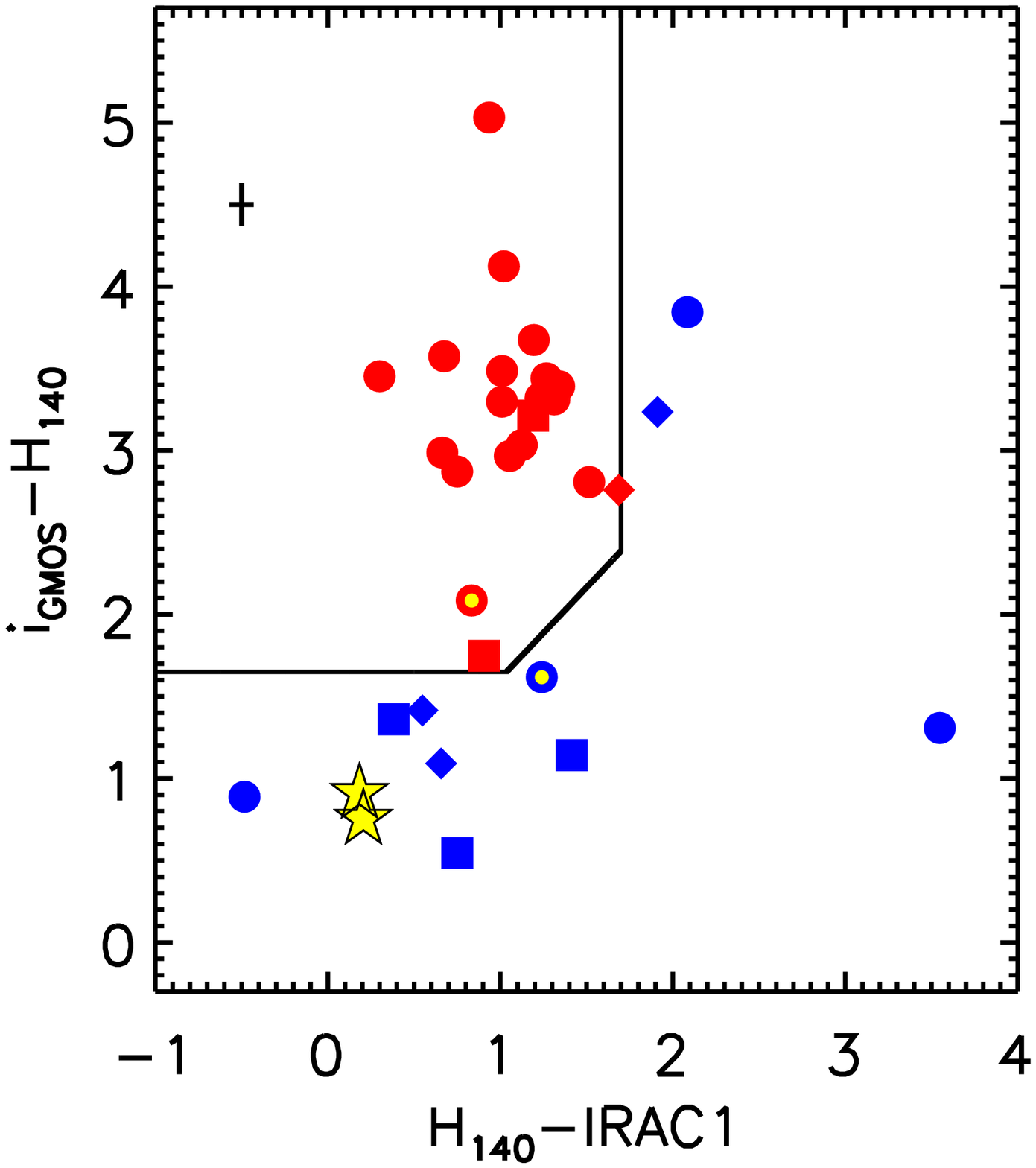}
\includegraphics[width=0.329\hsize]{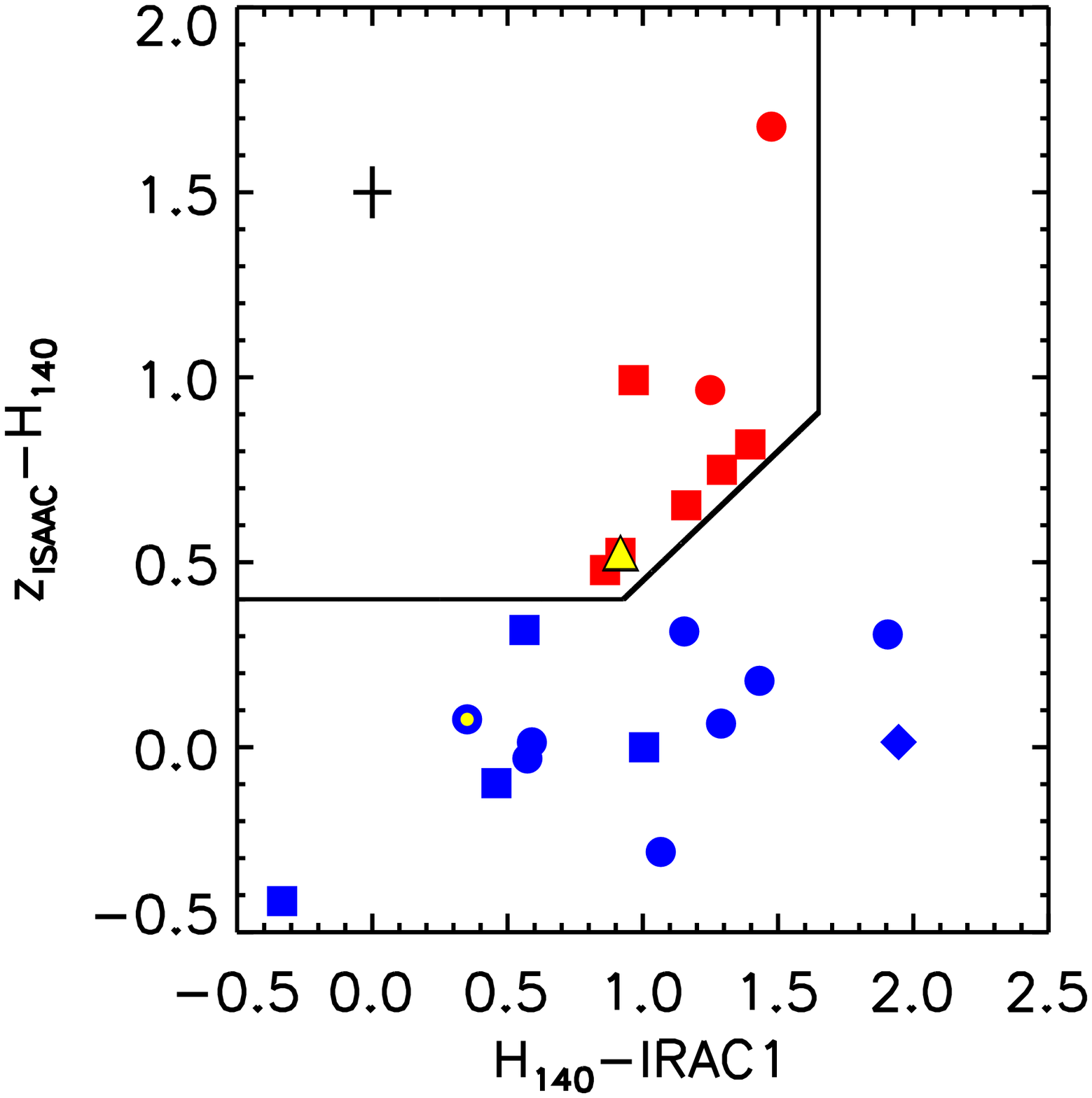}
\caption{Cluster color-color diagrams. Left : all clusters observed with ACAM. Middle: all clusters observed with GMOS. Right: CARLAJ2039-2514, observed with ISAAC. The continuous lines limit the regions that separate passive from active galaxies. Passive and active galaxies are shown in red and blue, respectively.  ETG are shown as circles. Disks and Irregular galaxies are shown as squares and diamonds, respectively. The cluster spectroscopic members show a yellow dot at the center. The RLQ and HzRG are shown as yellow stars and triangles, respectively, and have spectroscopic redshift even if we do not mark them with a dot. The other AGN have colors that are not included in the plots. The crosses on the upper left are the average color errors. \label{colcol-clus}}
\end{figure*}

\begin{table*}[ht]
\center
\caption{CARLA selected spectroscopically confirmed member classification as passive and active galaxies in this paper. We include only the spectroscopically confirmed members included in the galaxy selection used in this paper. In the column SFR flux, we show which line emission flux was used to estimate SFR.}  \label{tab:spectroclass}
\begin{tabular}{c c c c c c c c}
\hline \hline\\
Cluster&ID &tile&z&log(sSFR)&SFR Flux& Classification \\
\hline \hline \\
CARLA J0800+4029  & 436 &t1-1523&1.969$\pm$ 0.007&-9.4$\pm$0.5&OIII&Passive\\
CARLA J1129+0951&    240&t1-1222&1.546 $\pm$ 0.004&-8.7$\pm$0.5& H$_{\alpha}$&Active\\
CARLA J1129+0951&    261&t1-1263&1.537 $\pm$ 0.006&-9.1$\pm$0.5& H$_{\alpha}$&Passive\\
CARLA J1358+5752  &   414&t1-1462&1.377 $\pm$ 0.006&-8.9$\pm$0.5& H$_{\alpha}$&Active\\
CARLA J1358+5752  &   681&t1-1794&1.376 $\pm$ 0.005&-9.4$\pm$0.5& H$_{\alpha}$&Passive\\
CARLA J1358+5752  &   694&t1-1795&1.375 $\pm$ 0.006&-9.3$\pm$0.5& H$_{\alpha}$&Active\\
CARLA J2355-0002    &   762&t1-1374&1.492 $\pm$ 0.005&-9.2$\pm$0.6& H$_{\alpha}$&Passive\\
CARLA J2355-0002    &   334&t1-1405&1.485 $\pm$ 0.007&-9.1$\pm$0.5& H$_{\alpha}$&Passive\\
CARLA J2355-0002    &   337&t1-1410&1.500 $\pm$ 0.006&-9.6$\pm$0.3& H$_{\alpha}$&Passive\\
CARLA J2039-2514   & 174 &t1-1236&2.000$\pm$ 0.007 &-8.4$\pm$0.5&OIII&Active\\
\hline \hline \\
\end{tabular}
\end{table*}

\subsection{Passive and active galaxy selection} \label{sec:color}

Nine of our CARLA clusters have been observed in the {i} or {z}-band from the ground (Table~\ref{tab:ground}).
For these clusters, we use colour-colour diagrams, based on the {\it UVJ} diagrams commonly used in the literature, to separate passive from active galaxies (e.g. Labb\'e et al. 2005; Wuyts et al. 2007; Williams et al. 2009; Whitaker et al. 2011; Fang et al. 2018) in clusters with ground-based observations. Unfortunately, our limited spectral energy coverage, of only 4 bands per cluster, means we are unable to perform spectral energy distribution fitting and derive restframe $UVJ$ colours. Instead we use the colours of the observed bands since, at the redshift range of our clusters, the {\it i}-band and {\it z}-band observations correspond approximately to the rest-frame {\it U}-band, whilst $H_{140}$ and $IRAC1$ correspond to approximately the rest-frame {\it V} and  {\it J}-band, respectively. 
When the U-band rest-frame is not available, the FUV rest-frame can substitute it, and even lead to more precise specific star formation rate (sSFR) measurements \citep{13arn,19leja}. 

We use apparent magnitude colours to define passive regions of $i-H_{140}~IRAC1$ and $z-H_{140}~IRAC1$ colour space as shown in Figure~\ref{colcol-clus}, which shows the 271 galaxies selected for this paper analysis.  We define as quiescent galaxies with  log(sSFR)$<-9.5 \ {\rm [yr^{-1}]}$, which characterizes the quiescent region at our cluster redshifts (e.g., Whitaker et al. 2014; Leja et al. 2019). The method used to obtain these colours is based on a calibration on CANDELS apparent magnitudes, and is described in appendix \ref{app-pa}. 

Among the 271 galaxies selected for the study in this paper, only 10 are spectroscopically confirmed members and have sSFR measurements (Table~\ref{tab:spectroclass}). Of those, 6 ($\sim 20\%$ of the total selected passive population) are classified as passive galaxies and have log(sSFR)$<-9.5  \ {\rm [yr^{-1}]}$ at $\sim 1\sigma$. The other 4 are classified as active, and 3 of them have log(sSFR)$<-9.5  \ {\rm [yr^{-1}]}$ at $\sim 2 \sigma$. This confirms that our selection includes recently quenched star-forming galaxies (see appendix), and that many of star-forming galaxies in the CARLA clusters show SFR that are $\sim 3 \sigma$ below the field main sequence \citep{18noirot}.

All the AGN have ${\rm log(sSFR)}>-9.5 \ {\rm [yr^{-1}]}$ and are found in the active galaxy region, except the CARLAJ1753+6310 AGN that is found in the passive galaxy selection and has a log(sSFR)=$-8.6\pm0.3 \ {\rm [yr^{-1}]}$. The AGN that are not included in the plots have colors that are not in the figure range.

\begin{table*}
\center
\caption {\small{Corrected fractions of passive galaxies, ETG and mergers, and the minimum, maximum and median $\Sigma_N$ and galaxy stellar mass $log_{10}(\frac{M_*}{M_{\odot}})$. The different overdensities are labeled O1 and O2 for overdensity 1 and 2, respectively. The uncertainties are the average of the upper and lower limit uncertainty. } \label{tab:fraction}}
\resizebox{!}{4.3cm}{
\begin{tabular}{lccccccccc}
\hline \hline\\
Cluster/Overdensity &$z_{clus}$&Passive&ETG&Merger &$\Sigma_N$ &$log_{10}(\frac{M_*}{M_{\odot}})$ \\
&&&&&(${\rm gal  \  Mpc^{-2}}$)& \\ 
\hline \hline \\
CARLAJ0116-2052 &1.425 & - &0.59$\pm$0.15&0.41$\pm$0.15&220-750/390& 9.6-11.1/10.4\\
CARLAJ1018+0530/O1&1.952& 0.16$\pm$0.16& 0.22$\pm$0.16&0.33$\pm$0.18&120-280/175&9.9-12.0/10.3\\
CARLAJ1018+0530/O2&1.952& 0.62$\pm$0.20&0.64$\pm$0.19&0.26$\pm$0.18&120-1270/534&9.8-11.3/10.4\\
CARLAJ0800+4029  &1.986 &0.57$\pm$0.15&0.20$\pm$0.12&0.21$\pm$0.13&150-1370/ 660&9.8-12.2/10.3\\
CARLAJ0958-2904 & 1.392& - &0.41$\pm$0.25&0.44$\pm$0.26&130-260/170&9.7-11.2/10.0\\
CARLAJ1017+6116 & 2.801& - &0.39$\pm$0.16&0.47$\pm$0.16&180-620/ 380&10.1-12.5/10.8\\
CARLAJ1052+0806  &1.646&0.97$\pm$0.30 &0.76$\pm$0.14&0.4$\pm$0.3&170-1000/ 340&9.6-11.8/10.2\\
CARLAJ1103+3449  & 1.442&  0.62$\pm$0.20 & 0.74$\pm$0.18&0.06$\pm$0.30&110-210/150&9.7-11.0/10.1\\
CARLAJ1129+0951  &1.528& 0.51$\pm$0.17&0.60$\pm$0.16&0.33$\pm$0.15&130-850/400& 9.6-12.4/10.3\\
CARLAJ1131-2705 & 1.446& - &0.35$\pm$0.23&0.66$\pm$0.23&110-400/ 270&9.8-10.8/10.3\\
CARLAJ1300+4009  &1.675& - &0.50$\pm$0.16&0.04$\pm$0.30&180-560/280&9.6-12.0/10.4\\
CARLAJ1358+5752/O1& 1.368& 0.51$\pm$0.16&0.34$\pm$0.14&0.54$\pm$0.15&180-1500/300&9.6-12.6/10.0\\
 CARLAJ1358+5752/O2& 1.368& 0.29$\pm$0.14 &0.08$\pm$0.10&0.21$\pm$0.13&150-410/250&9.6-10.7/10.0\\
CARLAJ1510+5958  & 1.725& - &0.22$\pm$0.16&0.24$\pm$0.17&140-390/280&9.6-12.0/10.4\\
CARLAJ1753+6310  &1.582& 0.66$\pm$0.12&0.76$\pm$0.11&0.25$\pm$0.11&190-2200/640& 9.6-11.1/10.2\\
CARLAJ2039-2514/O1& 1.999& 0.29$\pm$0.19&0.34$\pm$0.19&0.56$\pm$0.19&157-403/200&9.9-11.1/10.4\\
CARLAJ2039-2514/O2& 1.999& 0.45$\pm$0.22&0.68$\pm$0.21&0.33$\pm$0.21&100-770/300&9.9-11.4/10.2\\
CARLAJ2355-0002 & 1.490& 0.68$\pm$0.15 &0.10$\pm$0.11&0.11$\pm$0.11&150-580/300&9.9-11.1/10.2\\
CARLAJ1017+6116 & 2.801& - &0.39$\pm$0.16&0.47$\pm$0.16&180-620/ 380&10.1-12.5/10.8\\
 \hline \hline\\
 \end{tabular}}
\end{table*}

\section{RESULTS} \label{sec:results}

We calculate galaxy type fractions as the ratio of galaxies in each category divided by the sum of total galaxies. The exception is the ETG fraction which we calculate  with respect to the total of the ETG plus LTG population (excluding the QSO). We use statistical background subtraction to correct for foreground and background galaxies that contaminate our cluster sample.

Following the method of \citet{05postman},we calculate the corrected number of galaxies in our clusters as $N_{corr}=N_{uncorr}+N_{miss}-N_{cont}$. $N_{uncorr}$ is the number of galaxies of each sub-type, without correcting for contamination. $N_{miss}$ are the galaxies that we miss by applying our selection criteria. $N_{cont}$ are the foreground and background galaxies that are not at our cluster redshift and are estimated to contaminate our sample, for each fraction category. 

We estimate $N_{miss}$ and $N_{cont}$ for each galaxy sample using the {\it CANDELS combined catalogue}. Within our colour, magnitude and mass selection, and in a circle of radius 0.5 arcminutes, we typically detect $2\pm1$ galaxies in the {\it CANDELS combined catalogue}. Of these galaxies,   28$\pm2$\% are ETG, 21$\pm1$\% are passive,  34$\pm2$\% are asymmetric, 1.7$\pm0.5$\% are mergers. We use these numbers to correct the cluster galaxy counts for fore- and background contaminants, i.e., $N_{cont}$, within each subsample of galaxies.  $N_{cont}$ is typically is in the range $3-15\%$. The fraction of missing galaxies is negligible ($<2\%$ for each category), and we approximate  $N_{miss} \sim 0$.

The percentage of CARLA active ETG with respect to the total number of ETG is $20 \pm 6$\%  at 1.35 < z <1.65, and $55 \pm 13\% $ at 1.65 < z < 2.05, compared to $6 \pm 5 \%$ in the {\it TPHOT-CANDELS catalogue} and $6 \pm 5 \%$ in our 'CANDELS combined catalog' for galaxies selected in the range $1.3<z<3$. In general, this shows a large percentage of active ETG in CARLA clusters with respect to the field, especially in the highest redshift range of our sample.

We list the corrected fraction of passive, ETG, asymmetric and interacting galaxies in each cluster in  Table~\ref{tab:fraction}.  Many clusters show enhanced fractions in all categories, but the uncertainties on these fractions are large because the number of galaxies in each cluster is small. We therefore combine all the cluster galaxies into a single sample to explore the effect of stellar mass and environment on the morphology and passivity of the cluster galaxies. In the appendix (Fig.~\ref{fig:af0}), we show the ETG and passive fractions as a function of cluster redshift. We find no correlation between these fractions and redshift, with Pearson coefficients of p=-0.03 and p=-0.003 for ETG and passive galaxies, respectively. This also justifies our decision to combine all the cluster galaxies into a larger sample.

\begin{figure*}
\includegraphics[angle=90,width=0.5\textwidth]{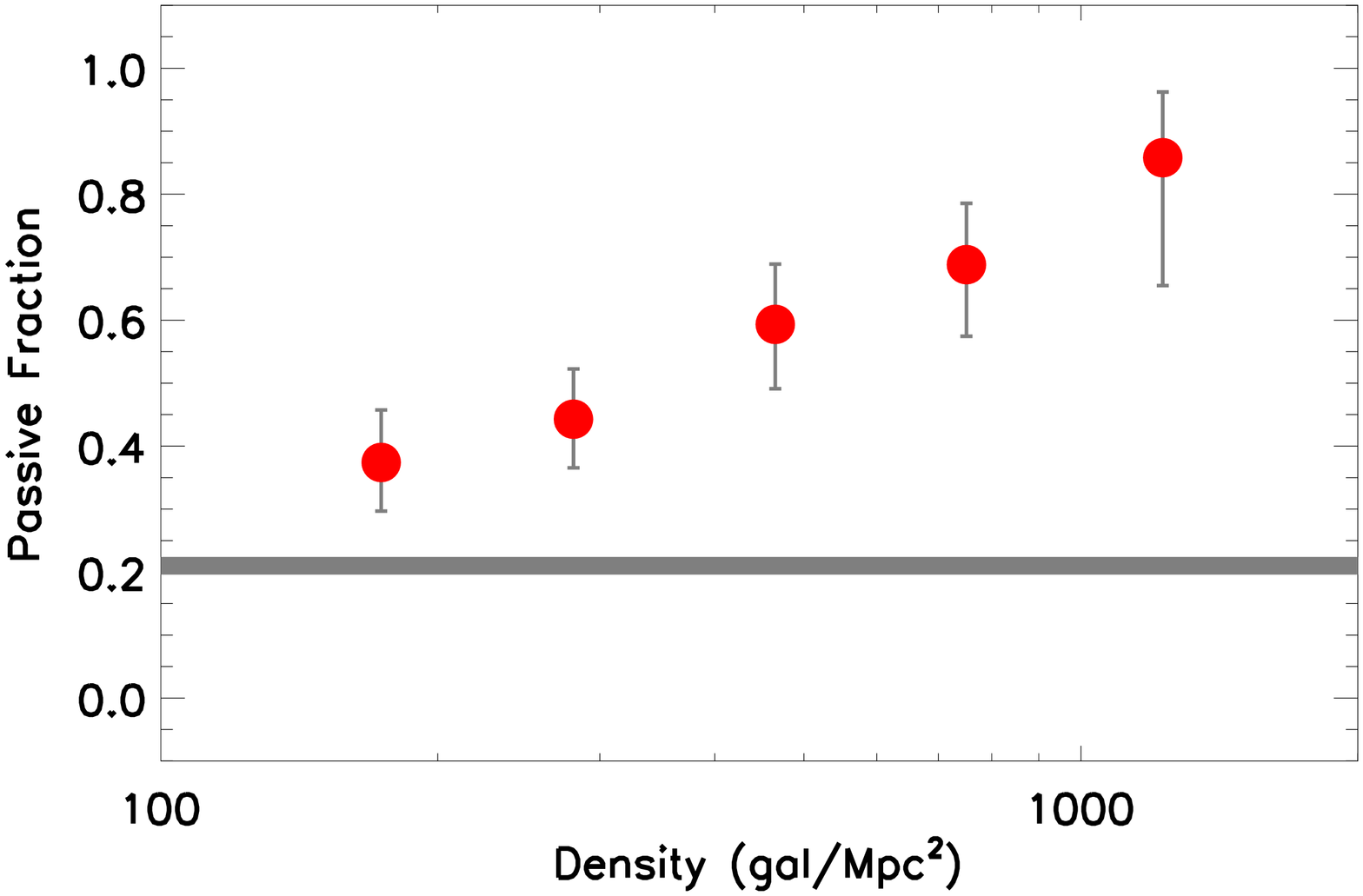}
\includegraphics[angle=90,width=0.5\textwidth]{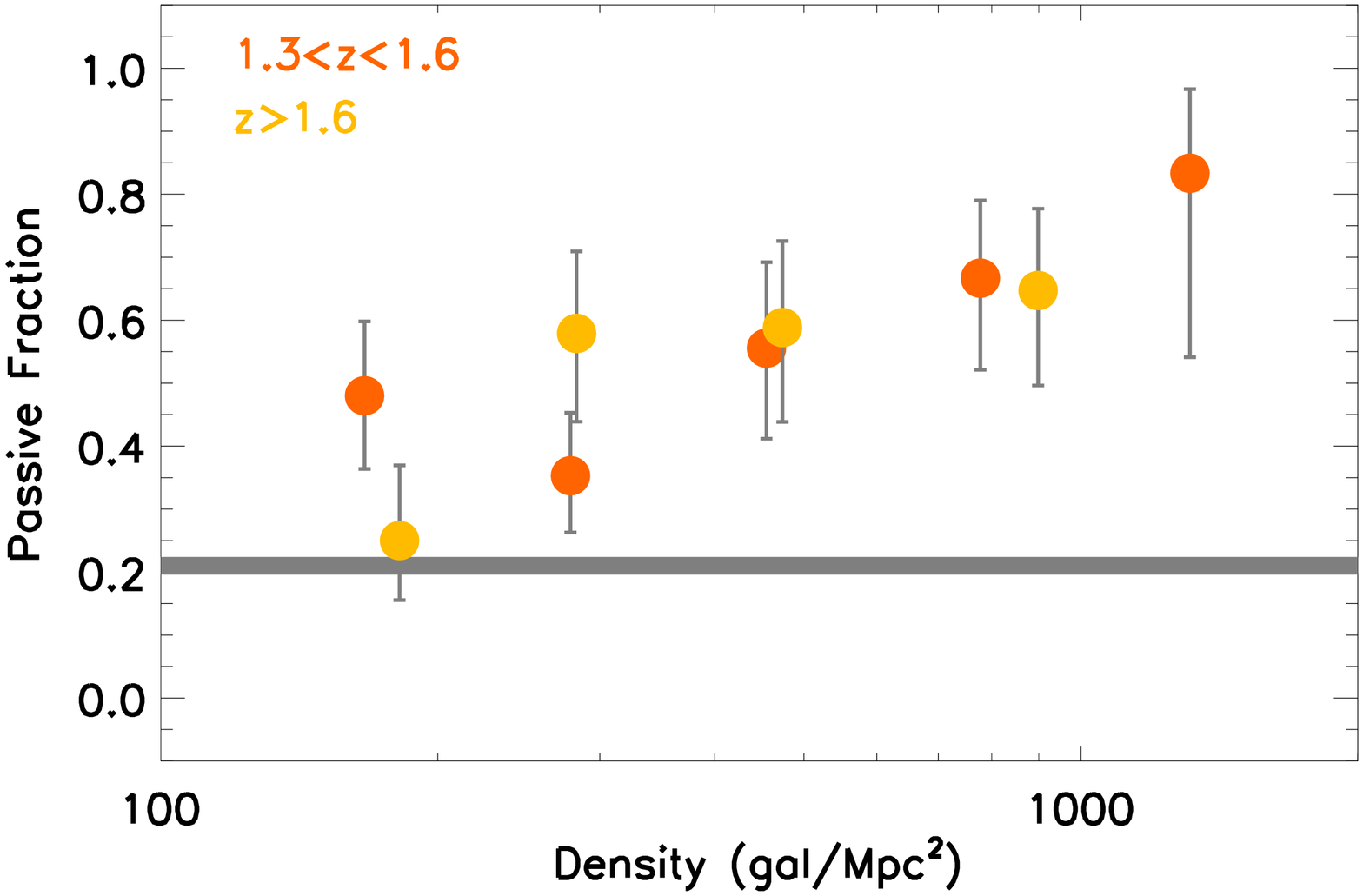}
\caption{Passive galaxy fractions as a function of environment. Left: our entire sample. Right: passive fractions in two redshift bins: $1.3<z<1.6$ (orange circles) and $1.6<z<2$ (yellow circles).  The grey region shows the $\pm1~\sigma$ range of the CANDELS passive fraction. We observe a significant passive-density relation up to $z \sim 2$.}\label{fig:f2}
\end{figure*}

\subsection{Influence of local environment on the morphology and passivity of cluster galaxies }

We first explore the fraction of passive galaxies within the clusters as a function of the local environment, shown in Fig~\ref{fig:f2}. We also split the cluster sample into a higher and lower redshift bin to explore whether there is any redshift evolution. We find there is a strong relation between the fraction of passive galaxies and the local environment of a cluster galaxy. Furthermore, we find no significant difference in this relationship between clusters at $1.3<z<1.6$ and $1.6<z<2.$ This result extends the work of \citet{19lemaux} that found the SFR-density relation is already in place at $z\sim 1.5$.

\begin{figure*}
\includegraphics[angle=90,width=0.5\textwidth]{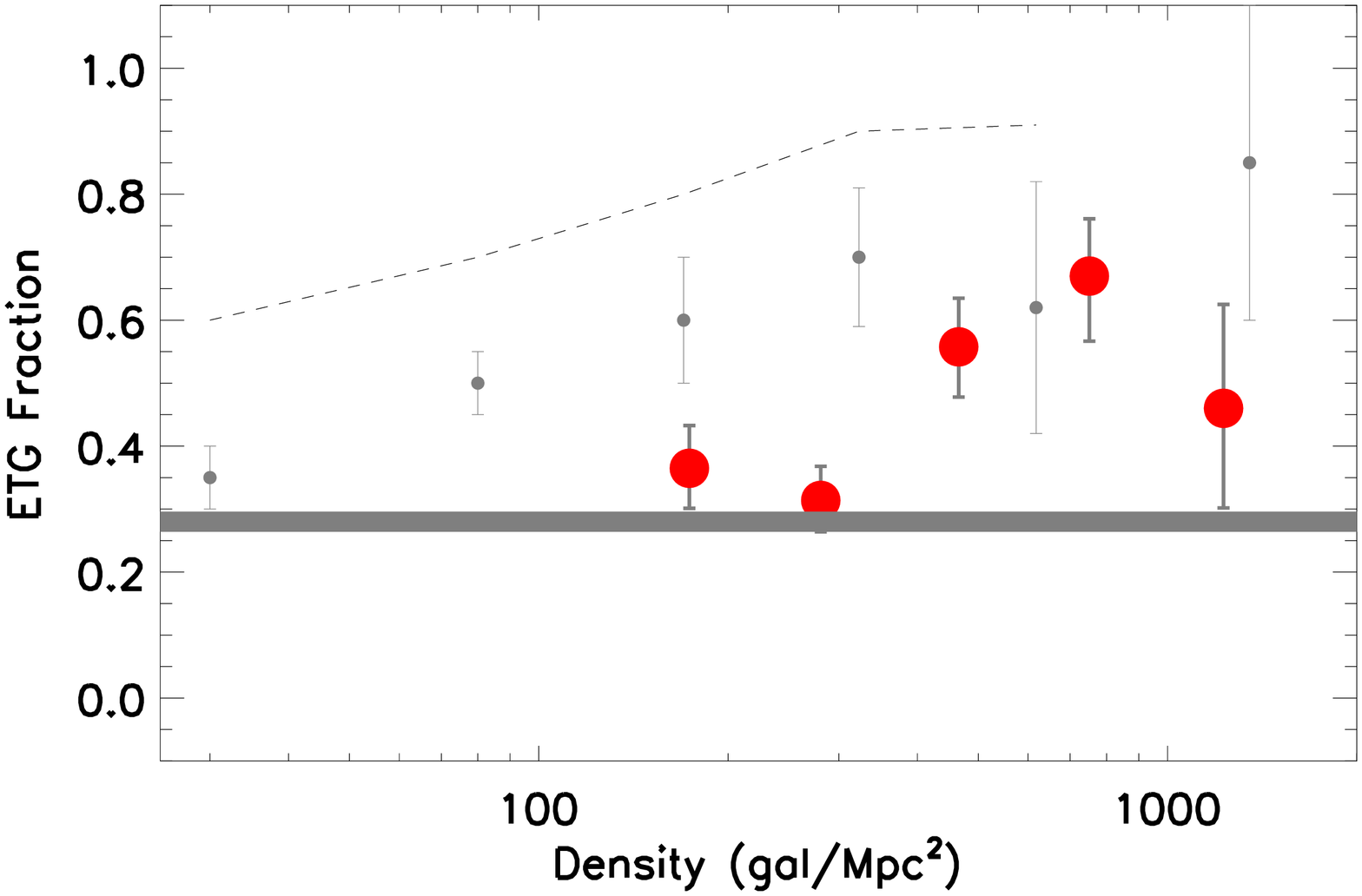}
\includegraphics[angle=90,width=0.5\textwidth]{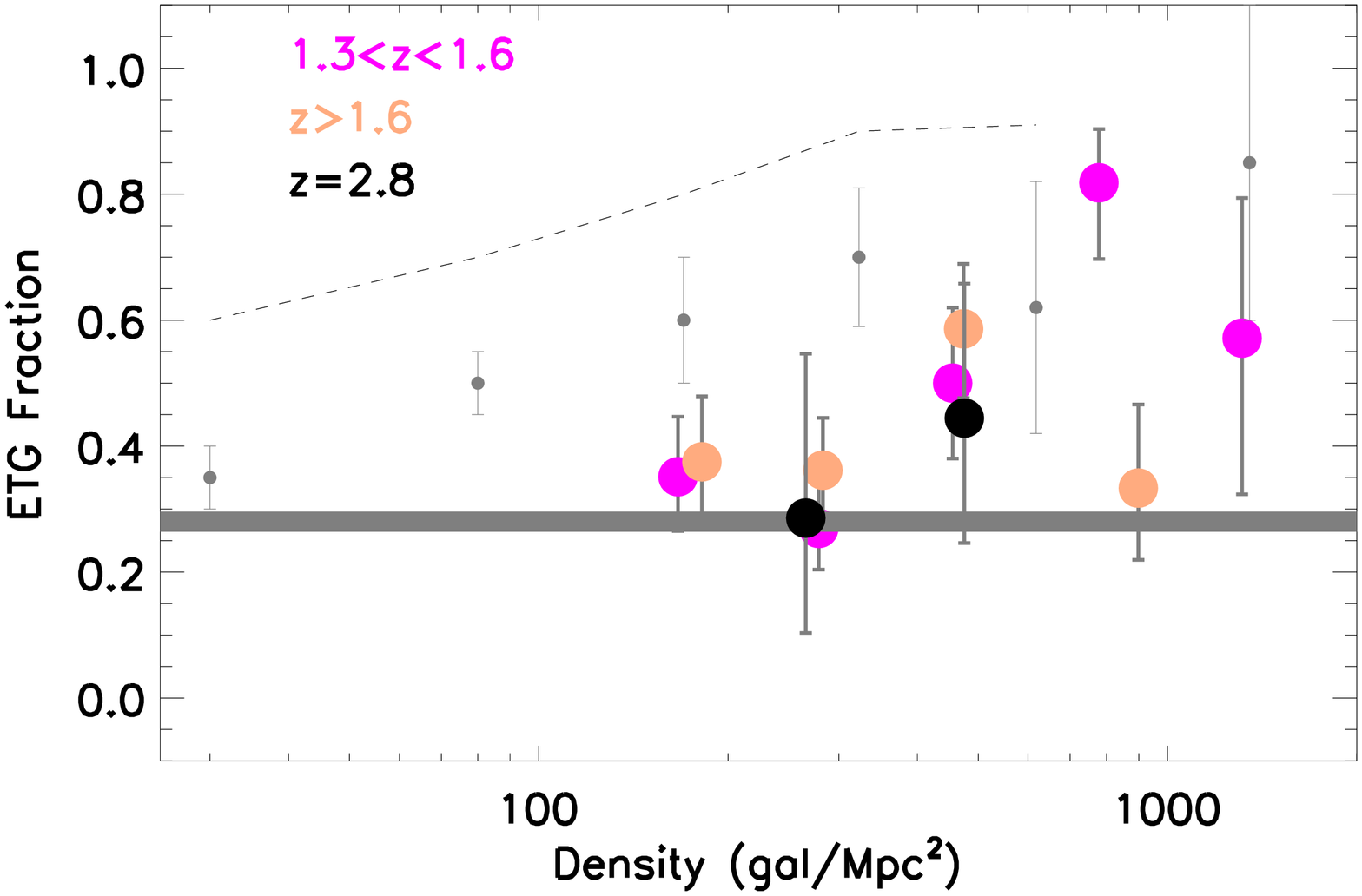}
\caption{ETG galaxy fractions as a function of environment. Left: our entire sample. Right: ETG fractions in two redshift bins: $1.3<z<1.6$ (magenta circles) and $1.6<z<2.8$ (pink circles). We also show with two black circles the fraction for CARLAJ1017+6116 at z=$2.8$. The small grey circles and error bars and the dashed line show the morphology-density relation at $z\sim1$ from \citet{05postman} and at lower redshift from \citet{80dressler}. The grey region shows the $\pm1~\sigma$ range of the CANDELS ETG fraction.  We observe a significant morphology-density relation up to $z\sim2$.  The cluster at z = 2.8 shows a similar fraction of ETG as in the other clusters, however one cluster does not provide enough statistics to confirm that the morphology-density relation is already in place at z$\sim$3. In fact, ETG fractions are correlated with environment, and increase from high to lower redshift in the intermediate dense regions.}\label{fig:f1}
\end{figure*}

\begin{figure*}
\includegraphics[angle=90,width=0.5\textwidth]{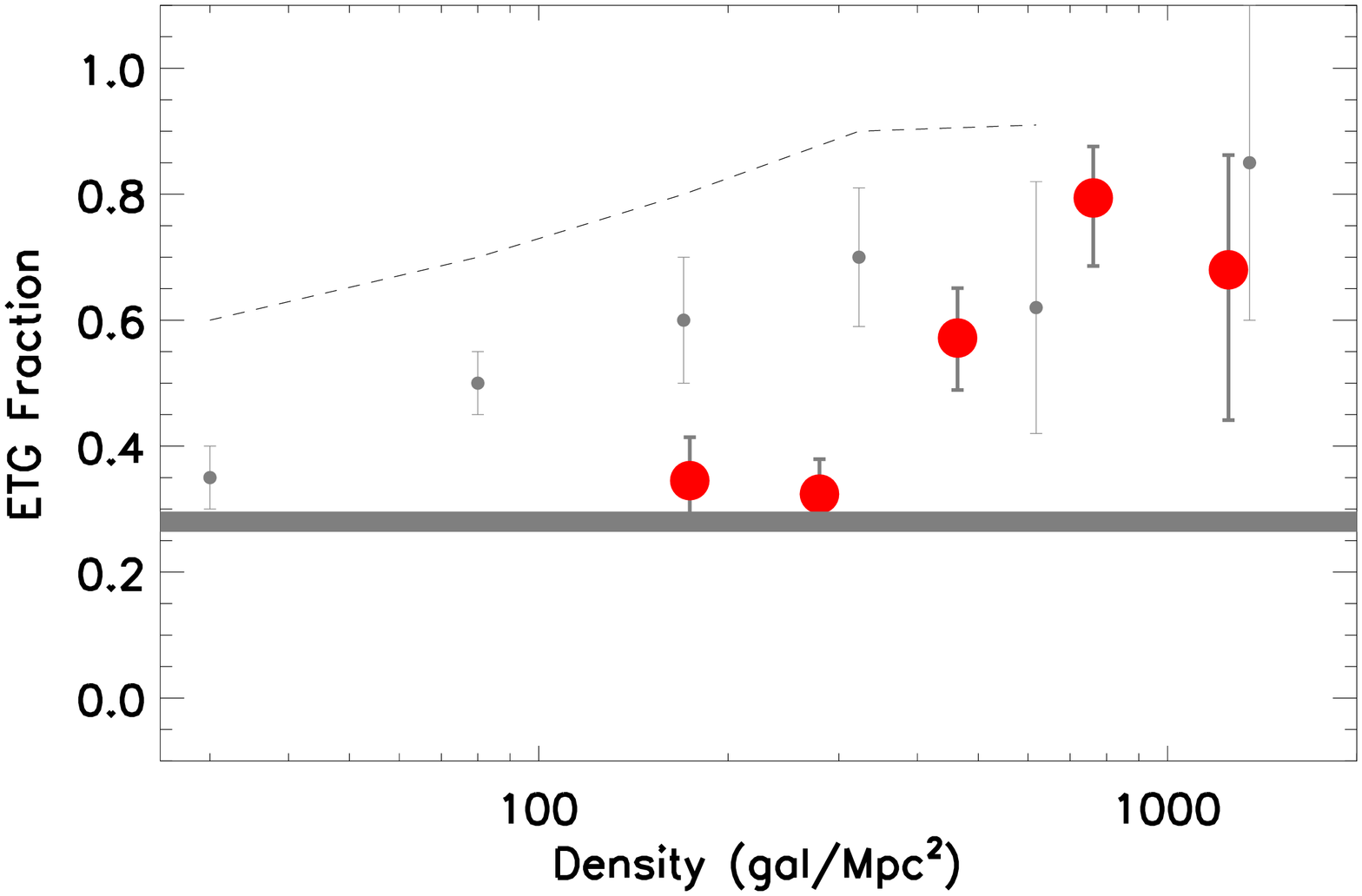}
\includegraphics[angle=90,width=0.5\textwidth]{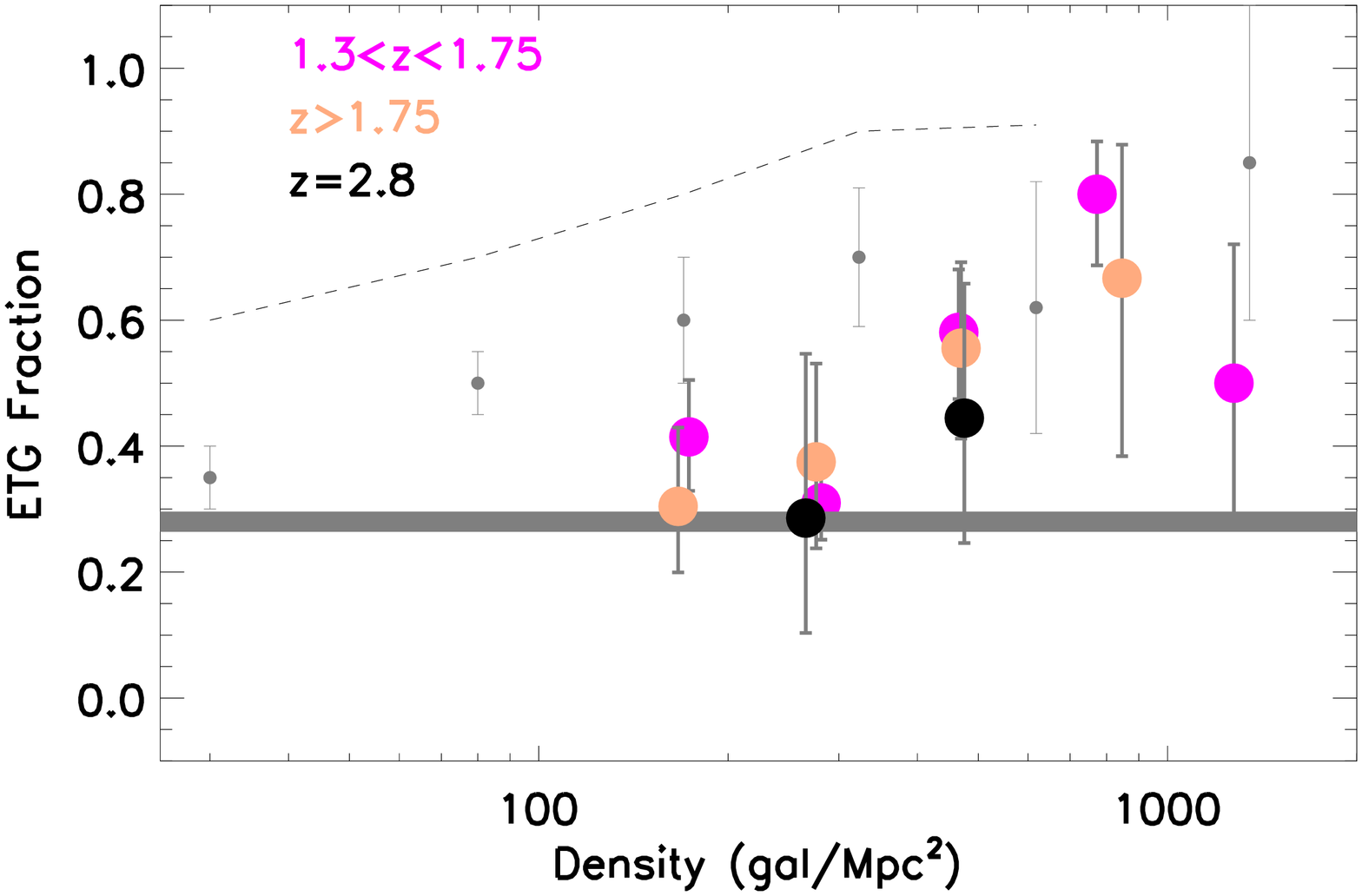}
\caption{The same as Fig.~\ref{fig:f1}, excluding the cluster CARLAJ0800+4029, which shows an high percentage of mergers in its densest regions. We observe a significant morphology-density relation up to $z\sim2$. The ETG fractions drop in the last bin, which is observed in the entire sample (see Fig.~\ref{fig:f1}), is mitigated. This means that the mergers in the high density regions in CARLAJ0800+4029 were dominating the ETG fraction drop observed when including this cluster in the analysis.}\label{fig:f1bis}
\end{figure*}

\begin{figure*}
\includegraphics[angle=90,width=0.5\textwidth]{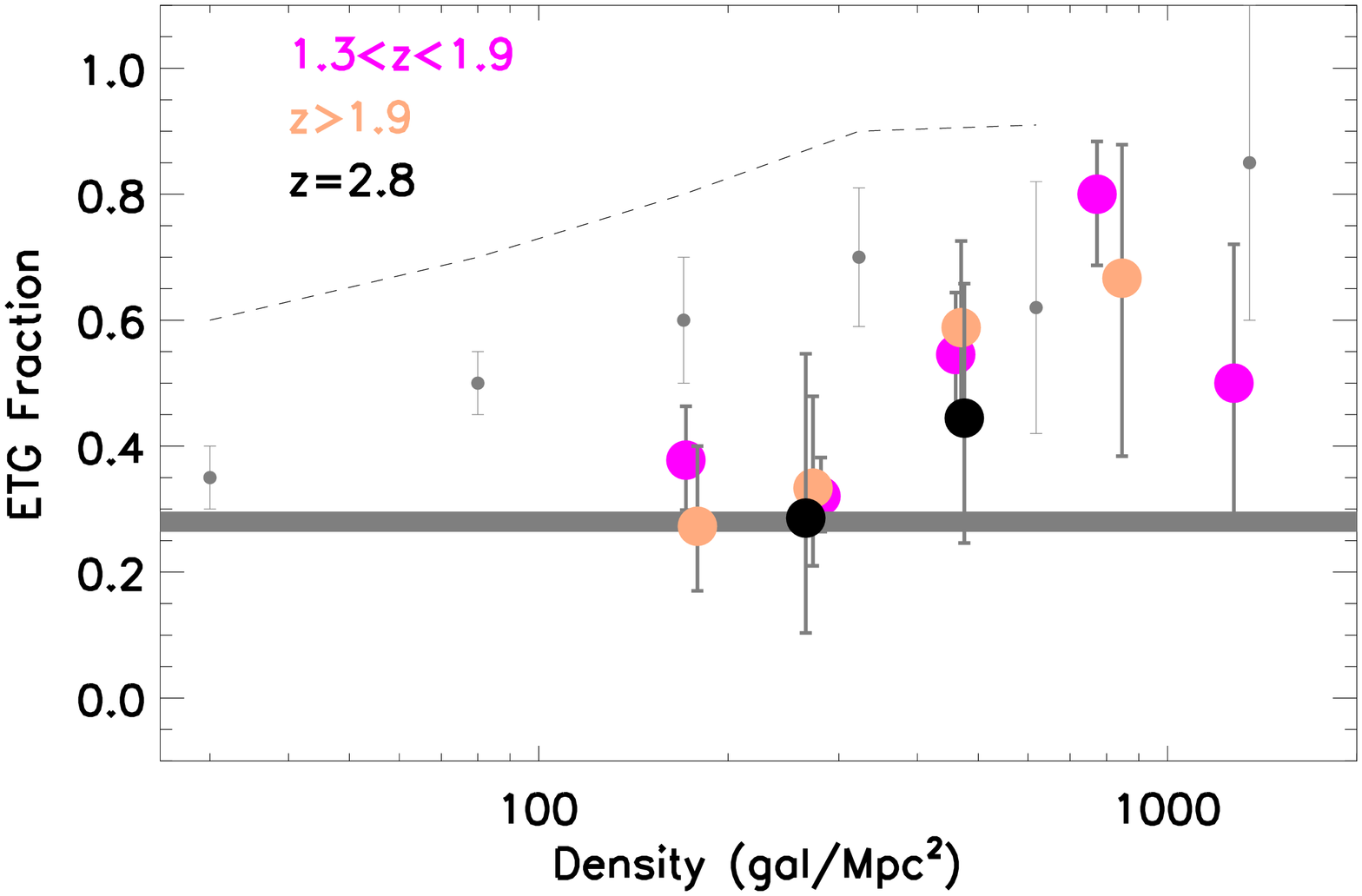}
\includegraphics[angle=90,width=0.5\textwidth]{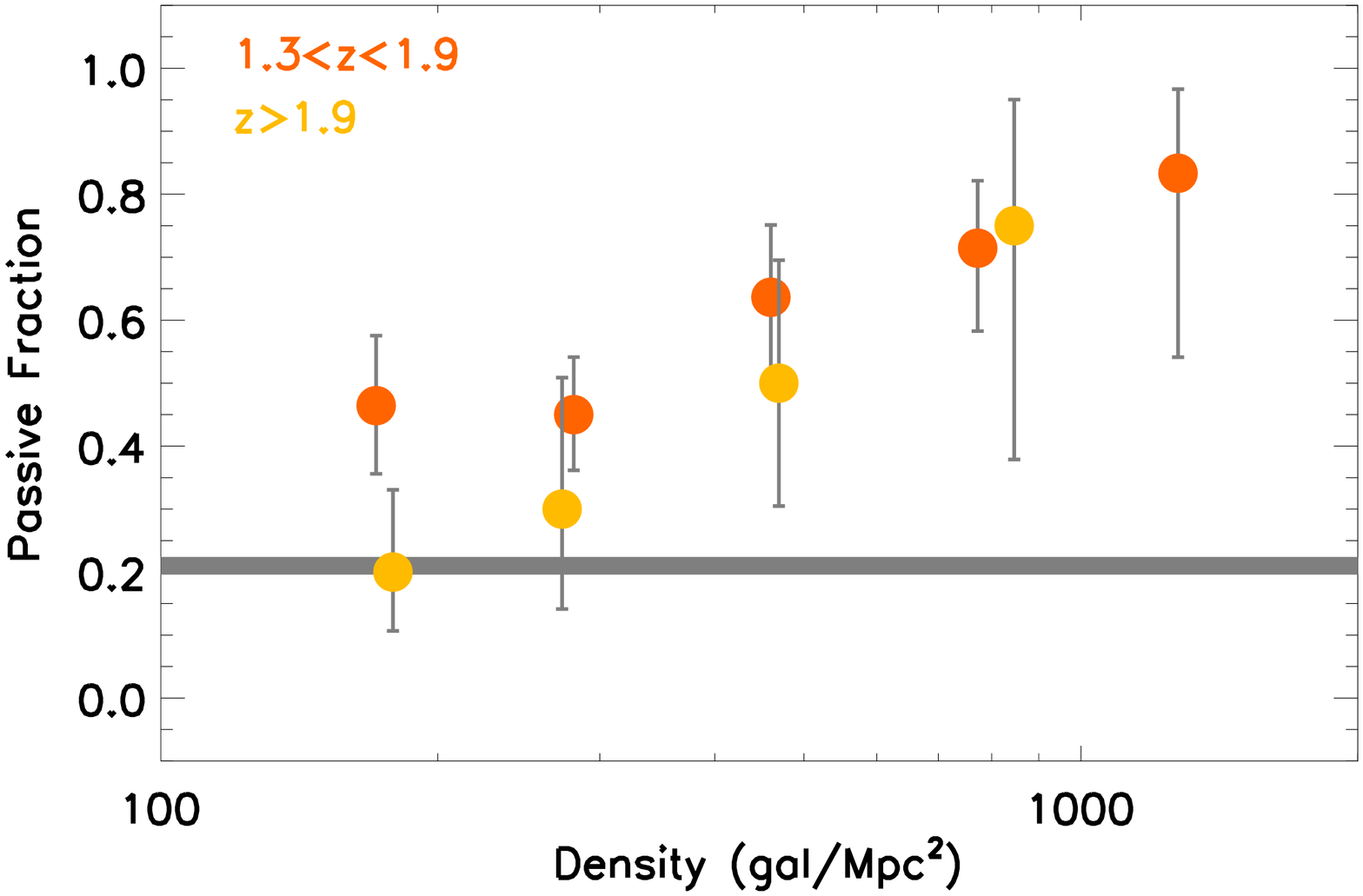}
\caption{The same as the right panels of Fig.~\ref{fig:f2} and Fig.~\ref{fig:f1bis}, excluding the cluster CARLAJ0800+4029, separating the $z\sim2$ clusters from the others at $z<1.9$. This shows that the relations holds at $z\sim2$.}\label{fig:fredbin}
\end{figure*}

We then explore the influence of local environment on the morphological properties of cluster galaxies. We start by examining the fraction of ETGs, also subdividing the analysis by redshift bins. 
 We find that the ETG fractions are correlated with local environment (Pearson coefficient p$\sim$0.93). The ETG fraction decreases in the last bin, however this drop is dominated by that lack of ETG and a high incidence of disk and irregular galaxy interactions in the high density regions of only one cluster, CARLAJ0800+402.  When this cluster is excluded from the analysis, the drop is mitigated, and the Pearson coefficient for the correlation between ETG fraction and local environment is p$\sim$0.98. These correlations shown in Figs~\ref{fig:f1}, \ref{fig:f1bis}, and~\ref{fig:fredbin} show that the morphology-density relation is present in high redshift clusters.  

The morphology-density relation is most clearly present in the lowest redshift bin, and holds up to $z=2$.  The cluster at $z=2.8$ only has galaxies in two local density bins, and the data are consistent with both the field values and with the morphology-density relation within uncertainties. We therefore conclude that the morphology-density relation is present in clusters up to $z=2$, but there is no strong evidence for or against this relationship persisting at higher redshifts. 

We compare our high-redshift cluster sample to data from the local Universe \citep{80dressler} and at $z\sim1$ \citep{05postman}. These are displayed in Figs~\ref{fig:f1} and \ref{fig:f1bis} as the dash line and the small grey circles. We find that the ETG fractions in high redshift clusters are consistent with those observed in clusters at $z\sim$1 by \citet{05postman} when the local density is $\Sigma_{N} > 700 \ {\rm gal/Mpc}^2$. Thus, there is already a strong morphological influence occurring in the densest local environment of clusters at $z=2$. But, at this high-local density, the ETG fraction of $z=1$ clusters and our $1.3<z<2$ CARLA clusters is lower than that of clusters in the local Universe. This suggests that further morphological transformation must still occur to the cluster population.    

At lower local densities, where  $\Sigma_{N} < 700 \ {\rm gal/Mpc}^2$, the CARLA clusters exhibit a similar ETG fraction as the field. This is in stark contradiction to clusters at $z=1$ that already exhibit  higher ETG fractions. It therefore seems that the densest of cluster environments influence the morphology of galaxies first, with lower density local environments either taking longer to influence galaxy morphology or only influencing galaxy morphology at later cosmological times.

\subsection{Influence of global environment on the morphology and passivity of cluster galaxies }

\begin{figure*}
\center
\includegraphics[angle=90,width=0.45\textwidth]{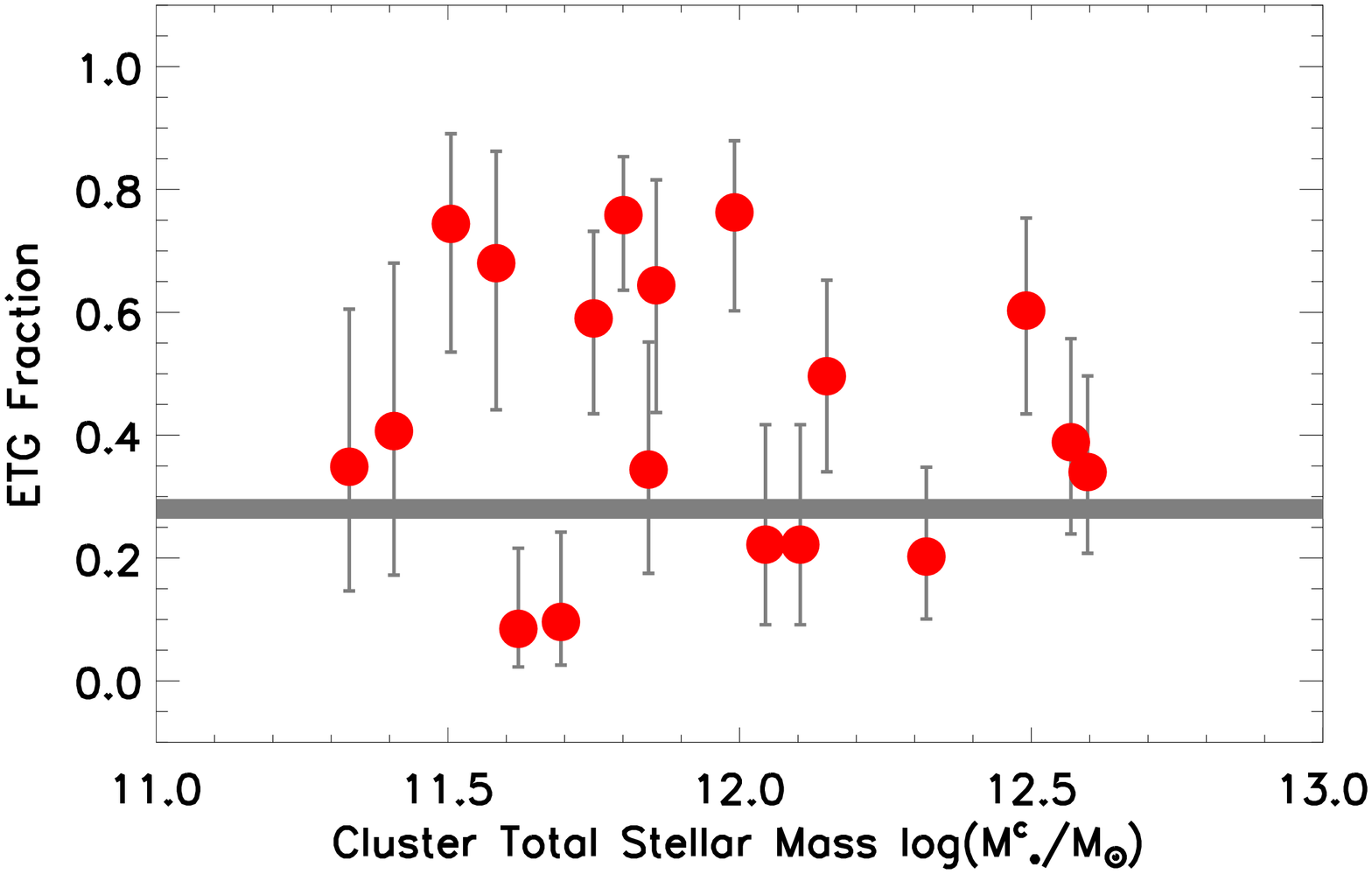}
\includegraphics[angle=90,width=0.45\textwidth]{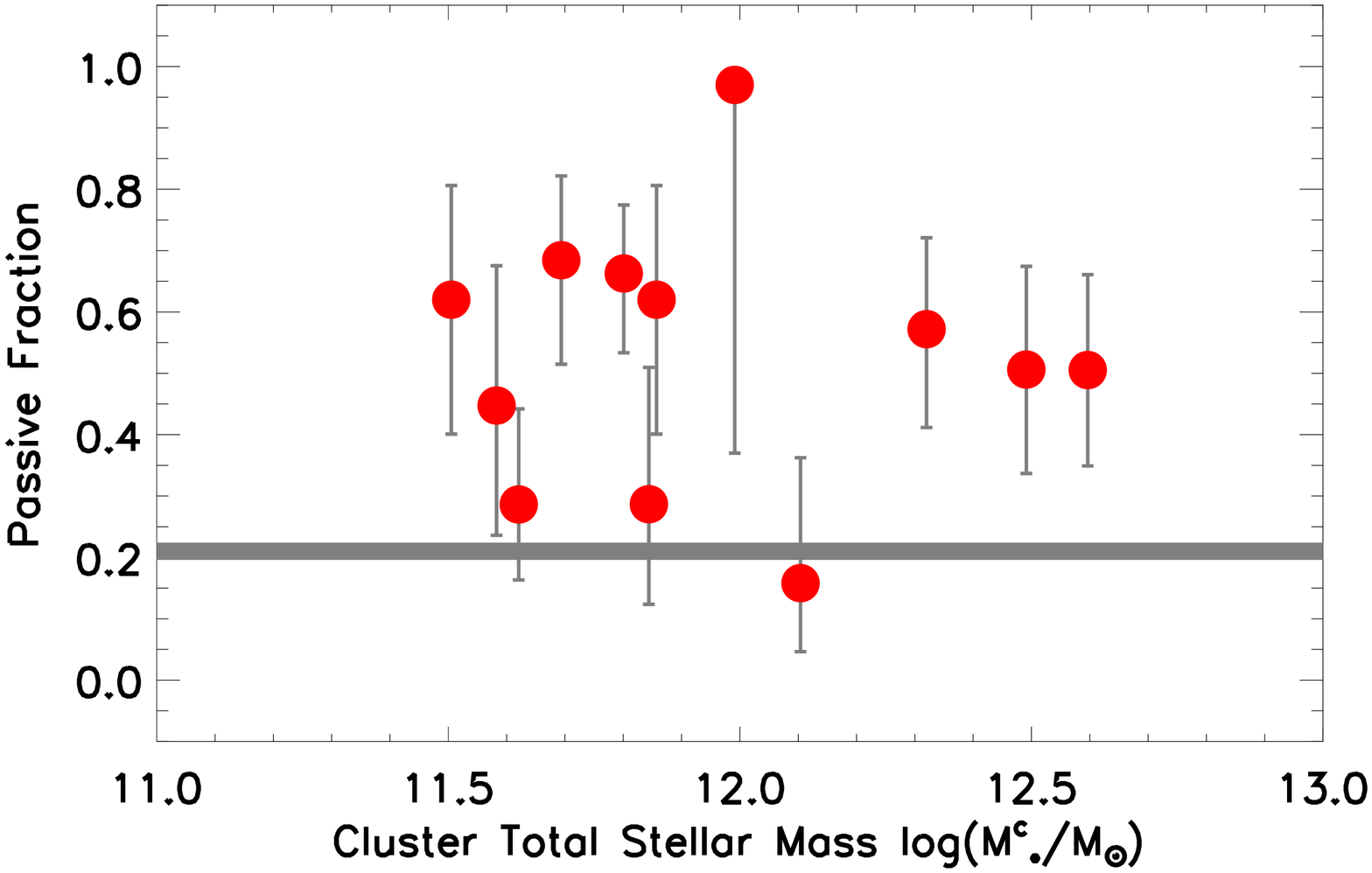}
\caption{ETG and passive galaxy fractions as a function of global environment defined as total galaxy stellar mass in the overdensity. The grey region shows the $\pm1~\sigma$ range of the CANDELS fractions. }\label{fig:global_environment}
\end{figure*}

Fig.~\ref{fig:global_environment} show ETG and passive galaxy fractions as a function of global environment defined as total galaxy stellar mass in the overdensity. The morphology and passivity of cluster galaxies do not depend on the cluster total stellar mass (the Pearson coefficient is p=-0.2 and p=-0.4 for the ETG and passive fractions, respectively). As shown in the appendix D, this lack of correlation is also observed with the overdensity SNR, as a proxy of density contrast.

\begin{figure*}
\includegraphics[angle=90,width=0.5\textwidth]{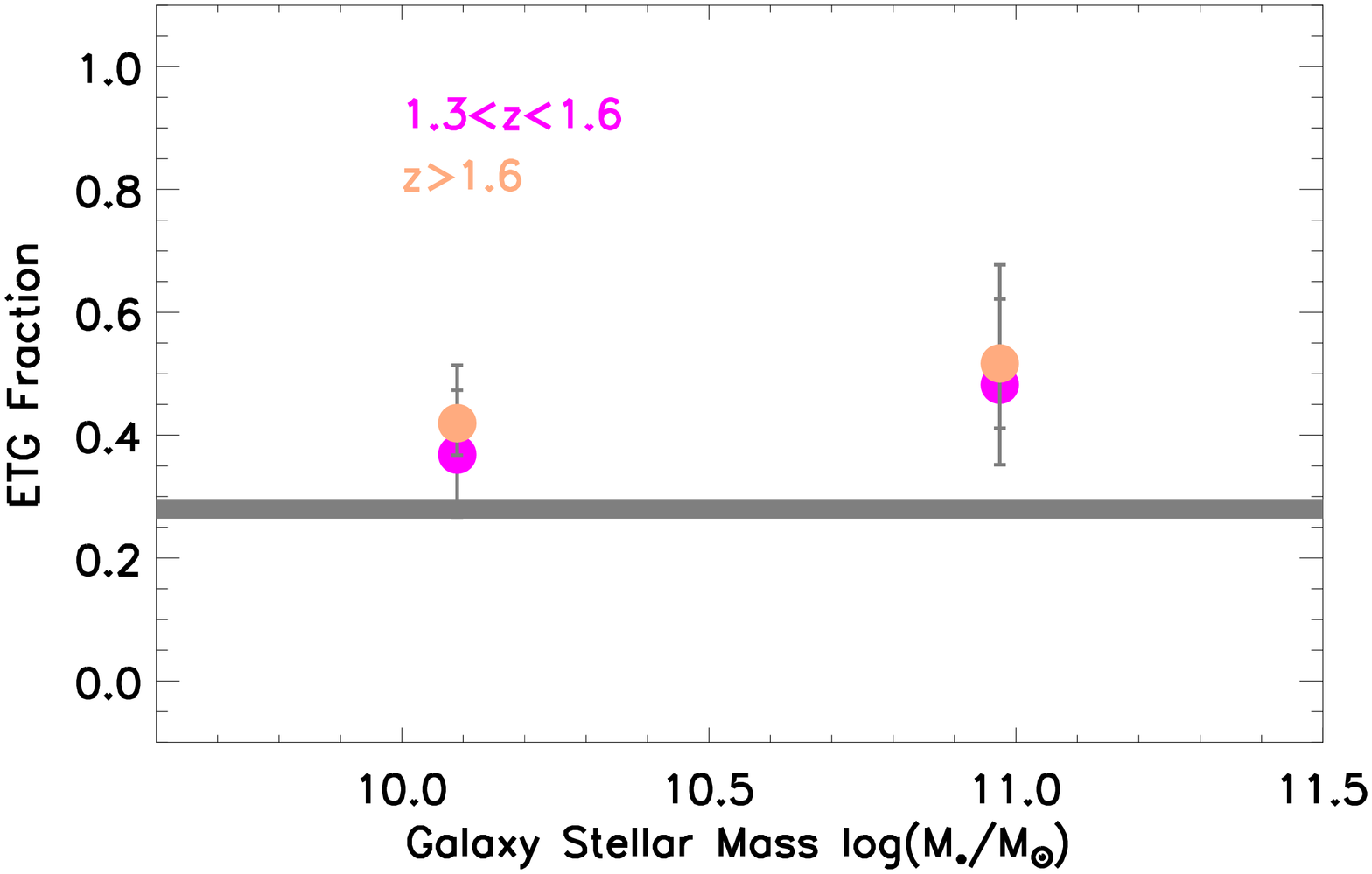}
\includegraphics[angle=90,width=0.5\textwidth]{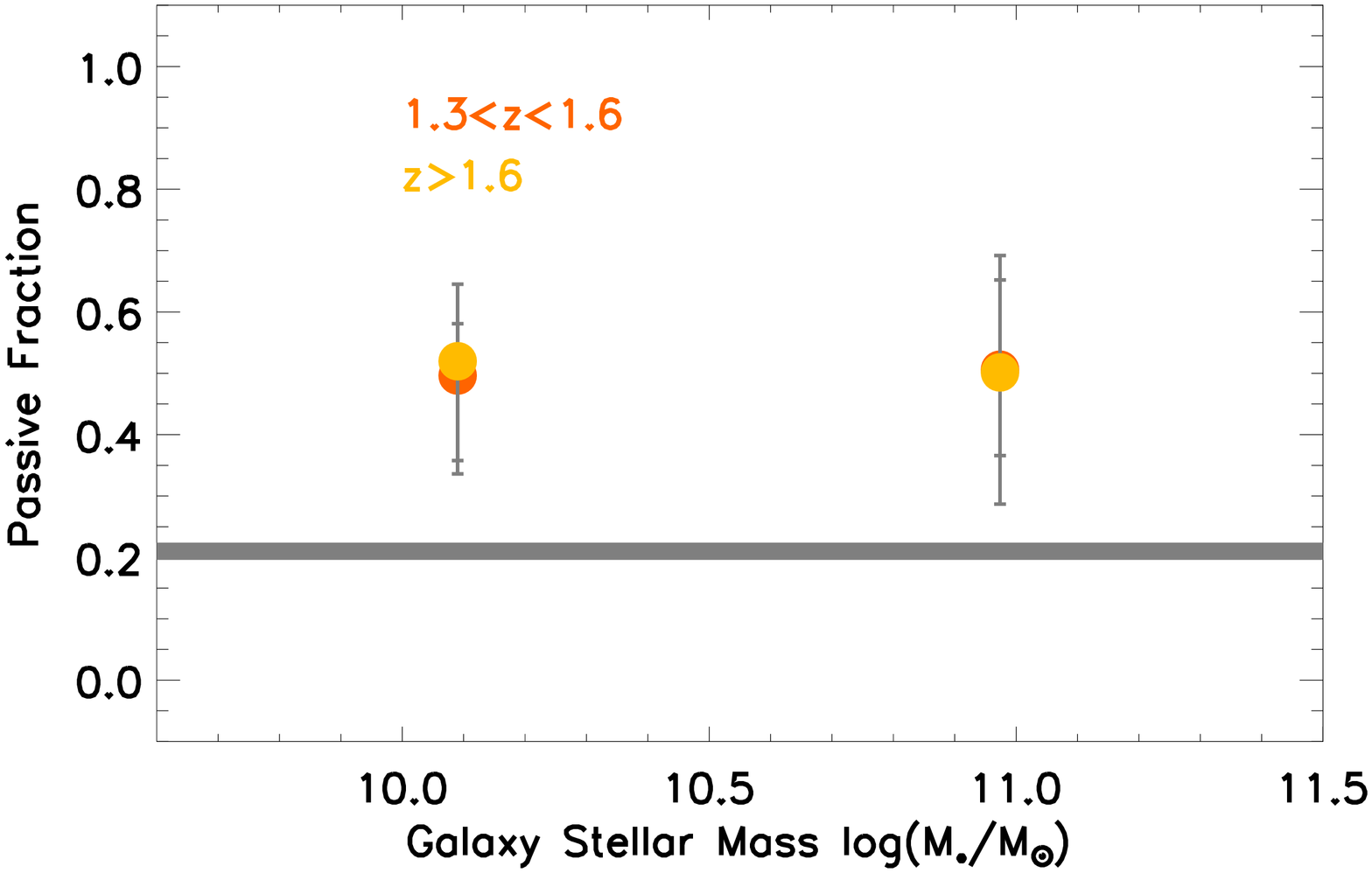}
\caption{ETG and passive galaxy fractions as a function of mass in two redshift bins: $1.3<z<1.6$ and $z>1.6$.  Symbol colors are the same as in Fig.~\ref{fig:f2}. The grey region shows the $\pm1~\sigma$ range of the CANDELS fractions.  ETG and passive fractions mildly increase with increasing mass.}\label{fig:f4}
\end{figure*}
\begin{figure*}
\includegraphics[angle=90,width=0.5\textwidth]{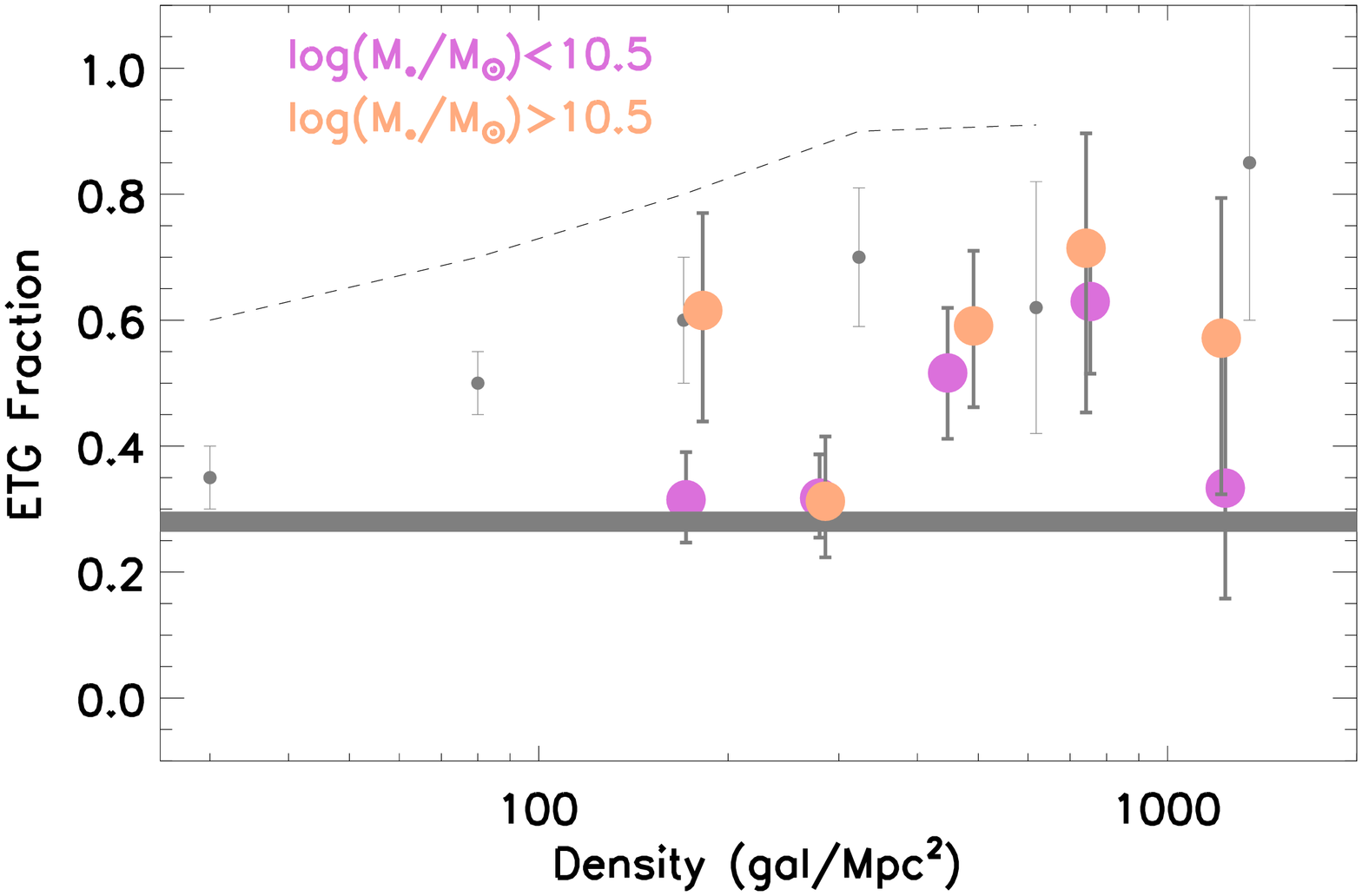}
\includegraphics[angle=90,width=0.5\textwidth]{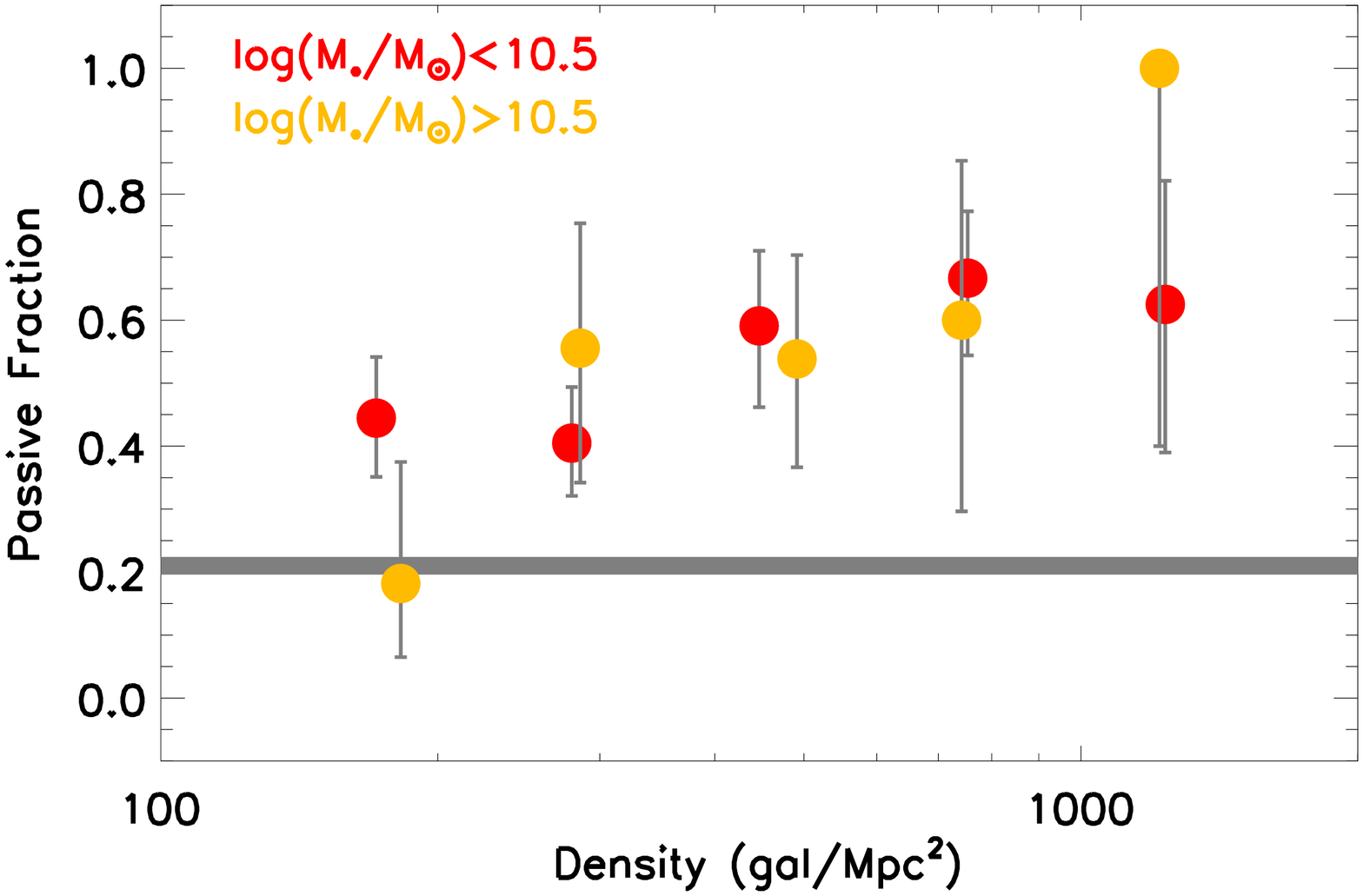}
\caption{ETG and galaxy fractions as a function of galaxy surface density in two mass bins: $log(\frac{M_*}{M_{\odot}})<10.5$  and $log(\frac{M_*}{M_{\odot}})>10.5$.  The small grey circles and error bars and the dashed line show the morphology-density relation at $z\sim1$ from \citet{05postman} and at lower redshift from \citet{80dressler}. The grey region shows the $\pm1~\sigma$ range of the CANDELS passive fraction. The mophology-density and passive density relation are already in place in both mass bins.}\label{fig:f3}
\end{figure*}

\subsection{Influence of stellar mass on the morphology and passivity of cluster galaxies }
\label{sec:mass_corr}
We examine how stellar mass influences the ETG and passive fractions in Fig~\ref{fig:f4}. The ETG and passive galaxy fractions are displayed in two mass bins and also subdivided into two redshift bins: $1.3<z<1.6$ and $z>1.6$. At both redshifts, more massive galaxies show higher ETG and passive fractions, but the uncertainties are large so we do not record a conclusive relationship in either case. This holds even when considering more mass bins, however the uncertainties increase because we select fewer galaxies in each bin, and the field contamination correction is also more uncertain.

Fig~\ref{fig:f2} and \ref{fig:f1} display a strong relationship between local environment and ETG \& passive fractions. To explore whether these relations are caused by an underlying  correlation between galaxy stellar mass and local environment, we plot in Fig~\ref{fig:f3} the ETG and passive fractions as a function of environment in two mass bins: $log(\frac{M_*}{M_{\odot}})<10.5$ and $log(\frac{M_*}{M_{\odot}})>10.5$. While the two separate samples are less statistically significant, the morphology and passive density relations are in place in both mass bins. For the ETG fractions, less massive galaxies have lower ETG fractions on average than the more massive galaxies, and in the densest and least dense bins they are consistent with the field. However, as in Fig~\ref{fig:f4}, the uncertainties are too large to make strong conclusions. We investigate whether there is a relationship between galaxy stellar mass and local environment, but we find that these parameters are not correlated in our cluster sample, with a Pearson coefficient p=0.11.  This implies that the morphology-density  and passive galaxy density relations are not driven by a correlation between galaxy mass and environment. 

We conclusion that the morphology-local density and passive-local density relations are already in place by $z=2$ for all galaxies with stellar masses above our mass limit, $log(\frac{M_*}{M_{\odot}}) \gtrsim 10$, but we cannot determine their dependence on stellar mass.

\begin{figure*}
\includegraphics[angle=90,width=0.5\textwidth]{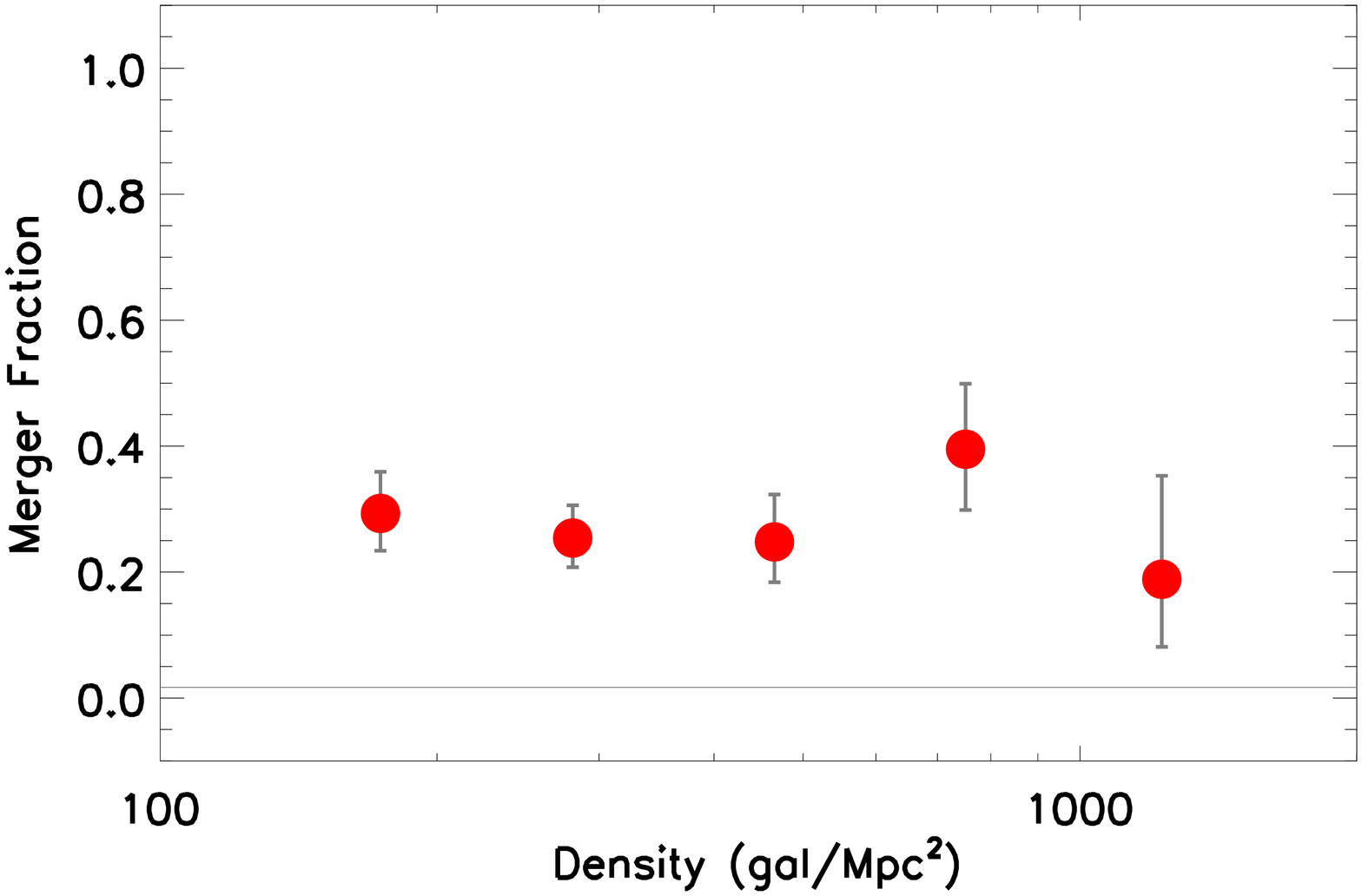}
\includegraphics[angle=90,width=0.5\textwidth]{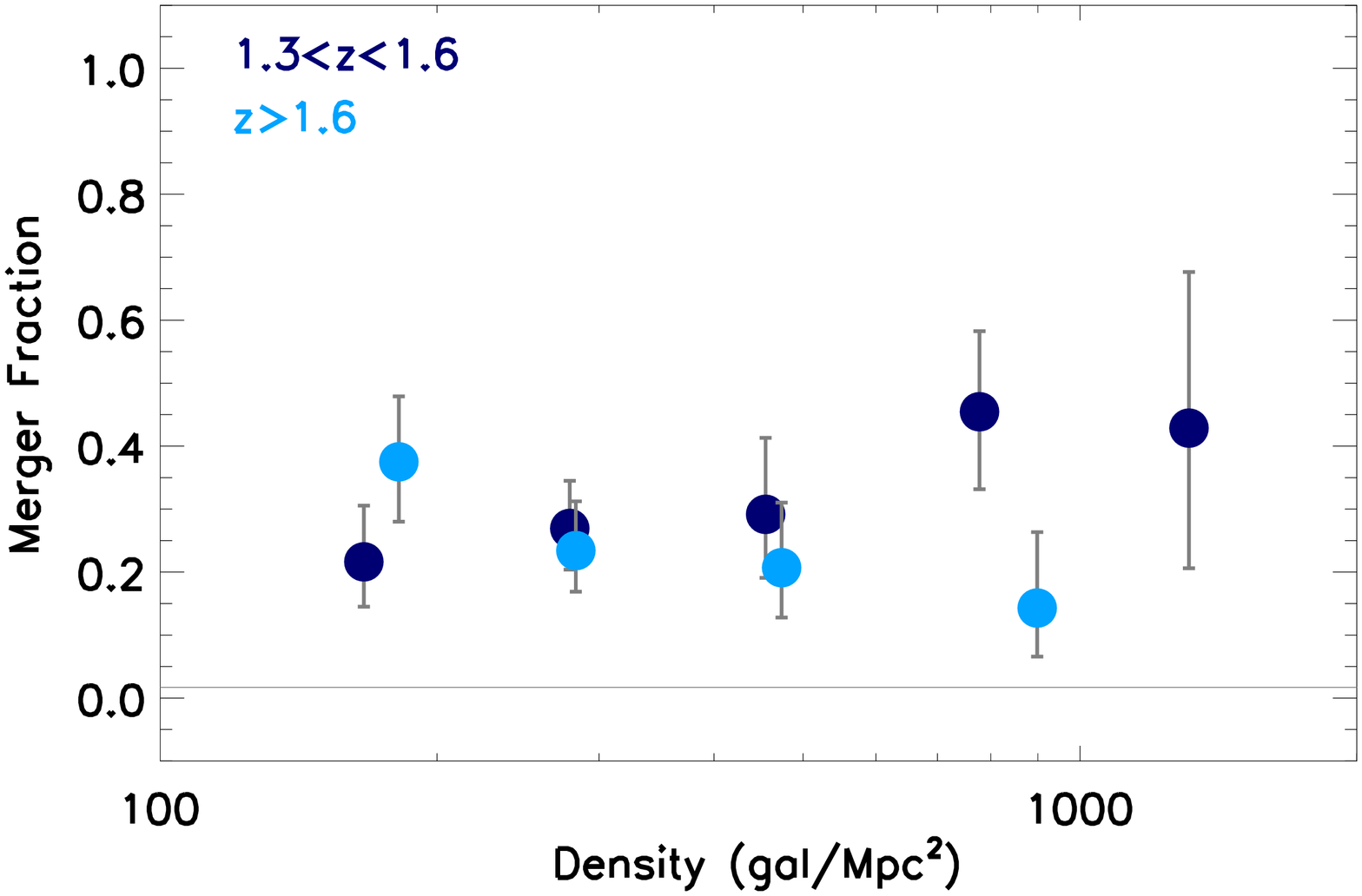}
\includegraphics[angle=90,width=0.45\textwidth]{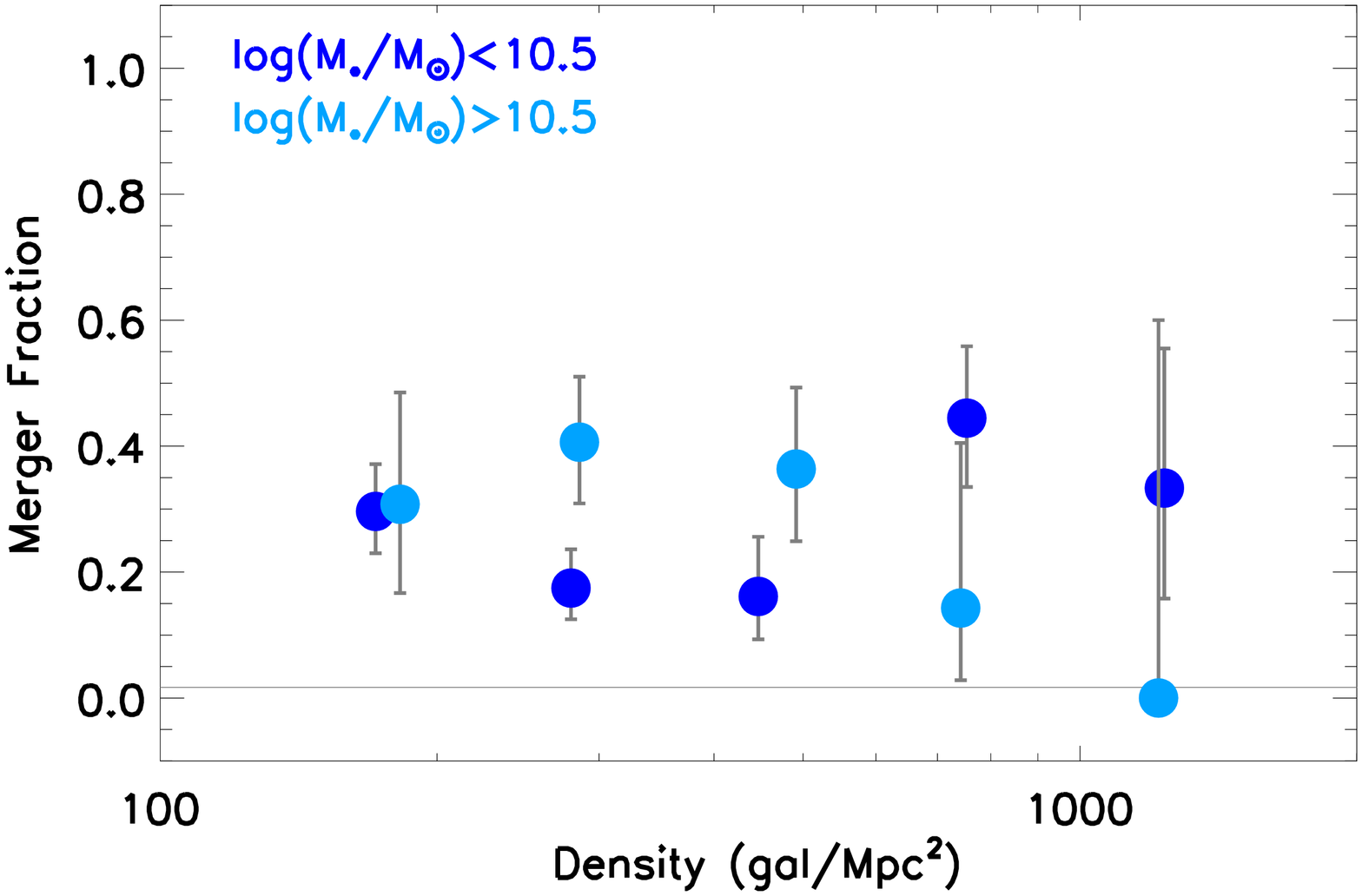}
\caption{Merger fractions as a function of environment. Top left: our entire sample. Top right: merger fractions in two redshift bins: $1.3<z<1.6$ (dark blue circles) and $1.6<z<2.8$ (light blue circles). Bottom: merger fractions as a function of galaxy stellar mass. The grey region shows the $\pm1~\sigma$ range of the CANDELS merger fraction.  The merger fraction in our cluster sample is significantly higher than in CANDELS, and shows a moderate correlation with environment and mass. }\label{fig:f5}
\end{figure*}


\subsection{Merger Fractions of $z>1.3$ cluster galaxies}

We concentrate on visually classified mergers rather than the number of close companions because the fractions of close companions can be biased by projection effects in clusters. The merger fraction for the entire sample of cluster galaxies is $26\pm3\%$, which is significantly higher than the CANDELS average merger fraction of $1.7\pm0.5$. In Fig~\ref{fig:f5} we display the merger fraction as a function of local environment and mass. The merger fraction is systematically higher ($\gtrsim 3\sigma$, except at high density where uncertainties are high) than the CANDELS merger fraction. Merger fractions show a moderate correlation with mass at all redshifts, with higher fractions for more massive galaxies and lower redshifts. However, similar to the results in Section~\ref{sec:mass_corr}, this result is not statistically significant because the differences in the observed fractions is consistent within $\sim 1 \sigma$. In contrast to the passive and ETG fractions, the merger fraction shows no significant correlation with local environment. We  explore the implications of this in the discussion below.

We additionally observe an average asymmetric galaxy fraction of $50\pm3\%$, which is significantly larger than the CANDELS asymmetric fraction of $34\pm2\%$. This provides supporting evidence that merger rates are enhanced in the cluster environment.
Our asymmetric galaxy fraction has no dependency on stellar mass, local or global environment, or redshift.

\subsection{AGN and spectroscopically confirmed member morphology} \label{sec:conf}

\begin{figure*}[t!]
\center
\includegraphics[width=1\hsize]{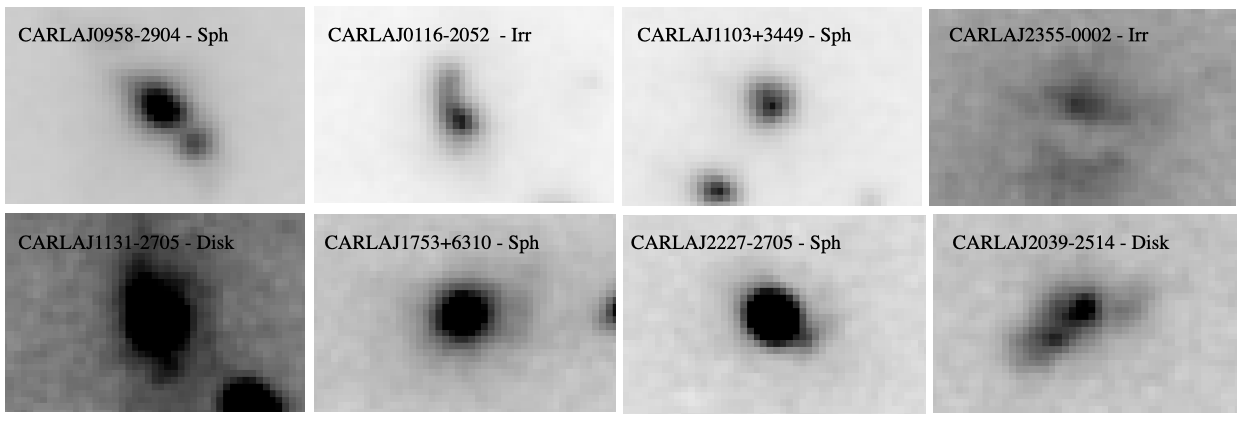}
\caption{Morphology of the eight central HzRG. They all show possible interactions and tidal arms. The radio source in CARLAJ2355-0002 has been discussed in \citet{18noirot}.}\label{fig:agnmorph}
\end{figure*}

The CARLA central AGN are eight high redshift radio galaxies (HzRG) and eight quasars (RLQ). 
We also confirmed two more quasars, one in the CARLAJ1510+5958 area, but at a different redshift (z=1.838) and one spectroscopically confirmed member in CARLAJ0800+4029 \citep{18noirot}. 
The RLQ all have unresolved morphology, often saturated and/or showing diffraction spikes. The HzRG morphologies are shown in Fig.~\ref{fig:agnmorph}: they are spheroids, disks and irregulars. They  are often found with a companion and tidal tails, and some are asymmetric, evidence for possible interactions or mergers.
The radio source in CARLAJ2355-0002 has been discussed in \citet{18noirot}. \citet{15collet} reported a peculiar radio-jet and gas properties in this radio source, which they found similar to the properties of the brightest cluster galaxies in low-redshift cool core clusters. \citet{18noirot} already discussed the morphology of this object in detail, and our HST image shows two sources and a complex morphology for this object. We do not use this galaxy pair in the galaxy morphology and color analysis in this paper. Its colors are consistent with being an active galaxy, and its morphology does not have a clear classification.

\begin{figure*}
\center
\includegraphics[width=0.35\hsize]{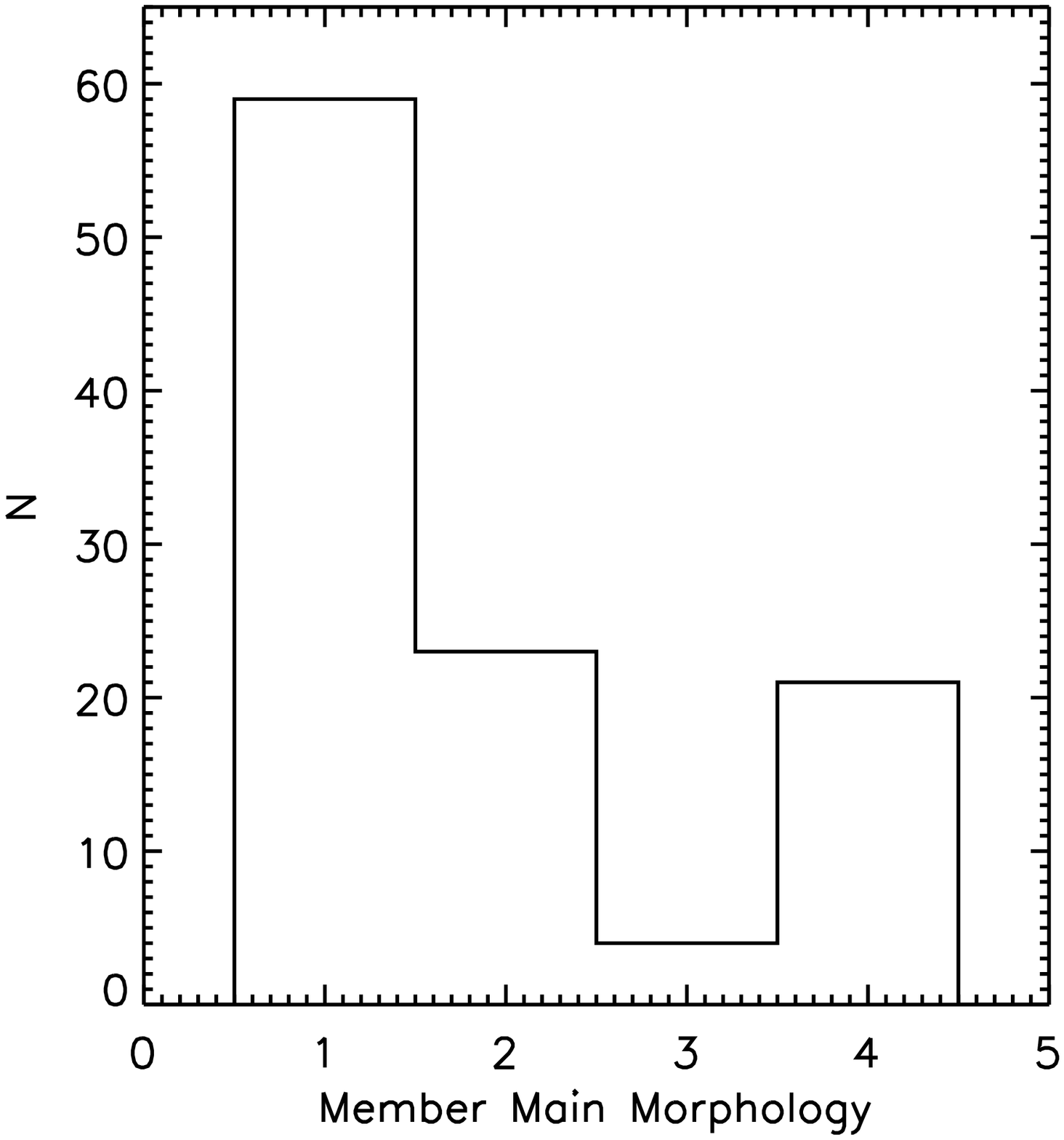}
\includegraphics[width=0.35\hsize]{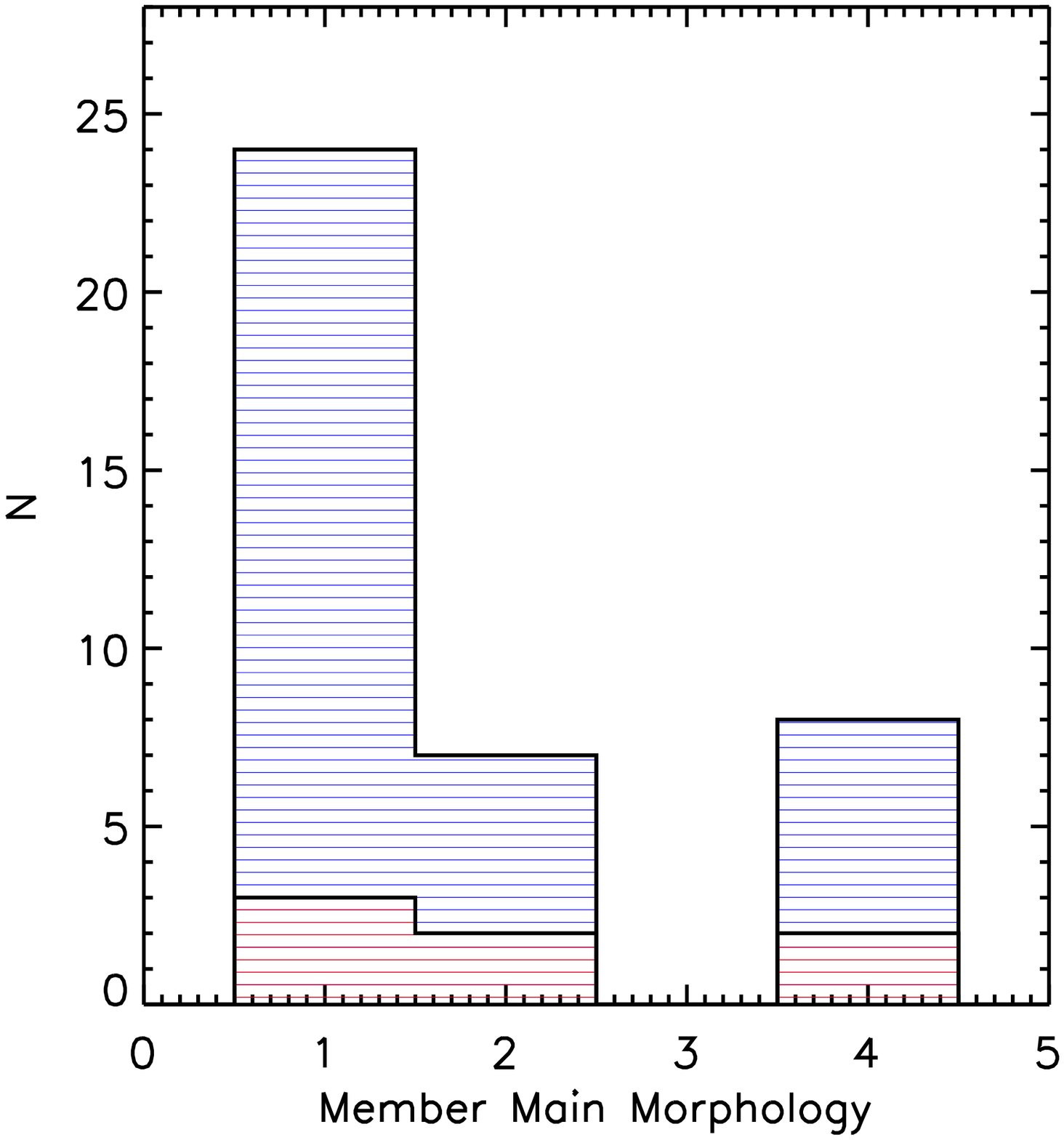}
\caption{Left: Morphology class distribution of the 107 spectroscopically confirmed cluster members from \citet{18noirot}. The classes are:  Disk (1), Spheroid (2), Irregular (3), Compact (4), Unclassifiable (5). Most of the members are disks, however there are large percentages of star-forming spheroids and compact galaxies. Right: Morphology class distribution of the members that have ground-based photometry. The blue and red lined histograms correspond to active and passive galaxies, respectively. The passive LTG are galaxies with star formation rates below the field main sequence. }\label{fig:memmo}
\end{figure*}

The CARLA spectroscopically confirmed members with a morphological classification (107 galaxies), are mostly disks (55\%), even if a large percentage of confirmed members are star-forming spheroids ($\sim$22\%; 4 of 23 are HzRG) and visually compact galaxies ($\sim$20\%). Only $\sim 4\%$ are irregular galaxies. Fig.~\ref{fig:memmo} shows the member morphology distribution. 
Excluding the AGN, the members that have sSFR 3$\sigma$ higher than the field main sequence (with log(sSFR)$\gtrsim -8.2 \ {\rm [yr^{-1}]}$ at $z<$1.5, and  log(sSFR)$\gtrsim -7.8 \ {\rm [yr^{-1}]}$ at $1.5<z<$2; from \citealp{14whita}) are 9 disks and 1 spheroid at $z<$1.5. Three disks show interaction (2 also show tidal tails), and 8 are asymmetric, indicating a possible connection between recent merger and starburst activity.
 At $1.5<z<$2, we find one visually compact galaxy in CARLAJ1129+0951 with log(sSFR)$\gtrsim -7.8 \ {\rm [yr^{-1}]}$ that does not show interactions, tidal tails or asymmetry.

\section{Discussion} \label{sec:discussion}

\subsection{Structure and heterogeneity of protoclusters}

Cosmological models predict that the clusters observed in the local Universe are built by the accretion of groups and filaments. At the redshifts of our sample, the main cluster progenitors are predicted to have halo mass in the range $13 \lesssim$log$_{10} (\frac{M^c_h}{M_{\odot}}) \lesssim 14$, and are expected to be surrounded by groups of similar or lower mass, spread out over scales of $\approx 35~h^{-1} Mpc^2$ comoving at $z\sim2$, which will accrete to assembly the clusters observed in the local Universe  \citep{13chiang, 18mul}. 

Our CARLA clusters have estimated halo masses of $13.6 \lesssim$log$_{10} (\frac{M^c_h}{M_{\odot}}) \lesssim 14.6$ and so are consistent with these model predictions. Nine of our clusters have expected total halo masses of log$_{10} (\frac{M^c_h}{M_{\odot}}) \gtrsim 14$, which defines a galaxy cluster\footnote{In fact, 90\% of the dark matter halos more massive than  log$_{10} (\frac{M^c_h}{M_{\odot}}) \sim 14$ are a very regular virialized cluster population up to redshift z$\sim$3 in simulations \citep{08ev,13chiang,18mul}}. All others have an expected  halo mass consistent with massive groups, which potentially are the main progenitors of local clusters.  

Our images do not cover large enough scales to observe the entire infall regions, but we do observe some structure. We observe three double structures (CARLAJ1018+0530, CARLAJ1358+5752, CARLAJ2039-2514), and multiple less significant overdensities in all our clusters except  CARLAJ1753+6310. In all double structures, one of the two overdensities reaches $\Sigma_N \gtrsim 600 {\rm \ gal \ Mpc^{-2}}$, and median  $\Sigma_N>300 {\rm \ gal \ Mpc^{-2}}$, typical of cluster core environments at $z \sim 1$ \citeg{05postman,19lemaux}, and the other has densities $\Sigma_N \lesssim 600 {\rm \ gal \ Mpc^{-2}}$,  typical of groups infalling in clusters \citeg{19lemaux}.

This results is consistent with the CARLA overdensities being clusters and groups/proto-clusters in the epoch of assembling. The exact characteristics of the assembly regions can only be better analyzed using a large imaging and spectroscopy coverage of these field.

Our local density analysis reveals that the galaxy populations within the cores of these high redshift clusters and protoclusters are heterogeneous: some consistent with the field, others with lower redshift groups and clusters. Around $25\%$ of our structures clearly show environments and galaxy fractions consistent with groups and filaments. For example, CARLAJ0958-2904 does not show properties much different from the field.  Whereas, around $30\%$ of our clusters have very dense local environments with $\Sigma_N > 1000 {\rm \ gal \ Mpc^{-2}}$, and here the galaxy population has ETG and/or passive fractions $>50\%$. Two clusters, CARLAJ1052+0806 and CARLAJ1753+6310 (both at $z\sim 1.6$), have ETG and passive fractions consistent with being mature clusters as those at $z<1.3$. 

The other structures are a more heterogeneous sample, mostly showing ETG and passive fractions and expected halo mass characteristic of groups and clusters, consistent with what is expected for cluster and proto-clusters at these redshifts. One exception to the general trends is a structure with low density environment $110 \lesssim \Sigma_N \lesssim 220  {\rm \ gal \ Mpc^{-2}}$, CARLAJ1103+3449, with expected halo mass that corresponds to a group, in which galaxy pre-processing has already happened. The central AGN is a star-forming ETG and its central region hosts a massive molecular gas reservoir \citep{20markov}, which suggests massive cool gas flows similar at those observed in the local Universe.

These results are consistent with a large body of literature on individual clusters. Clusters and proto-clusters at $z>1.5$ show a large variety in their galaxy populations. Some samples show large quiescent population and already quenched star-forming galaxies \citep{11andreon, 15cooke, 16stra, 18noirot, 20markov, 21shi,Sazonova2020}, others enhanced star-formation in their cores \citep{10tran, 11fass, 11haya, 12tada, 12zei,  13brodwin, 15mei, 16alberts, 16hayashi,18shima,21aoyama, 21koyama, 21polletta,21zheng}, others both populations \citep{16wang,17kubo,18stra}.

At higher redshift ($z>2$), some dense structures (clusters and protoclusters) present starbursts \citep{15casey,16casey,16wang}. 
Our findings are in contrast with the enhanced star-formation and star-burst activity in these samples, since we find an increasing fraction of passive galaxies in the densest local environments. This differences with other samples is probably due to the different sample selection, where some overdensities have been detected using star-forming and star-burst galaxy color selection, others by red galaxies overdensities and X-ray emission.

From our results, the CARLA overdensities selected around RLAGN have a more passive and ETG dominated galaxy population than the overdensities selected using star-forming and star-burst galaxy color selection, while they show similar populations as those found in X-ray and SZ selected clusters at $z\lesssim1.5$ \citeg{05postman,11andreon,16stra,19lemaux,19stra,20burg}. This might indicate a different evolution stage in the overdensities selected by different methods, with those selected using star-forming and star-burst galaxy color selection being still in a stage in which morphological transformations and quenching are still happening, in contrast with the galaxy population in CARLA, X-ray and SZ selected clusters that show large fractions of ETG and passive galaxies up to $z=2-3$. 

Our sample shows different percentages of ETG and passive galaxies in cluster regions that are otherwise similar in terms of global environment, dynamical state, and redshift. We therefore warn that ETG and passive fractions in individual high-redshift clusters or protoclusters should not be used to conclude whether or not morphological transformations or quenching happened first. This suggests that physical processes that lead to morphological transformations and quenching are often correlated but might not be exactly the same and can happen independently one from the other. Some models \citep{12delucia,18laigle} predict that these differences might be due to the fact that galaxies have different pre-processing histories that depend on the environments they were hosted by before their accretion in clusters, and tides in filaments that impact galaxy mass and quenching. In fact, some of the galaxies observed in our denser regions might have been accreted in much earlier epochs and others more recently, with different levels of pre-processing in the groups and filaments that hosted them in their previous history. 

The unique prospective from the CARLA survey is that by combining our large sample of clusters and protoclusters at $z>1.3$ we are able to average over a range of accretion histories and examine trends with both stellar mass and environment.

\subsection{The morphology and SFR-density relations of CARLA clusters in the context of previous results}  \label{compa}
 
Our main result is that galaxy morphology and passivity  are strongly correlated to local environment by $z\sim2$. We find the ETG and passive galaxy fractions are higher in denser environments. These fractions also mildly depend on galaxy stellar mass, with ETG and passive galaxies also being the most massive. But we find no correlation with  global environment (assuming that the central total stellar mass is correlated with total halo mass using \citealp{19beh}).

These results are reminiscent of the well defined morphology-density relation that is in place at  $z\lesssim 1$ \citeg{80dressler,05postman}. The densest regions in the Universe, found in galaxy clusters, host high percentages ($\gtrsim 80\%$) of passive ETG. Dense cluster environments can change galaxy morphologies and quench star formation via interactions between the galaxy and the intracluster medium or gravitational potential, or by galaxy-galaxy interactions which are frequent in the dense clusters. Quenching and morphological transformations are also observed to be correlated with galaxy mass, with more massive galaxies being more passive and more likely to possess ETG morphologies \citep{10peng}. In this case, quenching and morphology transformations are consequences of mass accretion and the formation of a bulge.

These studies have been extended to higher redshifts using field surveys. These works find that passive fractions are higher for massive galaxies up to $z\sim 2-3$ \citep{13ilbert, 13muzzin,14tom}.
\citet{15darvish,16darvish,18darvish} have studied galaxy properties as a function of environment in the COSMOS field up to $z \sim 3$. They found that the passive galaxy fraction depends on the environment at $z \lesssim1$ and on stellar mass out to $z\sim 3$. They suggested that at $z\sim1$ the denser environment could quench massive galaxies in a more efficient way because of the higher merger rate of massive galaxies in denser environments. In their field sample, less massive galaxies ($log_{10} \frac{M_*}{M_{\odot}} \lesssim 10.5-11$) are quenching, and more massive galaxies show star-bursts. 

With the CARLA cluster galaxies we find that both ETG and passive fractions depend on environment and only mildly on galaxy stellar mass. The explanation for this discrepancy is that we probe a very different environment to the earlier field studies. COSMOS covers an area of only few degrees and does not host a  significant cluster population. Most COSMOS galaxies are hosted in environments with a projected  galaxy surface density of $\Sigma \lesssim 300$. 

At $\Sigma \lesssim 300$, the CARLA clusters also exhibit ETG and passive fractions that are similar to the field. Thus our results are in complete agreement with the results of \citet{15darvish,16darvish,18darvish}. We only observe a significant enhancement in the ETG and passive fractions at $\Sigma > 700$.  We therefore should be wary of studies that look for environmental trends without exploring cluster environments.

One study that explored dense environments at high redshift is the  Observations of Redshift Evolution in Large-Scale Environments (ORELSE; \citealp{09lubin}; \citealp{17tom}). This collaboration studied a large range of environments, including groups around clusters at a scale of $\sim10-15 h^{-1}$ proper Mpc in the redshift range $0.55 < z < 1.4$. Using this sample, \citet{19lemaux} found that the average passive fraction is $\sim 10\%, \sim 40\%$ and $\sim 60\%$ in three environments: lower galaxy surface density $\log(1+\delta)<0.3$, intermediate $0.3<\log(1+\delta)<0.7$ and higher $\log(1+\delta)>0.7$, respectively. Their definition of local density  is $\log(1+\delta) \equiv \log \left(1+\frac{\Sigma_{VMC}-\Sigma^{bkg}_{VMC}}{\Sigma^{bkg}_{VMC}} \right)$, where $\Sigma_{VMC}$ is calculated with a Voronoi Monte-Carlo (VMC) method, and $\Sigma^{bkg}_{VMC}$ is the background local density. 

The three surface density bins considered in ORELSE  roughly correspond to our $\Sigma_N<140  {\rm \ gal \ Mpc^{-2}}$ ($\log(1+\delta)<0.3$),  $140<\Sigma_N<350  {\rm \ gal \ Mpc^{-2}}$ ($0.3<\log(1+\delta)<0.7$) and  $\Sigma_N>350  {\rm \ gal \ Mpc^{-2}}$ ($\log(1+\delta)>0.7$). To derive this correspondence, we have calculated our background surface density $\Sigma^{bkg}_N=70\pm30  {\rm \ gal \ Mpc^{-2}}$ from the {\it combined CANDELS catalog}, and have relied on results from \citet{15darvish}, that have shown that the density fields measured with $\Sigma_{VMC}$ and $\Sigma_N$ are in good agreement. Using this conversion, we see that our results for the passive fraction of $z>1.3$ cluster galaxies agrees with ORELSE results at $z<1.4$. The CARLA sample extends the relation between local density and SFR \& morphology up to $z=2$.

It is interesting that we find average passive fractions that are consistent with those found in \citet{19lemaux}, showing a lack of significant passive fraction redshift evolution at fixed environment from the ORELSE survey ($0.55 < z < 1.4$) to our passive galaxy sample ($1.3 < z < 2$).  

 \citet{20burg}  studied passive galaxy fractions in eleven clusters from the GOGREEN (Galaxies in Rich Early Environments; \citealp{17balogh}) survey at $ 1<z<1.4$. They find that the passive fraction strongly depends on galaxy mass, ranging from $\sim 0.3$ to $\sim 1$ at masses $\log_{10} \frac{M_*}{M_{\odot}} \sim 9.5$ and $\log_{10} \frac{M_*}{M_{\odot}}\sim 11.5$, respectively. A similar results is observed  in the ZFOURGE survey \citep{Kawi2017}, and in galaxy groups at $0.1\lesssim z \lesssim 2.3$ detected in deep near-infrared surveys by \citet{21sarron}. Furthermore, in a sample of five clusters at $1.4<z<1.7$ identified in the South Pole Telescope Sunyaev Zel'dovich effect (SPT-SZ) survey, \citet{19stra} found that the passive galaxy fraction depends on both mass and environment with efficient suppression of star formation occurring at $z \gtrsim 1.5$.
 
 In contrast to these works, the CARLA sample does not exhibit a strong stellar mass dependency for either the ETG fraction or the passive galaxy fraction. We stress that the uncertainties on our results on the mass-dependency are large and our lack of evidence of a dependency is not in disagreement with previous works within uncertainties. Our method, using statistical background subtraction, and estimating stellar masses using only IRAC1 photometry means that our mass estimates are more uncertain than the works mentioned above.

\subsection{The cause of morphological transformations in high-redshift clusters}

Morphological transformations of galaxies can occur through multiple routes. Rapid mass accretion resulting in a star-burst can form a bulge and transform LTGs into ETGs. Another route to forming an ETG morphology is through mergers and galaxy-galaxy interactions. In fact, ETG quenching can happen in different ways and at different epochs. While spheroids are formed as a consequence of star-bursts and dissipative collapse, they can also form from gas-rich disk mergers that do not cause  star-bursts \citeg{10mo}. If these galaxies do not accrete further gas, the spheroids resulting from the merger will become ETGs. If these merger remnants accrete substantial amounts of gas, then a disk will reform and the galaxy will present an LTG morphology. 

In our sample, the passive galaxy fraction increases in denser local environments, and we do not observe large fractions of star forming galaxies in dense environments. We also do not observe enhanced starburst activity in the cluster cores.  In fact, excluding the AGN, we only find 9 disk galaxies and 1 spheroid at $z<$1.5 that have sSFR 3$\sigma$ higher than the main sequence (log(sSFR)$\gtrsim -8.1 \ {\rm [yr^{-1}]}$ at $z<$1.5, and log(sSFR)$\gtrsim -7.8 \ {\rm [yr^{-1}]}$ at $1.5<z<$2), and only one visually compact galaxy at $z>$1.5 in CARLAJ1129+0951, that does not show interactions, tidal tails or asymmetry. On the other hand, three disks show interaction (2 also show tidal tails), and 8 are asymmetric, indicating a possible connection between recent merger and starburst activity. 

The lack of enhanced star-formation and star-burst activity in the CARLA sample suggests that morphological transformations are not caused primarily by rapid mass accretion that results in a starburst and formation of a bulge. Or alternatively, these starbursts have already occurred and caused the morphological transformations, or they are heavily dust-obscured and therefore not detectable in our optical/near-infrared analysis. However, an analysis of the AllWISE\footnote{http://wise2.ipac.caltech.edu/docs/release/allwise/} catalog sources in the area covered by our Spitzer observations shows that only our radio sources are detected in the 22$\mu$m channel, which is the Wide-field Infrared Survey Explorer (WISE, \citealp{Wright10}) W4 channel. This  also indicated the absence of sources with high star-formation rates. In fact, this channel is closer to the peak of the infrared emission by star-forming galaxies and the lack of significant detections means that there are not intense star-forming galaxies in the area \citeg{Hwang12, Chang15, Cluver17}.

On the other hand, previous studies concluded that a high fraction of mergers in clusters at $z>1.5$ would have been necessary to explain their high quiescent fractions   at $z\sim1-1.5$ \citep{15darvish, 16darvish, 18darvish,19lemaux}. We indeed observe merger fractions that are systematically higher than in the CANDELS fields in the CARLA clusters.  Our average merger fraction is $26\pm3\%$, significantly higher ($\gtrsim 3\sigma$) than the CANDELS average merger fraction of $1.7\pm0.5$.

At first look this seems to point towards galaxy interactions and mergers acting as the primary cause of transforming LTG into ETG. However, whilst the ETG and passive galaxy fractions strongly increase with local environment, the merger galaxy fraction is approximately constant across all local environments. In fact, the galaxy merger fraction in the CARLA clusters does not correlate with local or global environment, redshift or galaxy stellar mass. Therefore, the merger rate is enhanced to the same level in the low-density cluster environments (where the passive and ETG fractions are low) as in the high-density cluster environments (where the passive and ETG fractions are high). So mergers {\it alone} do not seem to be responsible for the morphological transformations and quenching. 

One possibility solution that is consistent with the data from the CARLA galaxy morphologies is the theory of that the LTG or ETG morphology results from the amount of gas accreted after a merger \citeg{Zavala2019,Harshan2021}. Let us take the merge fraction data at face value and presume that mergers occur at the same rate in all regions of the high redshift clusters. In the low-density regions, new disks may be able to re-form around these merger remnants through the inflow of substantial amounts of pristine gas through cold streams. However, if cold streams are disrupted in the densest regions of the clusters (as suggested by \citealt{09Dekel}), then the merger remnants in the densest regions will become ETGs. Such a scenario would account for our observations of a strong morphology-local density and SFR-local density relation within clusters, and also with a high merger rate in all cluster environments.

\subsection{Active ETG}

We find higher percentages of active ETG in CARLA clusters compared to the field, mainly within the highest redshift range of our sample. In fact, CARLA active ETG are $21 \pm 6$\% and $59 \pm 14\% $  of the total ETG population at 1.35 < z <1.65 and  1.65 < z < 2.05, respectively, which are $ \gtrsim 2.5 \sigma$ higher than the corresponding values of $6 \pm 5 \%$ and the $8 \pm 2 \%$  found in our field {\it TPHOT-CANDELS catalogue} and 'CANDELS combined catalog' for galaxies selected in the range $1.3<z<3$, after applying our CARLA cluster sample selection criteria. About half of the active ETG are mergers or asymmetric in both redshift bins. \citet{Afanasiev2022} also find that $45\pm18$\% and $42\pm17$\% of the CARLA active ETGs lie within $1\sigma$ of the passive MSR ($64^{+16}_{-20}$\%  and $58\pm17$\% for 2.5$\sigma$) at $1.35<z<1.65$ and $1.65<z<2.05$, respectively.

The presence of active ETGs in clusters has been observed at $z\sim 0$ \citeg{Sheen2016} and up to $z\sim 2$ \citeg{Ferreras2000,06meia,15mei,Jaffe2011,Mansheim2017}. 
These galaxies could be evidence that morphological transformation occurs before quenching \citeg{Barro13, Barro14}, but they could also be remnants of recent events that
triggered the star formation, such as mergers or feedback \citeg{Martig2009,Bournaud11,Kaviraj2011, Kaviraj2013}.

In the local Universe ($z<1$), active ETGs show
evidence for recent gas-rich minor merger events or interactions 
with neighbouring gas-rich galaxies, and they are thought to become 
quiescent when the gas acquired during the merger, and
which fuels star formation, has been exhausted \citeg{Lee2006, Huertas2010,George2015,George2017}. If this is true also at the higher redshifts that we
probe, at least part of these galaxies have most probably gone through a recent
merger or neighbour galaxy interaction, and would most probably quench 
at a later epoch, thereby increasing the fraction of
passive ETGs in the cluster population.

The larger percentage of blue ETG in our higher redshift clusters could point to an increase of 
mergers or neighbour galaxy interactions at high redshift; however we do not observe a significant increase in the fraction of mergers in the highest redshift bin that
we probe (with respect to the lower redshift bin), even if the merger fractions in our cluster sample are higher than in the field (see sec. 3.4). Observations at higher redshift are needed to test this hypothesis, as much as a larger statistical sample to assess if this result is universal.




\section{SUMMARY} \label{sec:summary}

We perform an in-depth study of galaxy population, morphology and quenching in the CARLA cluster sample \citep{18noirot}. The CARLA survey is a unique sample of spectroscopically confirmed clusters and proto-clusters at $1.3<z<2.8$, with HST, Spitzer and ground-based imaging and HST grism spectroscopy.  Our cluster total stellar mass spans the range $11.3 \lesssim$log$_{10} (\frac{M^c_*}{M_{\odot}}) \lesssim 12.6$, which corresponds to an approximative halo mass range $13.6 \lesssim$log$_{10} (\frac{M^c_h}{M_{\odot}}) \lesssim 14.6$.

\vskip 0.3truecm

This sample permits us to extend morphology and quenching studies in clusters at $1.3<z<2.8$, and study them as a function of local and global environment, and cluster redshift and density contrast. Our main results are:

\begin{itemize}

\item The morphology-density and passive density relations are already in place at $z\sim2$. The cluster ETG and passive fractions depend on local environment and mildly on galaxy mass. They do not depend on global environment (defined as the cluster core total stellar mass). This points to local environment as the main driver of morphological transformations and quenching at $1.3 \lesssim z \lesssim 2-3$.

\vskip 0.5truecm

\item At lower local densities, where $\Sigma_N < 700$ gal/Mpc$^2$, the
CARLA clusters exhibit a similar ETG fraction as the field, in contrast to clusters at z = 1 which
ready exhibit higher ETG fractions. This implies that the densest of cluster environments influence the morphology of galaxies first, with lower density  environments either taking longer to influence galaxy morphology or only
influencing galaxy morphology at later cosmological times.

\vskip 0.5truecm

\item The percentages of active ETG in CARLA clusters are higher compared to the field, mainly within the highest redshift range of our sample. We find that CARLA active ETG are $21 \pm 6$\% and $59 \pm 14\% $  of the total ETG population at 1.35 < z <1.65, and  1.65 < z < 2.05, respectively, $ \gtrsim 2.5 \sigma$ higher than the $6 \pm 5 \%$ and the $8 \pm 2 \%$ respectively found in our field {\it TPHOT-CANDELS catalogue} and 'CANDELS combined catalog' for galaxies selected in the range $1.3<z<3$, applying our CARLA cluster sample selection criteria. About half of the blue ETG are mergers or asymmetric in both redshift bins. The large percentages of blue ETG in our higher
redshift clusters could be evidence of an increase of mergers or neighbour galaxy interactions at high redshift. However, even if our merger fractions are higher than in the field at all redshift, we do not observe an increase of merger fraction in the highest redshift bin that
we probe.

\vskip 0.5truecm

\item  Merger fractions in clusters are ten time larger than in the CANDELS fields, as needed in previous studies to explain their high quiescent fractions at lower redshift \citep{15darvish, 16darvish, 18darvish,19lemaux}, and do not depend on environment. One cluster, CARLAJ0800+4029 does not host any ETG and shows a large fractions of disk and irregular mergers in its densest regions. This suggests that morphological transformations did not happen yet, and that ETG formation might result in later epochs from the mergers that we  that we are catching in the act. The constant rate of mergers as a function of environment, and the high fractions of ETG and passive galaxies in the densest local environments support a scenario in which cold gas streams still fuel star formation in merger remnants within the lower density environs, while streams are disrupted in merger remnants in the densest regions, allowing them to  become ETG.

\vskip 0.5truecm

\item  We do not find large fractions of star forming galaxies and star-bursts in dense environments. This suggests that morphological transformations are not caused primarily by rapid mass accretion that results in a starburst and formation of a bulge. This is different from the enhanced star-formation and star-burst activity found in other overdensity samples and consistent with other clusters that present a significant ETG and quenched population at $z \gtrsim 1.5$. These differences are most probably due to different sample selection functions. In fact, some overdensities have been detected using star-forming and star-burst galaxy color selection, others by red galaxy overdensities and X-ray or SZ emission, and might probe different environments that host galaxies with different accretion and pre-processing histories.

\vskip 0.5truecm

\item  We confirm that selecting dense regions around RLAGN is efficient to identify structures consistent with being clusters and proto-clusters. Our clusters and proto-clusters have estimated halo masses consistent with progenitors of local clusters, and three present multiple structures that are predicted to be found in cluster progenitors at $z=1.5-3$ \citep{13chiang,18mul}

\end{itemize}

\begin{acknowledgements}
This work is based on observations made with the NASA/ESA Hubble Space Telescope, obtained at the Space Telescope Science Institute, which is operated by the Association of Universities for Research in Astronomy Inc., under NASA contract NAS 5-26555. These observations are associated with program GO-13740. Support for program GO-13740 was provided by NASA through a grant from the Space Telescope Science Institute, which is operated by the Association of Universities for Research in Astronomy Inc., under NASA contract NAS 5-26555. This work is  based on observations made with the {\it Spitzer} Space Telescope, which is operated by the Jet Propulsion Laboratory, California Institute of Technology, under a contract with NASA. We thank Stefano Andreon, Veronique Buat, Adriano Fontana, Claudia Maraston and Alvio Renzini for useful comments and suggestions. We thank Sandra Faber and Dave Kocevski for helping with CANDELS templates. We thank Universit\'e de Paris, which founded Anton Afanasiev' s Ph.D. research. We thank the anonymous referee for her/his careful reading of the manuscript and useful suggestions that helped to improve the paper. The work of DS was carried out at the Jet Propulsion Laboratory, California Institute of Technology, under a contract with NASA.  GN acknowledges funding support from the Natural Sciences and Engineering Research Council (NSERC) of Canada through a Discovery Grant and Discovery Accelerator Supplement, and from the Canadian Space Agency through grant 18JWST-GTO1. This work was supported by the French Space Agency (CNES). 
\end{acknowledgements} 
\clearpage

%






\begin{appendix}

\section{APPENDIX A. SEXTRACTOR AND TPHOT PARAMETERS USED FOUR OUR PHOTOMETRY} 

\begin{table}[hp]
\caption{SExtractor parameters.}
\label{tab:sexpar}
\centering
\begin{tabular}{lcccccc}
\hline\hline
SExtractor & Cold Mode & Hot Mode \\
\hline
DETECT\_MINAREA & 5.0 & 10.0 \\
DETECT\_THRESH & 0.75 & 0.7 \\ 
ANALYSIS\_THRESH & 5.0 & 0.8 \\
FILTER\_NAME & tophat\_9.0\_9x9.conv & gauss\_4.0\_7x7.conv \\
DEBLEND\_NTHRESH & 16 & 64 \\
DEBLEND\_MINCONT & 0.0001 & 0.001 \\
BACK\_SIZE & 256 & 128 \\
BACK\_FILTERSIZE & 9 & 5 \\ 
BACKPHOTO\_THICK & 100 & 48 \\
\hline
\end{tabular}
\end{table}

\begin{table}[hp]
\caption{TPHOT parameters.}
\label{tab:tphotpar}
\centering
\begin{tabular}{lcccccc}
\hline\hline 
\multirow{2}{*}{Pipeline} & 1st pass & priors, convolve, fit, diags, dance \\
& 2nd pass & convolve, fit, diags, archive \\ 
\hline 
\multirow{3}{*}{Priors stage} & usereal & true \\
& usemodels & false \\
& useunresolved & false \\ 
\hline 
Convolution stage & FFTconv & true \\ 
\hline 
\multirow{7}{*}{Fitting stage} & fitting & coo \\
& cellmask & true \\
& maskfloor & 1e-9 \\
& fitbackground & false \\
& threshold & 0.0 \\
& linsyssolver & lu \\
& clip & true \\
\hline
\end{tabular}
\end{table}
\clearpage

\section{OVERDENSITIES AND SPATIAL DISTRIBUTION OF SELECTED GALAXIES} \label{overden}

\begin{figure*}[h!]
\center
\includegraphics[width=0.32\textwidth]{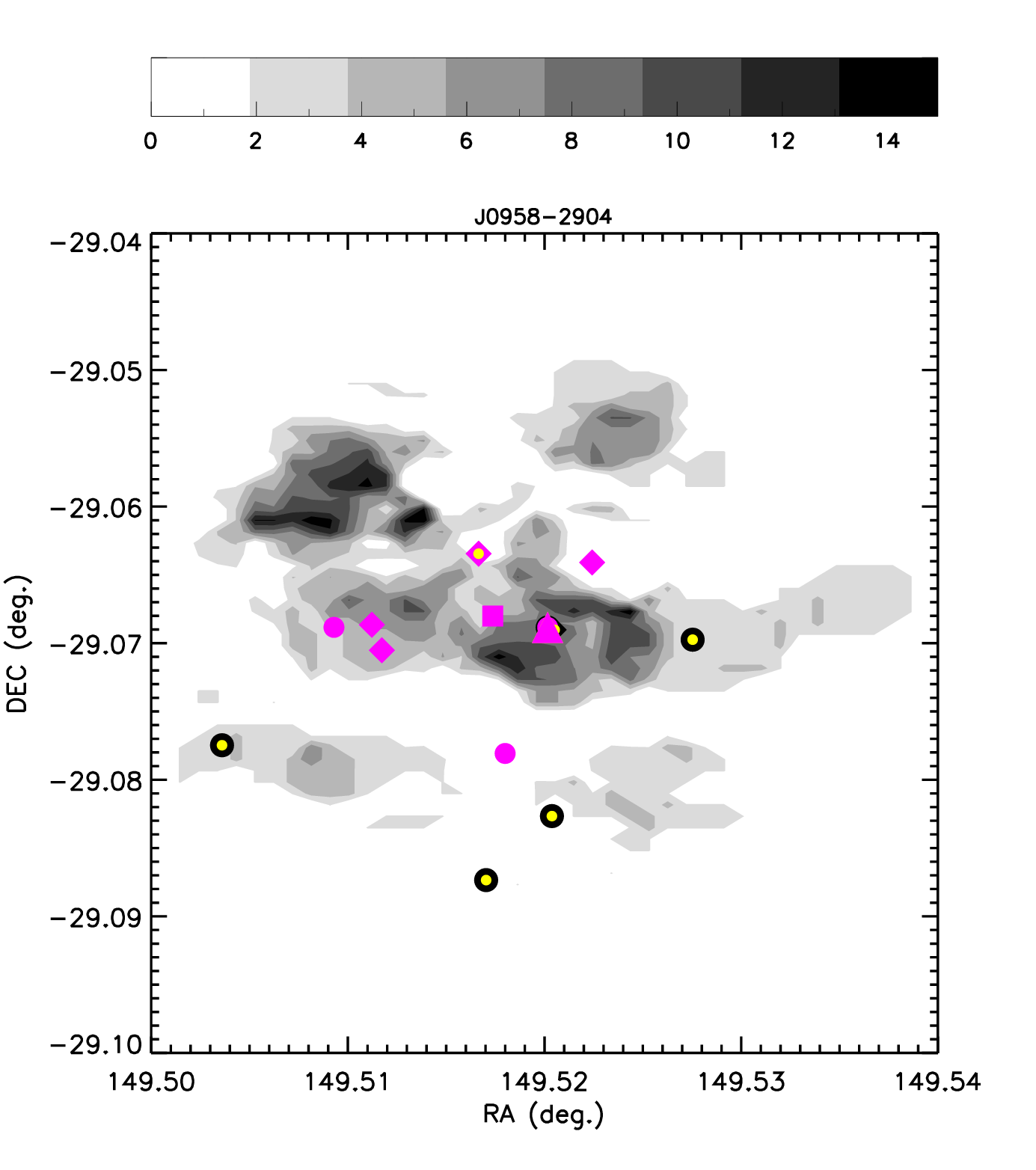}
\includegraphics[width=0.32\textwidth]{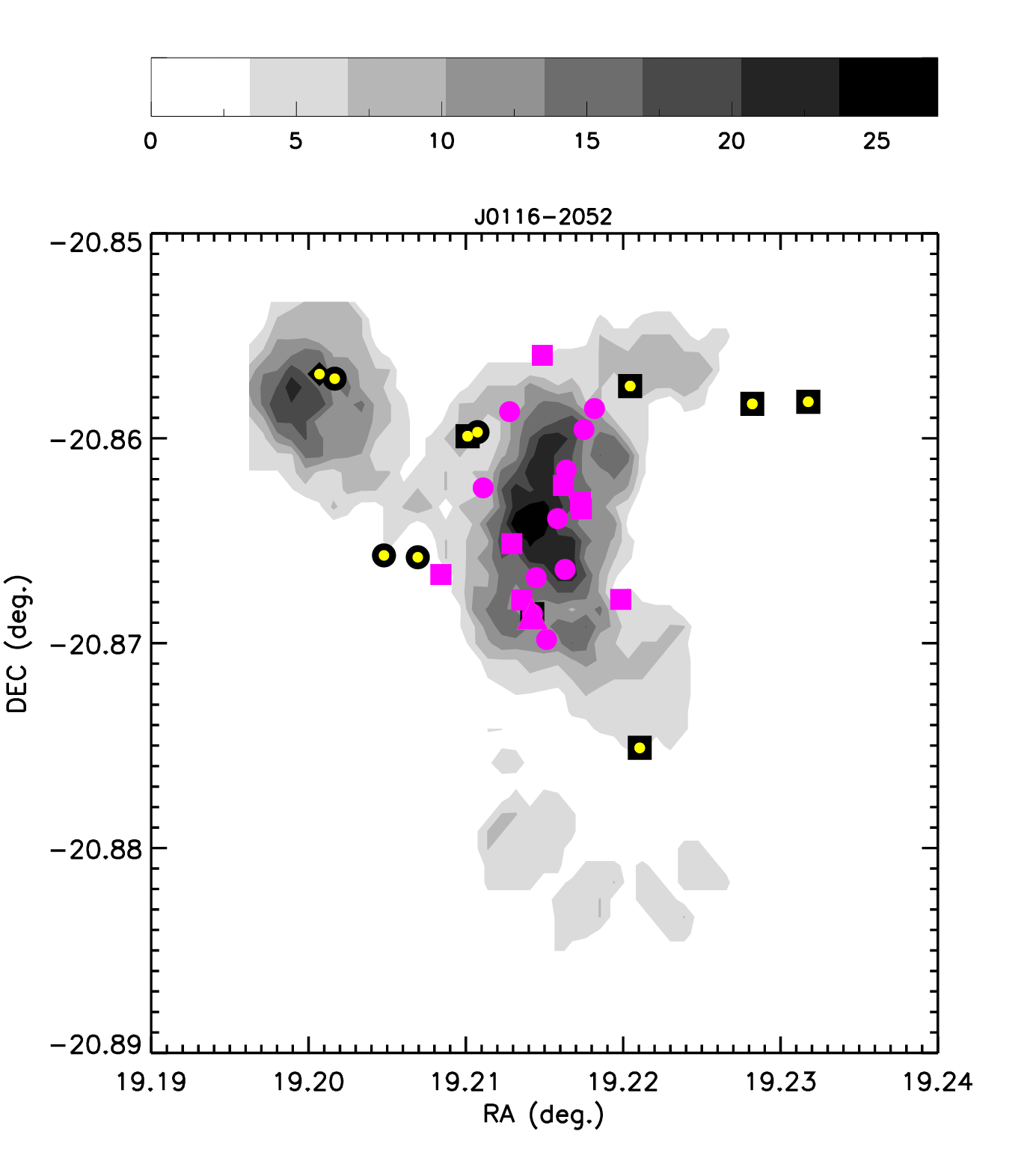}
\includegraphics[width=0.32\textwidth]{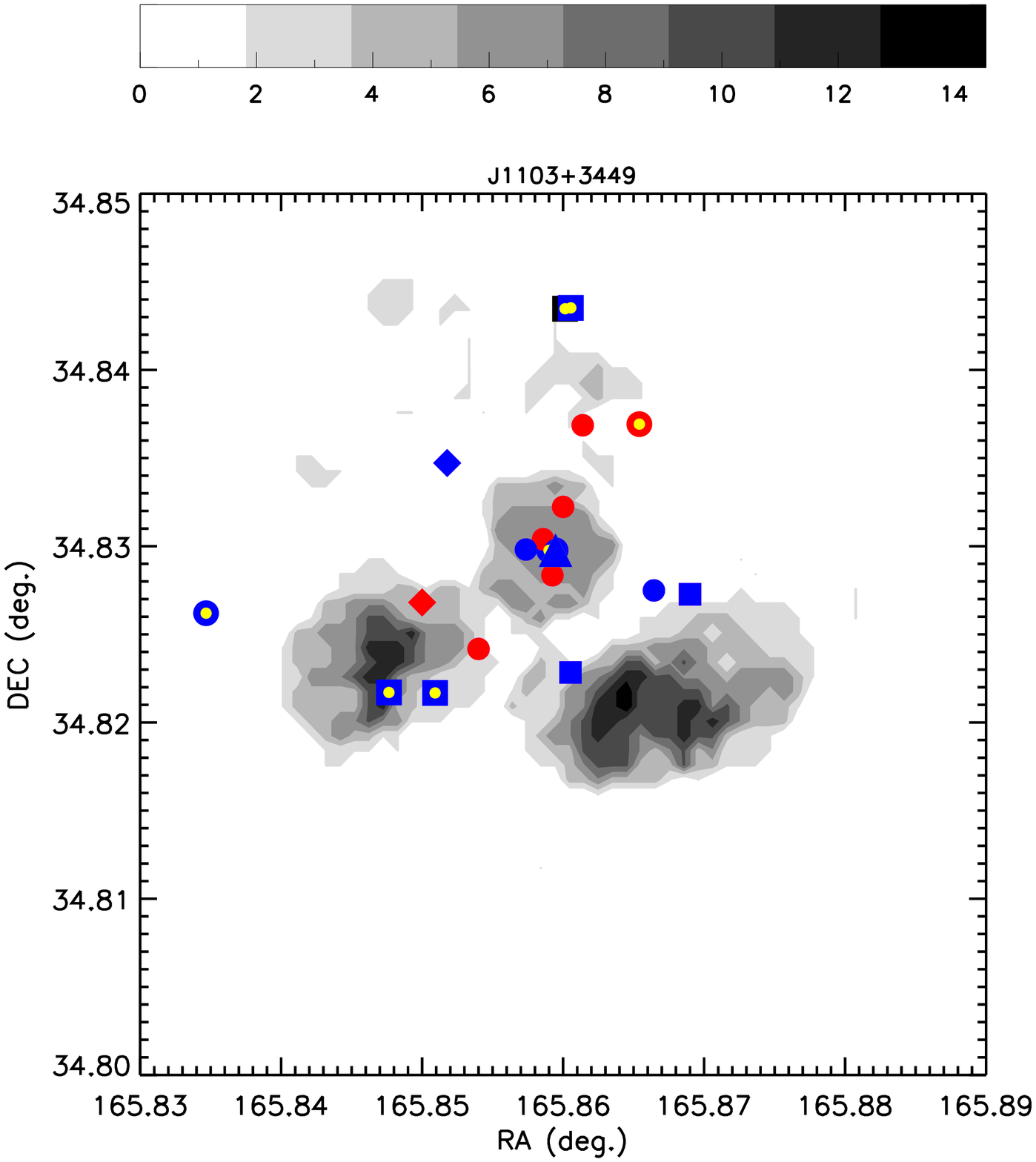}
\vskip 0.cm
\includegraphics[width=0.32\textwidth]{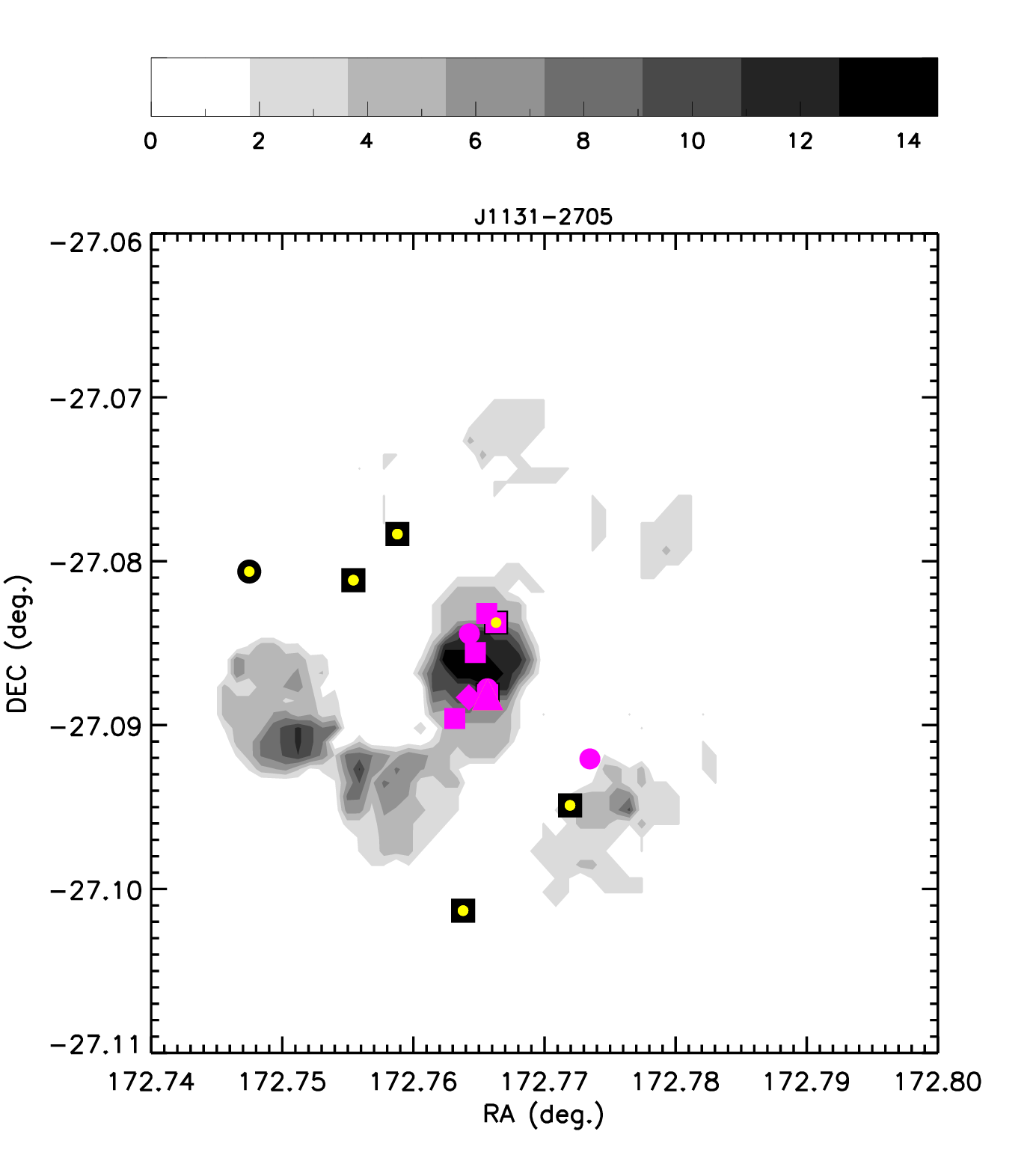}
\includegraphics[width=0.32\textwidth]{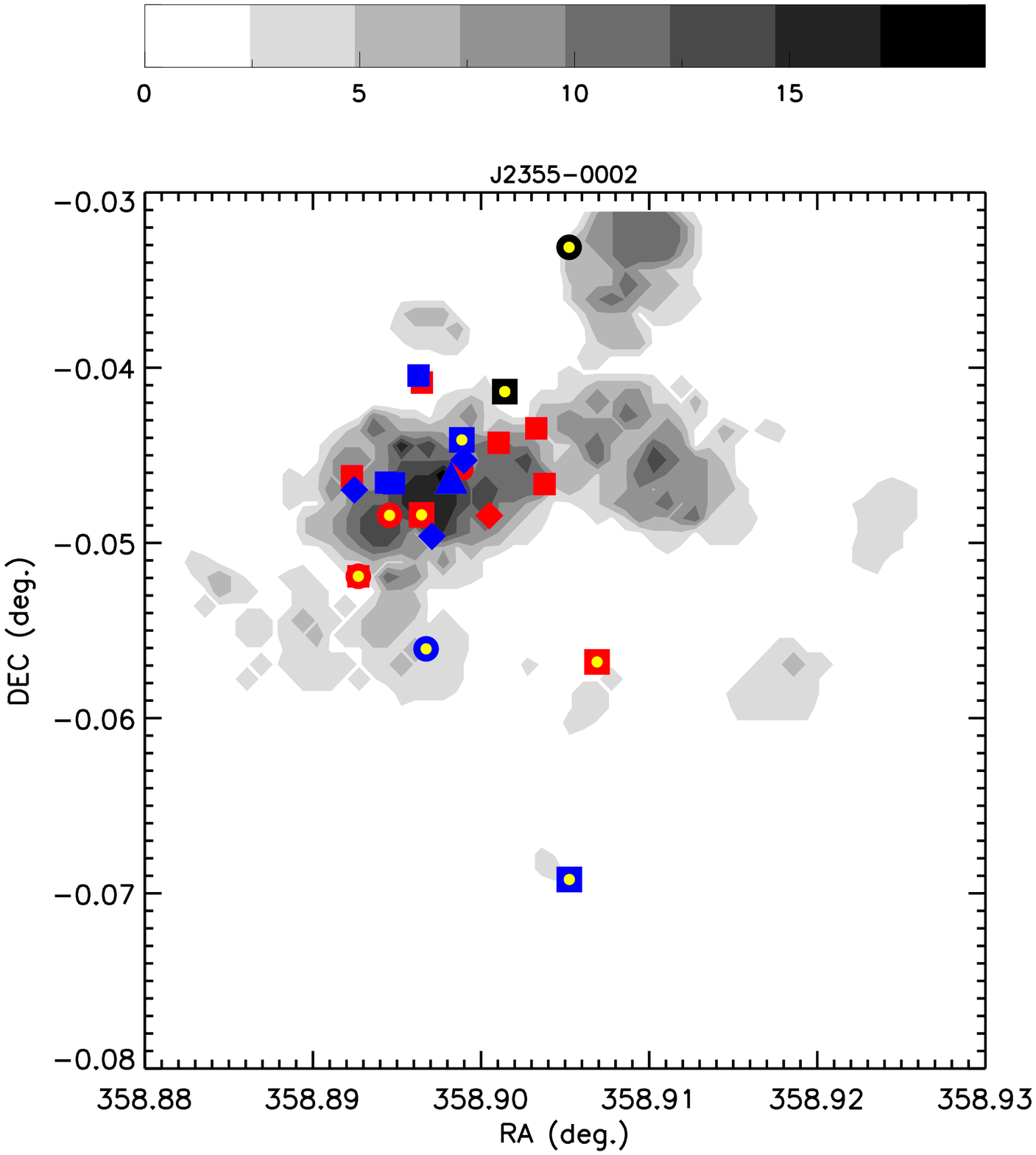}
\includegraphics[width=0.32\textwidth]{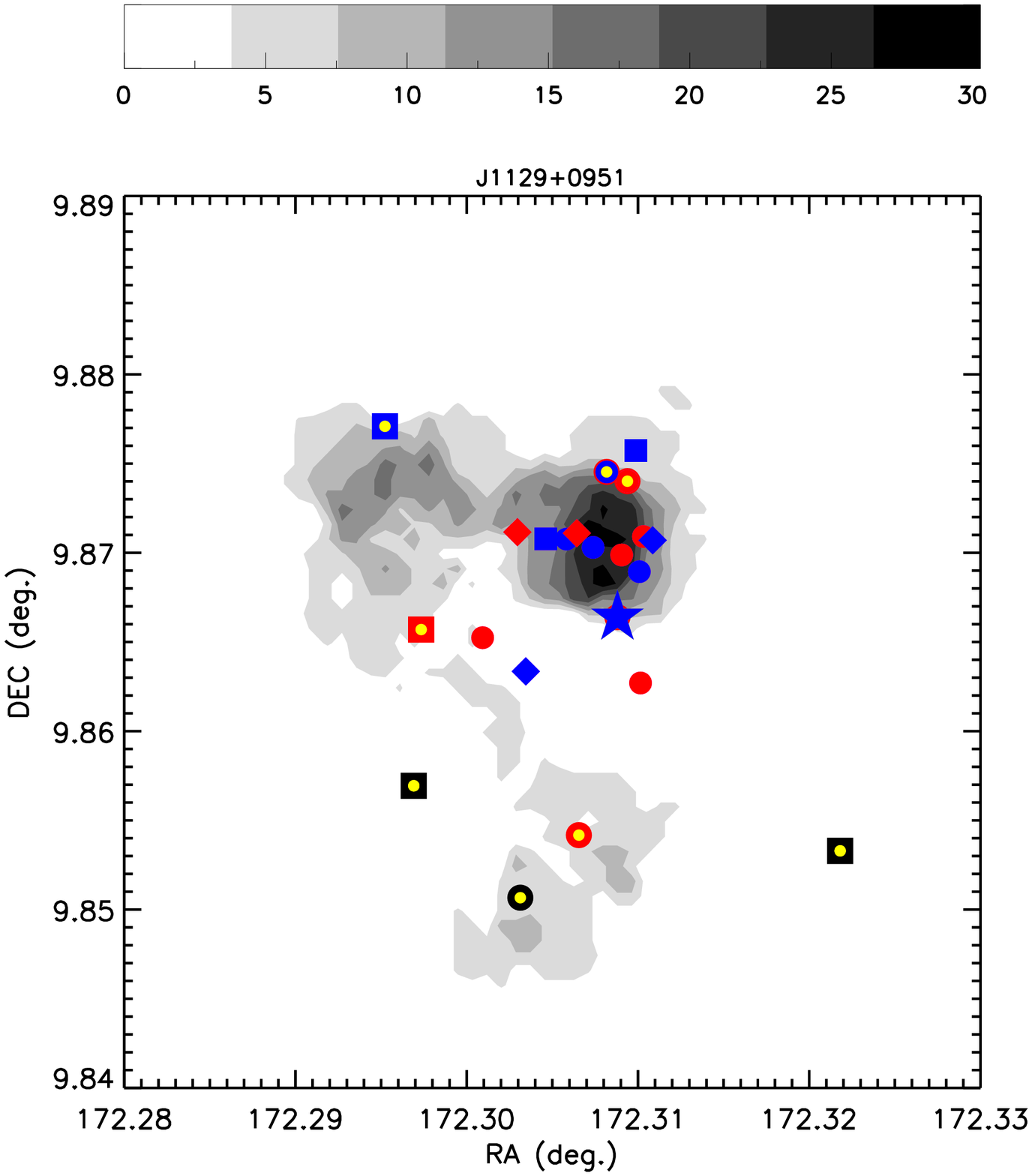}
\caption{Selected high density regions for CARLAJ0958-2904 at z=1.392, CARLAJ0116-2052 at z=1.425, CARLAJ1103+3449 at z=1.444, CARLAJ1131-2705 at z=1.446, CARLAJ2355-0002 at z=1.49.  and CARLAJ1129+0951 at z=1.528. Passive and active galaxies are shown in red and blue, respectively. We use pink when we cannot separate passive and active galaxies, either because the cluster was not observed in four bandpasses or because the galaxy was not detected in one of the four bandpasses. ETG are shown as circles. LTG are shown as squares (disks) and diamonds (irregulars). The cluster spectroscopic members are marked by a small yellow circle, when they are black they are not included in our magnitude, color and spatial selection but are shown to better explain our overdensity selection.  The RLQ and HzRG are shown as stars and triangles, respectively.}\label{fig:c1}
\end{figure*}

\begin{figure*}
\center

\includegraphics[width=0.32\textwidth]{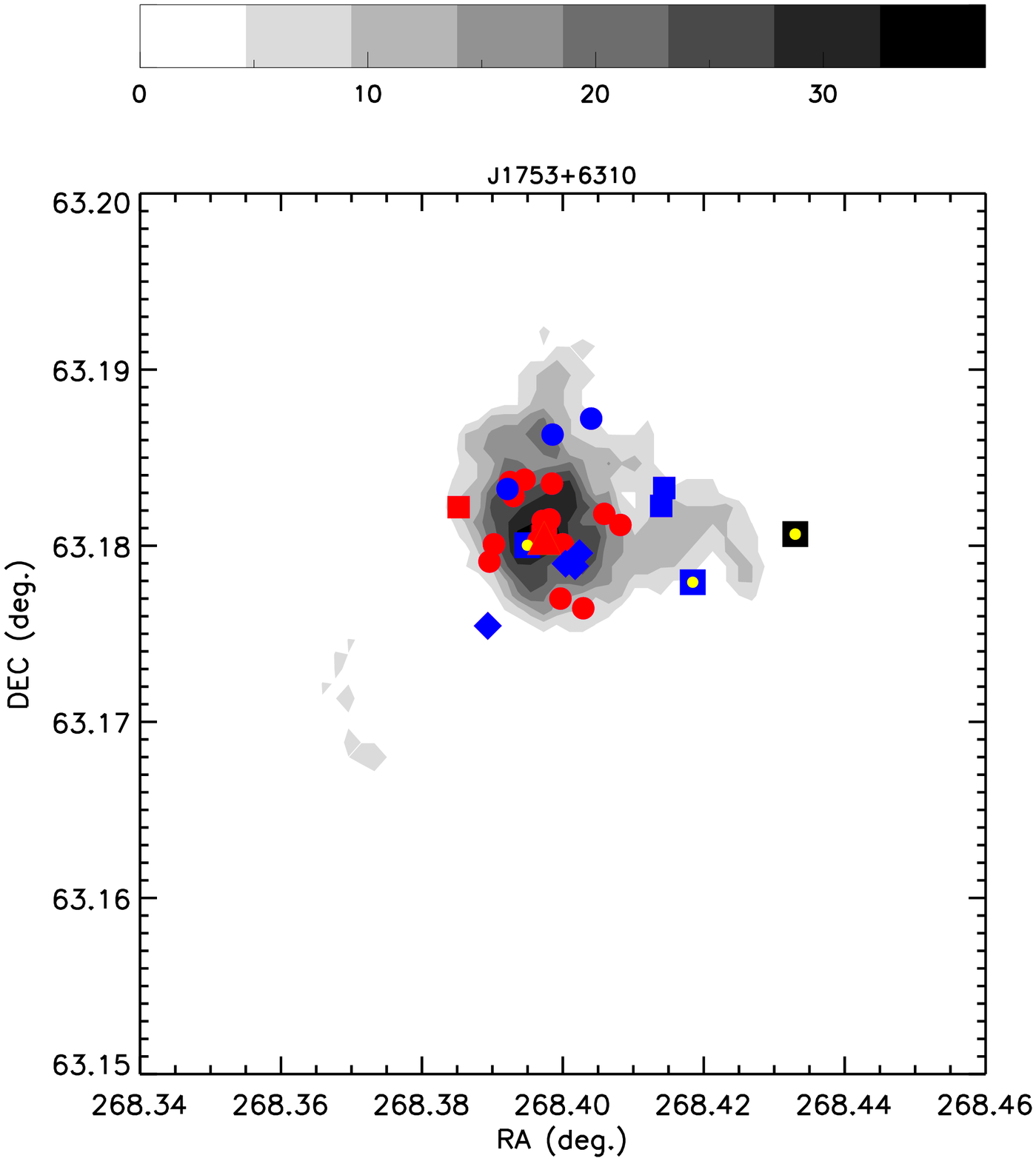}
\includegraphics[width=0.32\textwidth]{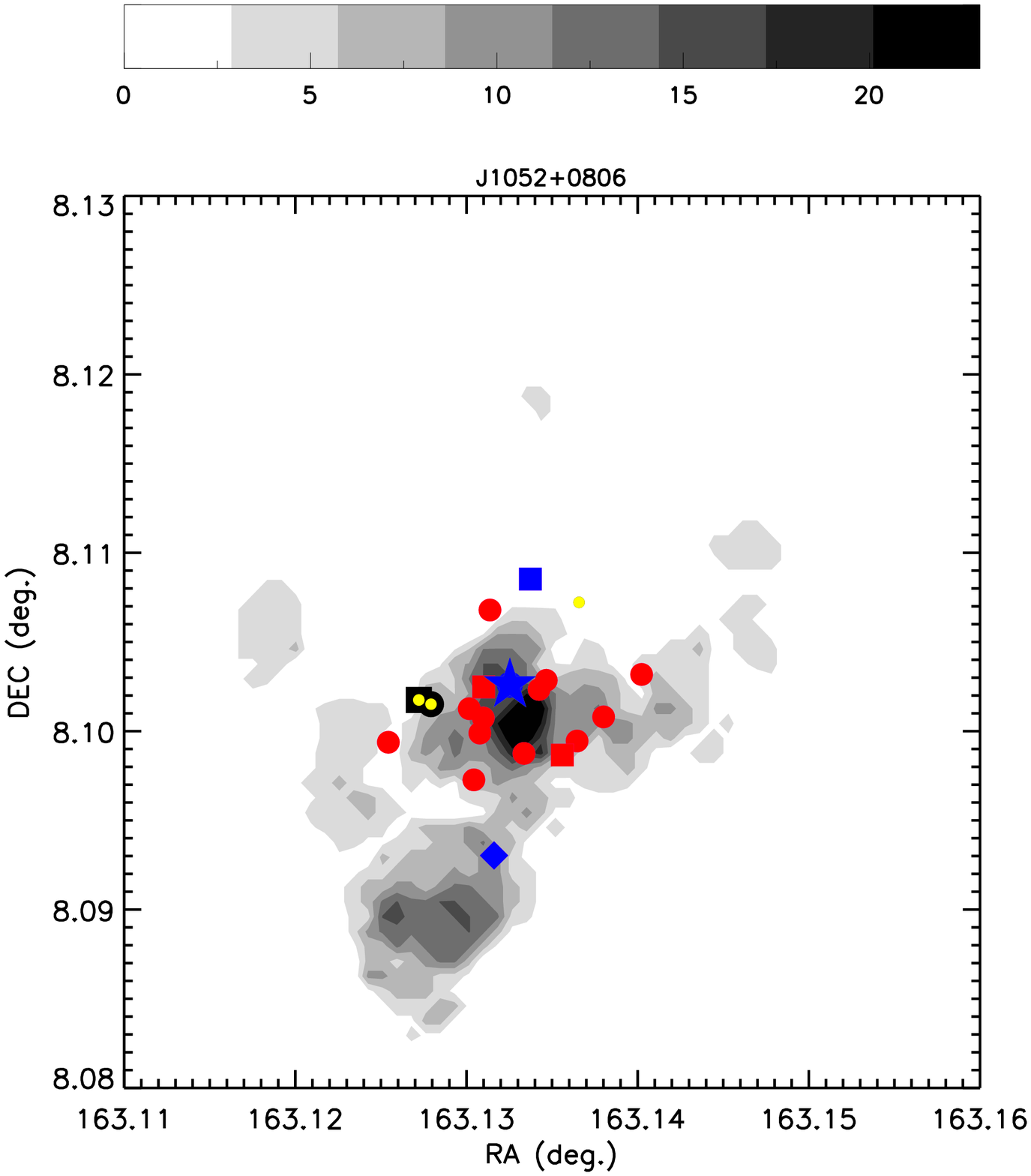}
\includegraphics[width=0.32\textwidth]{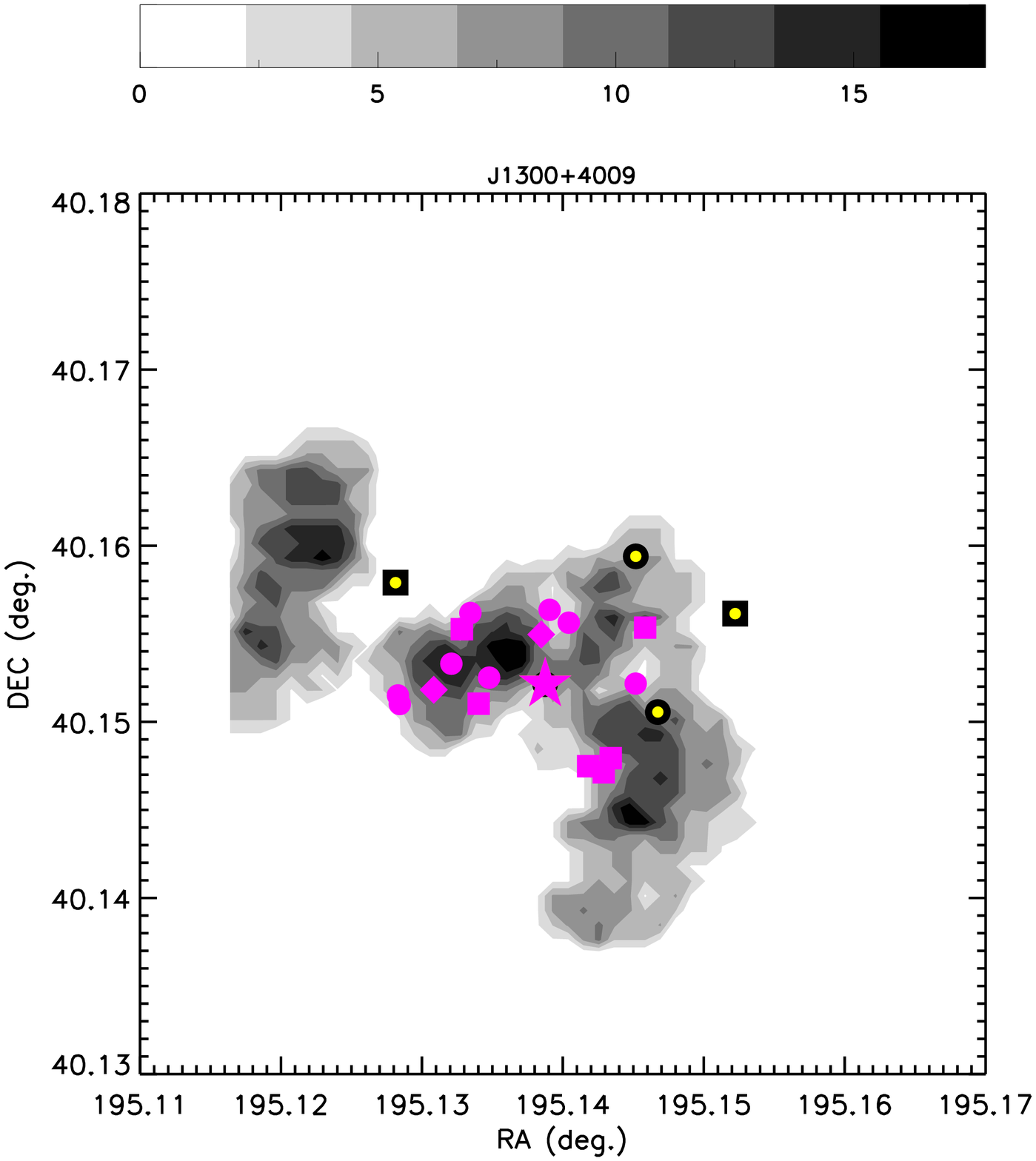}

\includegraphics[width=0.32\textwidth]{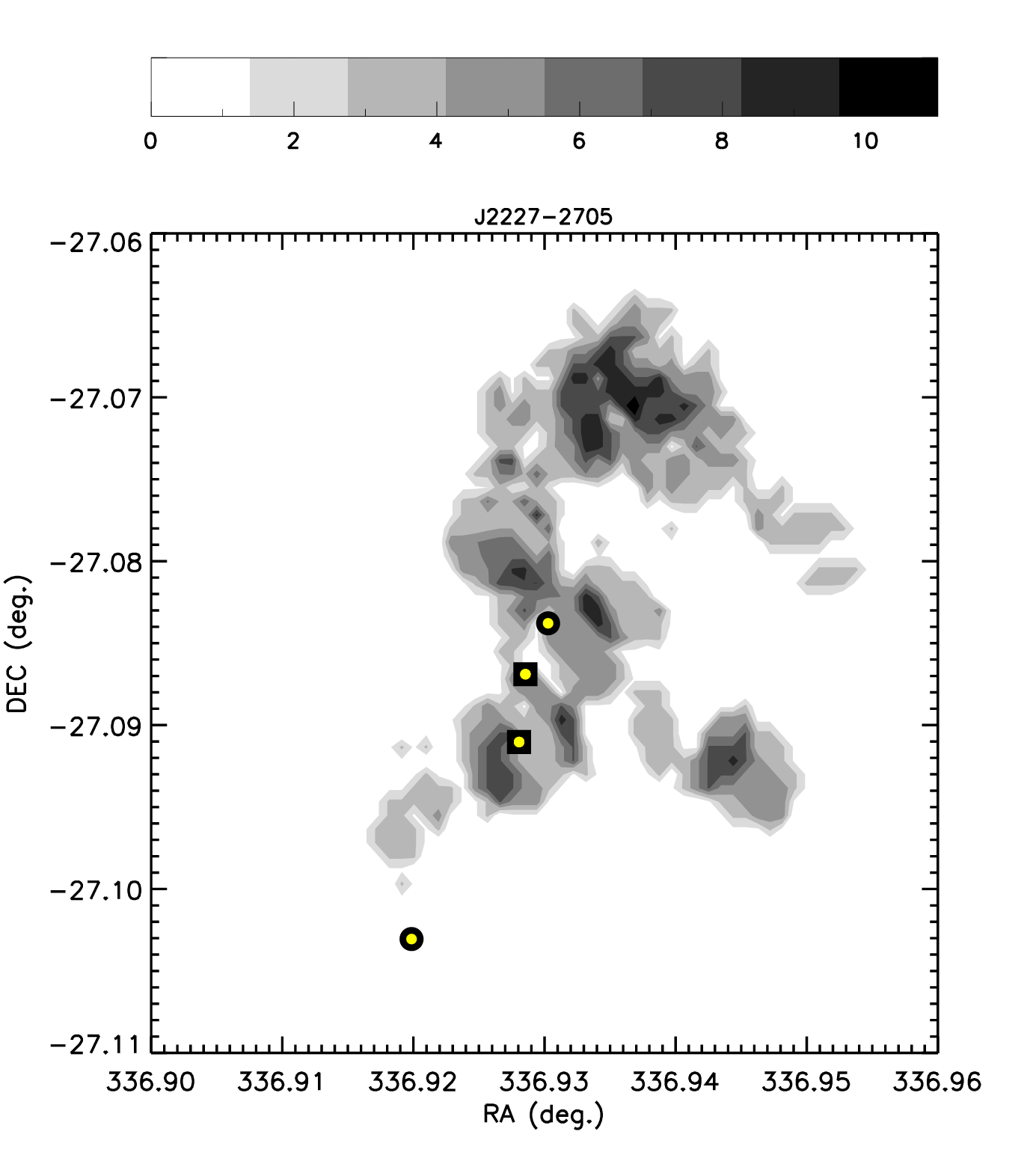}
\includegraphics[width=0.32\textwidth]{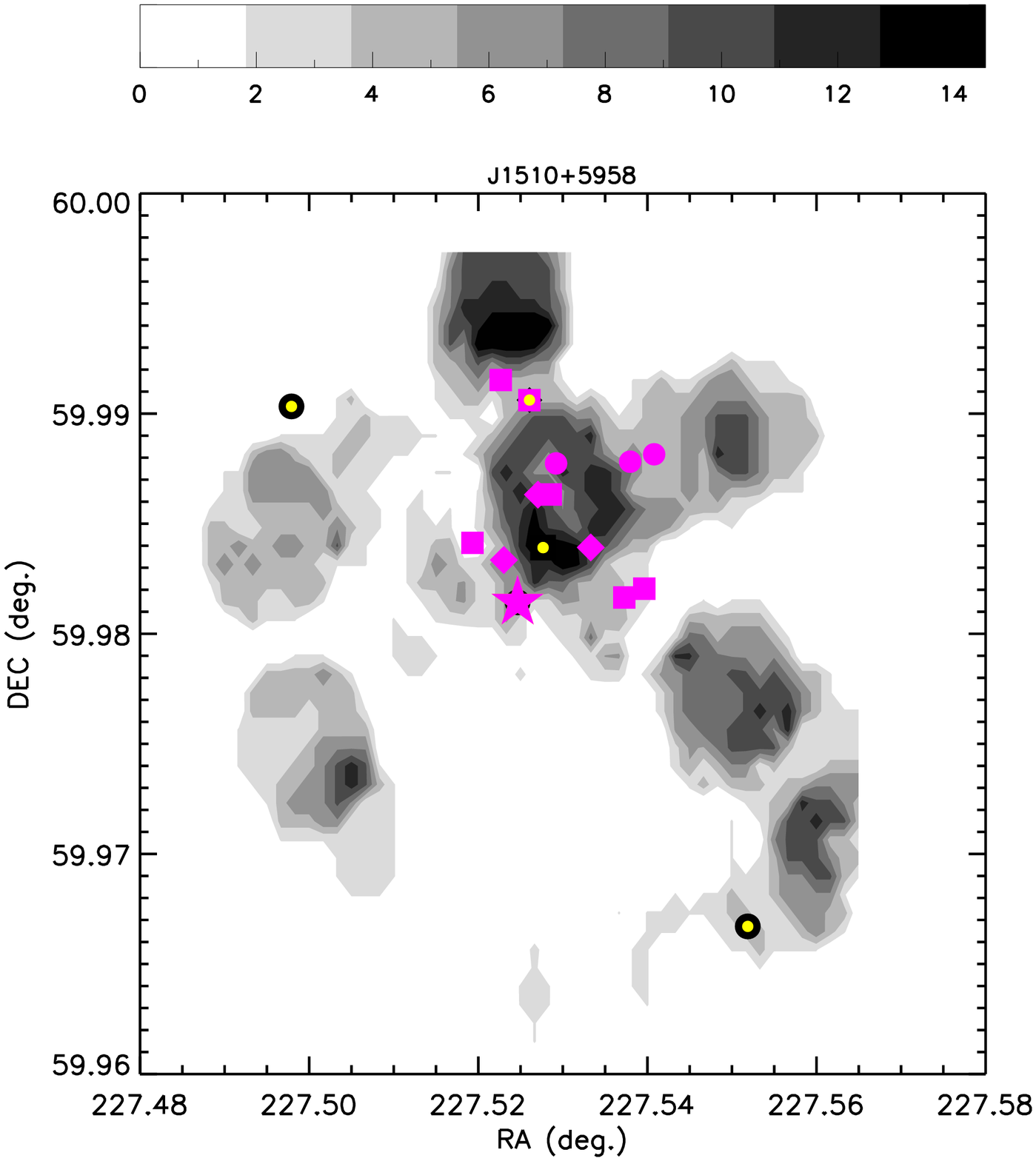}
\includegraphics[width=0.32\textwidth]{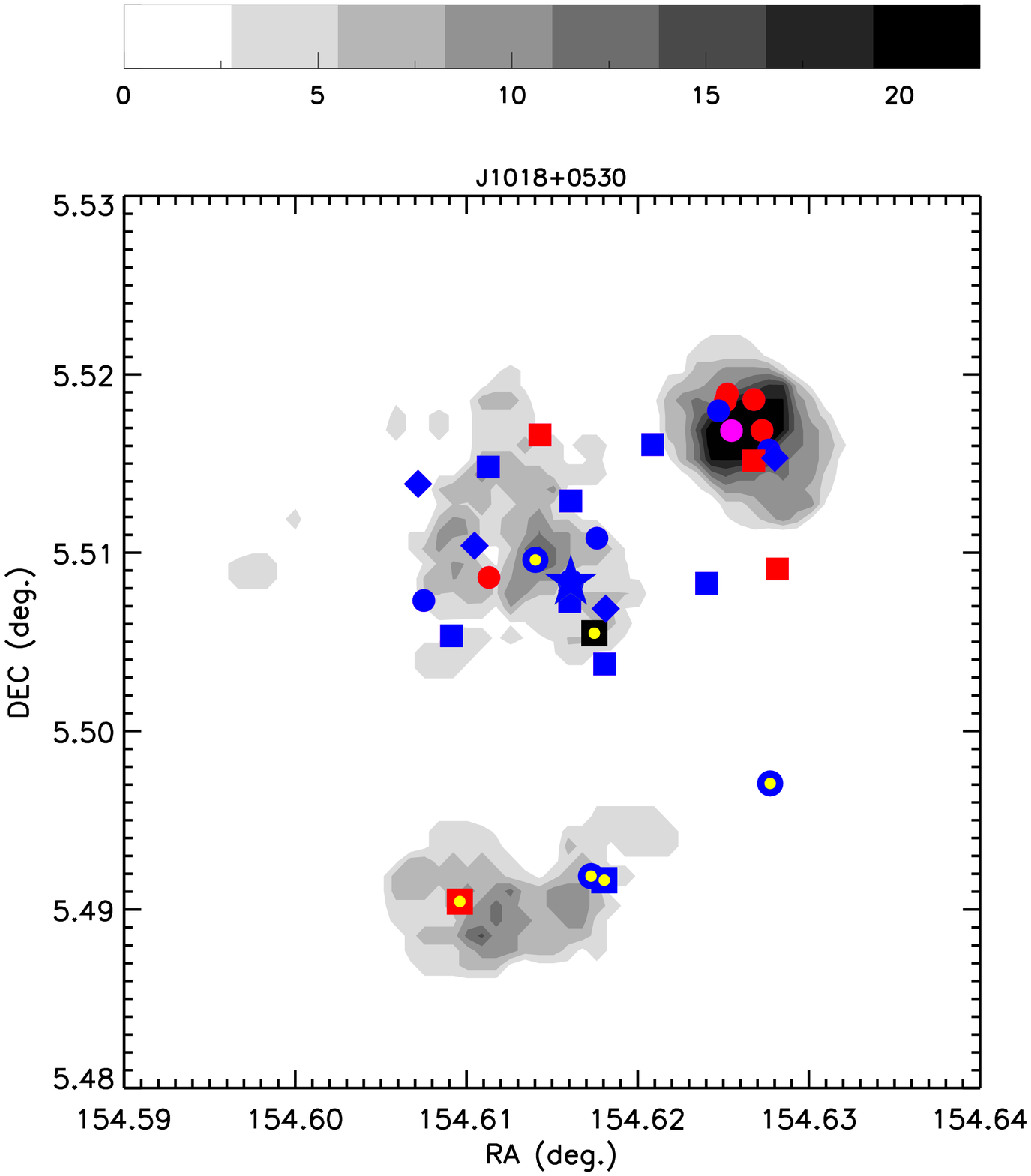}
\caption{Selected high density regions for CARLAJ1753+6310 at z=1.582, CARLAJ1052+0806 densitymap at z=1.646, CARLAJ1300+4009 at z=1.675, CARLAJ2227-2705 at z=1.692, CARLAJ1510+5958 at z=1.725, and CARLAJ1018+0530 at z=1.952 (two selected regions). The symbols are the same as in Fig.~\ref{fig:c1}.}\label{fig:c2}
\end{figure*}

\begin{figure*}
\center

\includegraphics[width=0.32\textwidth]{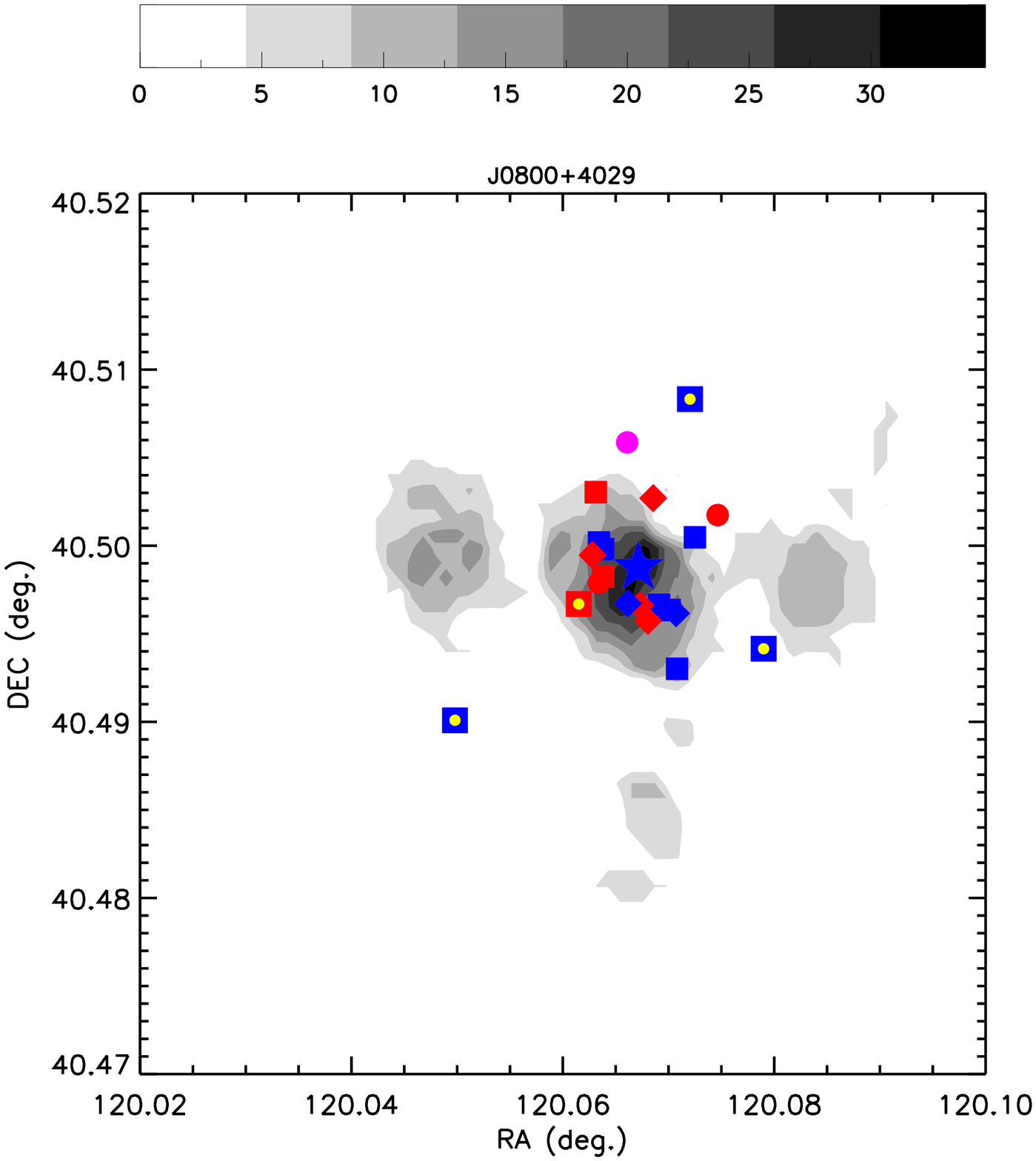}
\includegraphics[width=0.32\textwidth]{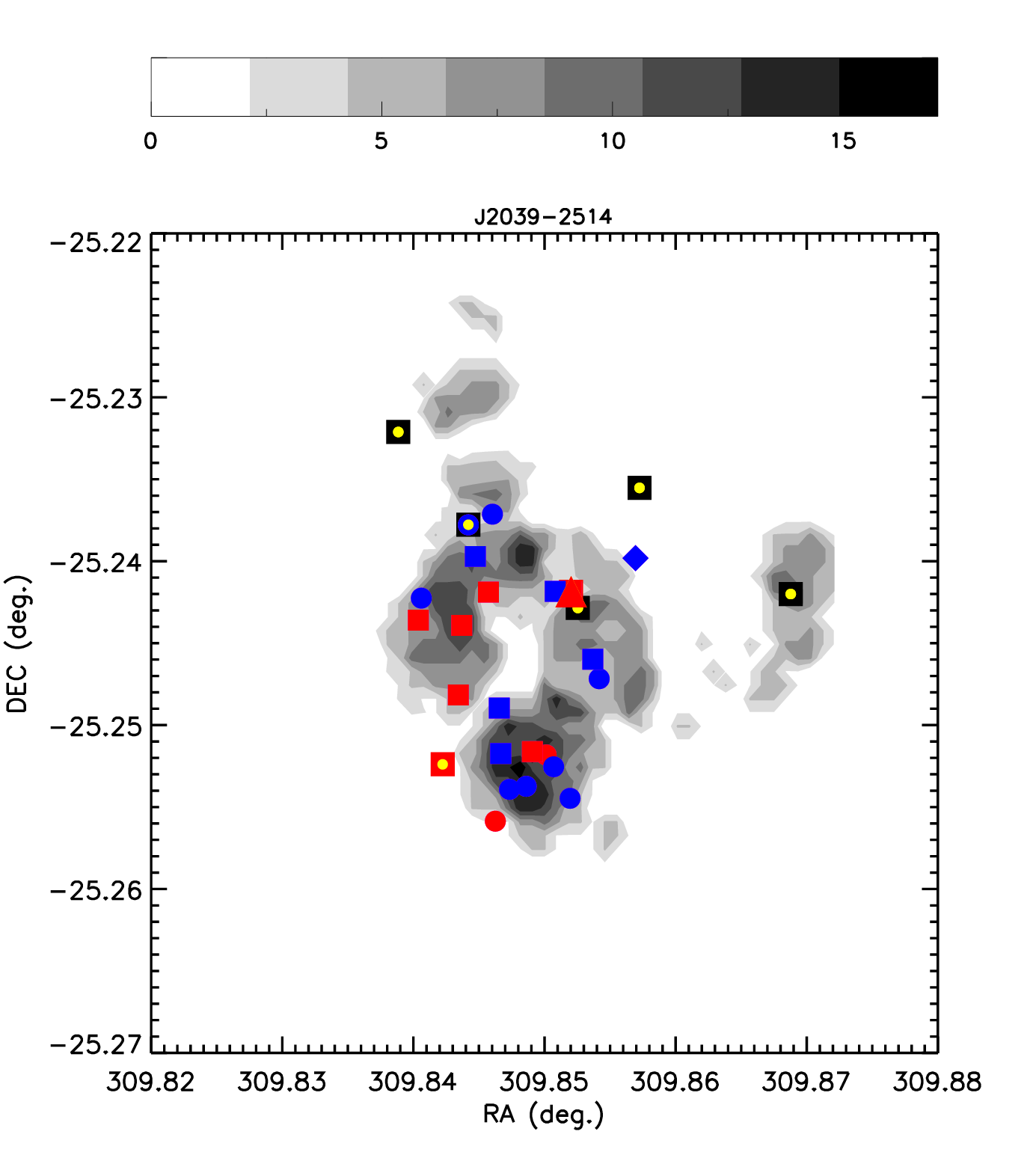}
\includegraphics[width=0.32\textwidth]{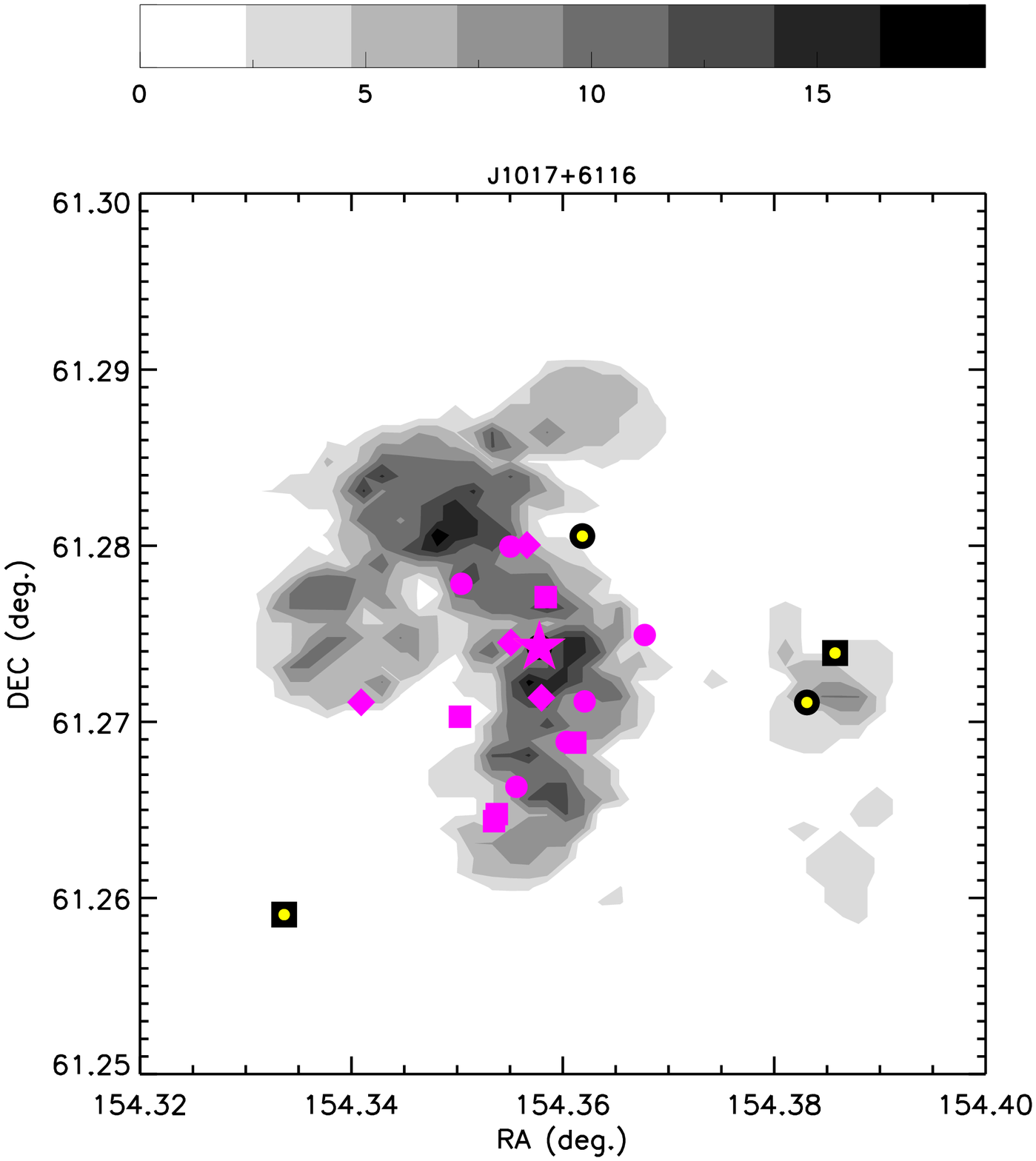}
\caption{Selected high density regions for  CARLAJ0800+4029 at z=1.986, CARLAJ2039-2514 at z=1.999 (two selected regions), and CARLAJ1017+6116 at z=2.801. The symbols are the same as in Fig.~\ref{fig:c1}.}\label{fig:c3}
\end{figure*}

In section \ref{sec:sample} we describe how we identify the densest regions in each CARLA cluster, where the contamination by interloper galaxies is $\lesssim 20\%$.  Figs.~\ref{fig:c1}~-~\ref{fig:c3} present the SNR maps for all CARLA clusters and in this section we explain our subjective choices.

 CARLAJ1358+5752 at z=1.368: the AGN is close to the highest peak. There are other three peaks. One of these last two is only partially covered by the HST image and  another visually seems to be an extension of the second peak. We focus our analysis on the first two peaks, at SNR$_c$=17 and 15.
 
CARLAJ0958-2904 at z=1.392:  the AGN is close to one of the three highest peaks, all with similar SNR,  SNR$_c$=6, and we center our region on the AGN. This is the structure in which we expect more contamination, with $\sim 28\%$ being field galaxies.

CARLAJ0116-2052 at z=1.425: the AGN is within 0.5\arcmin~from the highest SNR peak, SNR$_c$=16. We concentrate only on the region centered around the highest SNR, because  we cannot study in details the second highest SNR region, which is only partially covered by the HST image.

CARLAJ1103+3449 at z=1.442:  The highest SNR peak, is at $\sim0.6$\arcmin  \ south of the AGN (in the right bottom of the SNR map), and the spectroscopically confirmed members are found in other two highest peaks. We don't have enough information to know if the highest SNR peak is associated with the AGN, and we consider the region centered on the AGN in this analysis, with SNR$_c$=10. The AGN host is a star-forming ETG that lies on the field star formation main sequence at the cluster redshift, and we found a large reservoir of molecular gas tens of kiloparsecs south of it \citep{20markov}.

 CARLAJ1131-2705 at z=1.446: the AGN is close to the highest peak, with SNR$_c$=7. The second highest peak is a too much contaminated by field galaxies. We concentrate only on the region centered around the highest SNR peak.
    
  CARLAJ2355-0002 at z=1.490: the AGN is close to the center of the highest SNR peak at SNR$_c$=15.  The other peaks have much lower SNR.
 
CARLAJ1129+0951 at z=1.528: we observe two high peaks. The AGN is within 0.5\arcmin \ from the highest peak. Both regions host spectroscopically confirmed members. The lowest peak is only partially covered by the HST image and visually appears as an extension of the main overdensity. We focus our analysis on the highest peak, which has SNR$_c$=14.

 CARLAJ1753+6310 at 1.582: the AGN is close to the center of the highest peak at SNR$_c$=23. \citet{18noirot} discovered a serendipitous structure with 6 spectroscopically confirmed members at z=2.117, superposed to the highest peak. The overdensity around the highest SNR has a regular circular form, and CARLAJ1753+6310  is one of the three clusters that we classified as highly probable cluster \citep{18noirot}. 
 The caveat with this cluster is that we cannot separate the cluster that hosts the AGN from the z=2.117 structure. However, with this caveat in mind, we  make the hypothesis that most of the galaxies in the overdensity belong to the cluster associated to the AGN. \citet{16cooke} has shown that the core of t is dominated by passive galaxies.

CARLAJ1052+0806 at z=1.646: the AGN is at the center of the highest peak, SNR$_c$=15. The second highest peak is too much contaminated by field galaxies. We concentrate only on the region centered around the highest SNR peak.

 CARLAJ1300+4009 at z=1.675:   the AGN is close to the highest peak, SNR$_c$=15. The other two peaks are at lower SNR$_c$. One is an extension of the main peak, and the other does not host any spectroscopically confirmed members. We concentrate only on the region centered around the highest SNR peak.

   CARLAJ2227-2705 at z=1.692: the cluster shows several low SNR peaks. The highest peak is at $\sim$0.9\arcmin \ north of the AGN.  The region around this peak and other minor peaks hosts several spectroscopically confirmed members of the two serendipitous structures discovered by \citet{18noirot} at z=1.357 and z=1.478. We do not have enough information to associate neither this or the other overdensities to the AGN and the spectroscopically confirmed members. Even if the region centered around the AGN has a SNR$_c$=11, and includes several confirmed members, we consider that the contamination from the background and the serendipitous structures is too uncertain to keep this cluster in our analysis.
  
 CARLAJ1510+5958 at z=1.725: we observe three SNR peaks. The highest density peak (top of the SNR map) is one only partially covered by the HST image. This and another peak also host several spectroscopically confirmed members belonging to the serendipitous discoveries at z=0.875 and z=0.977 \citep{18noirot}. We concentrate on the third peak (SNR$_c$=11) that is also close to the AGN and around which are also found other three spectroscopically confirmed members.

CARLAJ0800+4029 at z=1.896: the AGN is at the center of the highest peak, SNR$_c$=17. There are other overdensities but they are too much contaminated by field galaxies. We concentrate only on the region centered around the highest SNR peak.

CARLAJ1018+0530 at z=1.952: we observe three high SNR peaks. The AGN is closest to the first highest peak, with SNR$_c$=12. We find cluster members in all of these three regions. One of these highest peak  (in the SNR map, the peak on the bottom) encompasses a region that is only partially covered by the HST image. We keep only the two highest peaks that are completely covered by HST (SNR$_c$=11 and SNR$_c$=12) for this cluster, which might be a multiple cluster structure. These regions correspond to the regions discussed in \citet{18noirot}, where we point out that there is a concentration of bright continuum-only sources in a region northeast of the AGN. This corresponds to our second overdensity, and it shows potential intra-cluster light. Narrow-band imaging available for this field permits us to derive photometric redshifts that are consistent with the AGN redshifts (Werner et al., in preparation). 

 CARLAJ2039-2514 at z=1.999: the region with the highest SNR is $\sim0.7$\arcmin \ north of the AGN. The first two highest peaks are within 30\arcsec, and the AGN is close to the third highest peak. We select two regions, centered on the highest peak (SNR$_c$=10), and the third peak (SNR$_c$=8). This cluster presents a clearly defined color-magnitude relation \citep{16noirot} and it might be an extended multi-peak structure. 
 
 CARLAJ1017+6116 at z=2.801: the region with the highest SNR$_c$ is centered at a distance of $\sim0.6$\arcmin  \ from the AGN.  \citet{18noirot} discovered a galaxy overdensity around this highest peak, with spectroscopical redshift $z=1.235$.  The second highest peak, with SNR$_c$=14, is close to the AGN, and closer to the cluster spectroscopically confirmed members. In this analysis, we exclude the overdensity at $z=1.235$, and consider the  second higher overdensity associated with the AGN.

\section{PASSIVE AND ACTIVE GALAXY SELECTION} \label{app-pa}

Here we detail how we arrived at the apparent magnitude colour-colour cuts we use in section~\ref{sec:color} to define passive cluster galaxies. We also show that rest--frame {\it UVJ} diagrams obtained with only our four TPHOT magnitudes are not accurate enough to separate passive galaxies at the redshift of the CARLA clusters. 

Throughout this analysis, we use the Santini et al.'s sSFR as \lq true\rq\ sSFR and hereafter define as sSFR--passive galaxies those with  ${\rm sSFR} < 10^{-9.5} M_{\odot}$/year, which characterizes the quiescent region at our cluster redshifts \citeg{14whita,19leja}.

We start with the {\it combined CANDELS catalogue}, described in $\sec~\ref{sec:sample}$, and select galaxies with $IRAC1 <22.6$~mag,  ($IRAC1-IRAC2)>-0.1$~mag and at $1<z<2$. This selection results in 646 galaxies, of which $26 \pm 3$\% have ${\rm sSFR} < 10^{-9.5} M_{\odot}$/year, which we define as passive. 

Applying the same magnitude, colour and redshift selection to the {\it TPHOT-CANDELS catalogue}, we obtain 147 galaxies, of which $20 \pm 4$\% are passive. Because of the limited statistics from the {\it TPHOT-CANDELS catalogue}, we consolidate our analysis by using the {\it combined CANDELS catalog}, and scale our results to the TPHOT photometry using empirical relations.

We use this sample to build {\it UVJ} diagrams using CANDELS rest-frame {\it U, V,} and {\it J} bands from our {\it combined CANDELS catalogue}.  The passive region defined by \citet{09will} includes $77\pm 4$\% of the sSFR--passive galaxies. The contamination of sSFR--active galaxies (hereafter defined as galaxies with ${\rm sSFR} \geq 10^{-9.5} M_{\odot}$/year) is $12\pm 3$\%. Hereafter, we define galaxies that are selected within the passive region defined by \citet{09will} as {\it UVJ}--passive. This means that the {\it UVJ}--passive sample includes a sample of CANDELS sSFR--passive galaxies that is $\sim 80$\%  complete and $\sim 90$\% pure. {\it UVJ}--passive galaxies are $24 \pm 2$\% of our total selected galaxies.

We then derive new rest-frame {\it U, V} and {\it J}-band magnitudes by running the software EAZY (Brammer et al. 2008) on our  {\it TPHOT-CANDELS catalogue} fixing the galaxy redshifts to CANDELS  spectroscopic redshifts, and using the same templates as those used by the CANDELS collaboration in \citet{18fang} (Dave Kocevski's private communication). When deriving the {\it U, V} and {\it J}-band in this way (i.e. with four TPHOT bandpasses), our {\it UVJ}--passive sample is $\sim$65\% complete and $\sim$80\% pure.  Hence, we consider that our TPHOT bandpasses do not permit us a reasonable estimation of the galaxy spectral energy distribution from which we can derive rest-frame {\it U, V} and {\it J}-band magnitudes. 

\begin{figure*}
\includegraphics[width=0.32\textwidth]{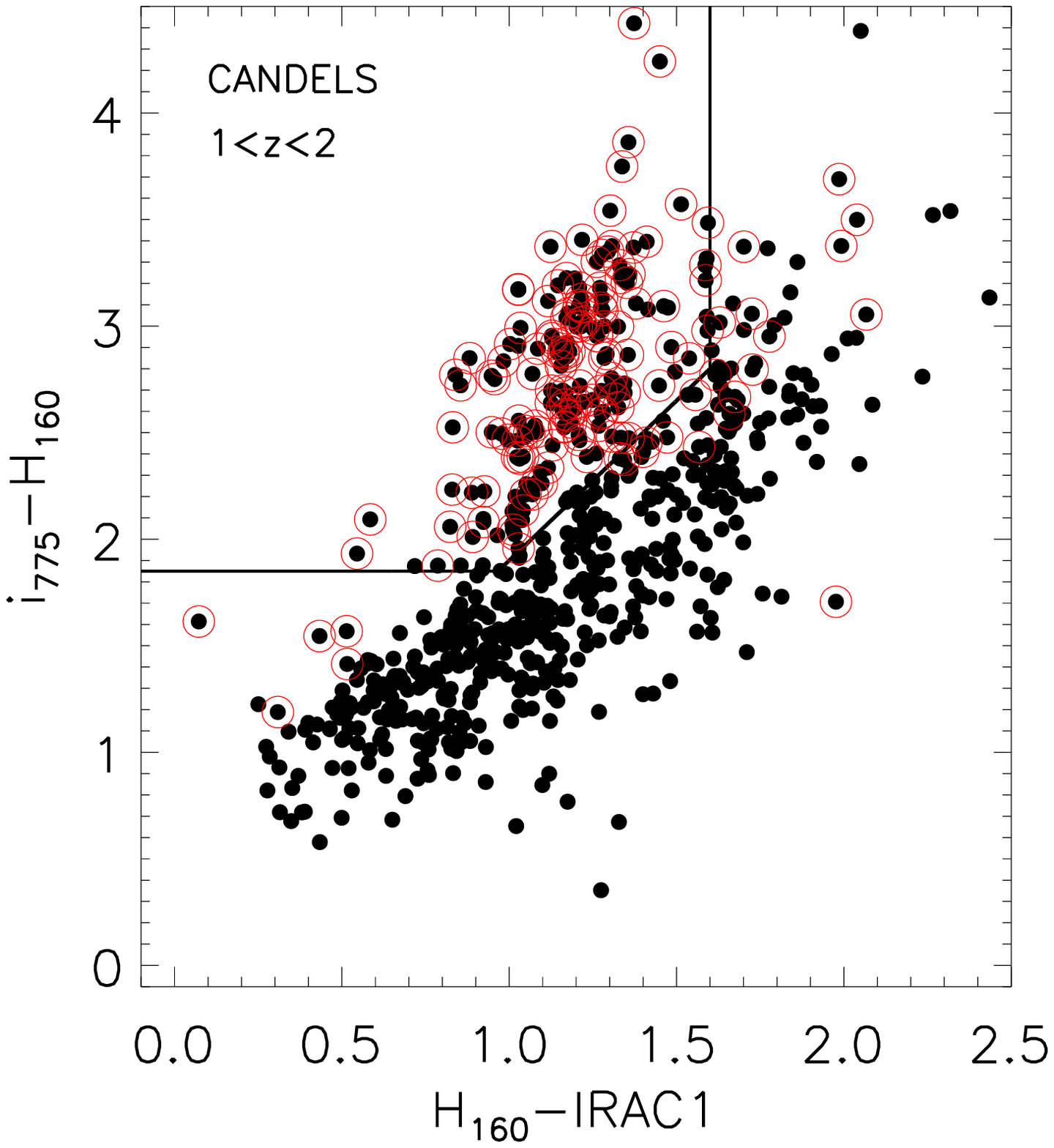}
\includegraphics[width=0.32\textwidth]{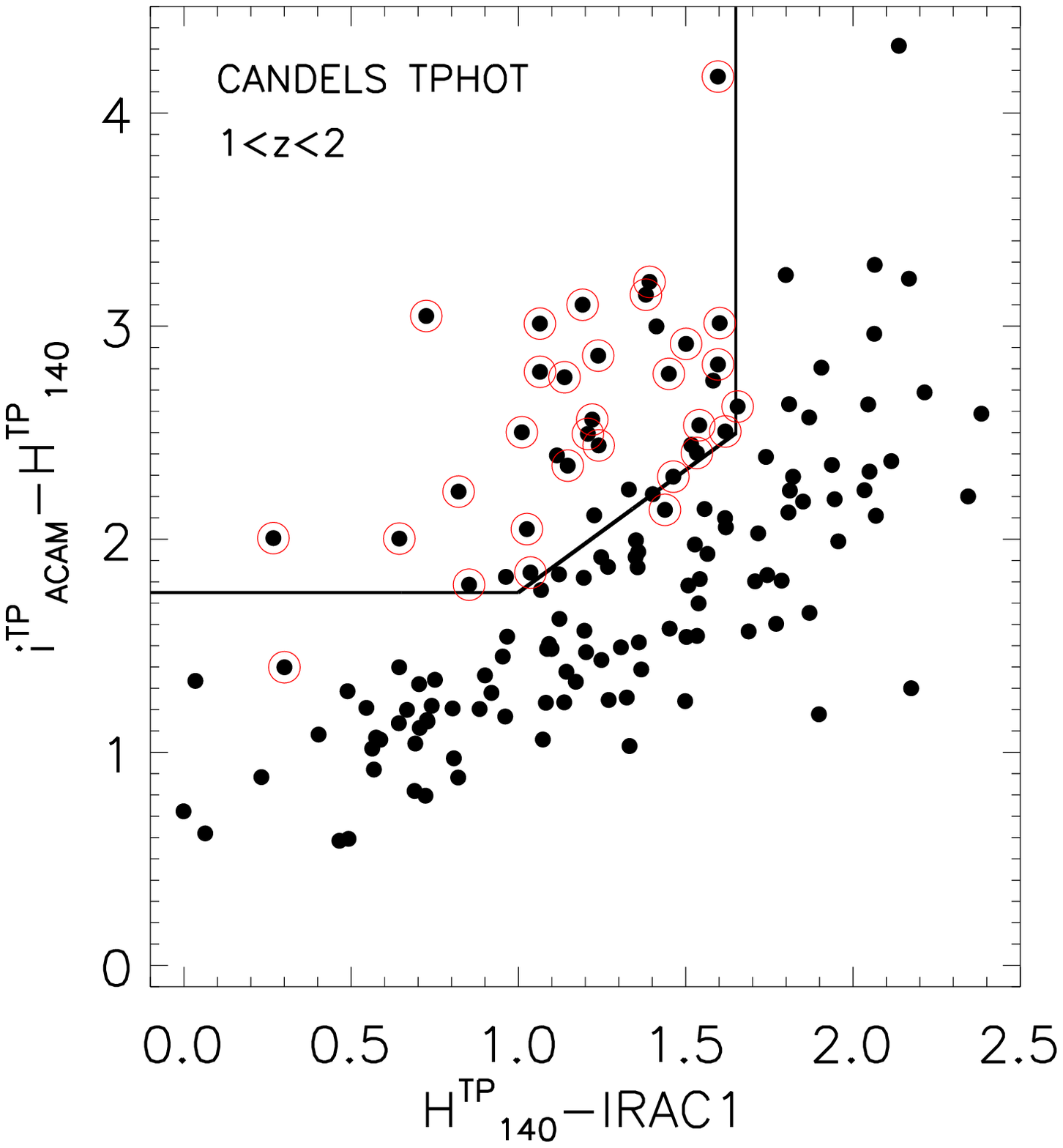}
\includegraphics[width=0.32\textwidth]{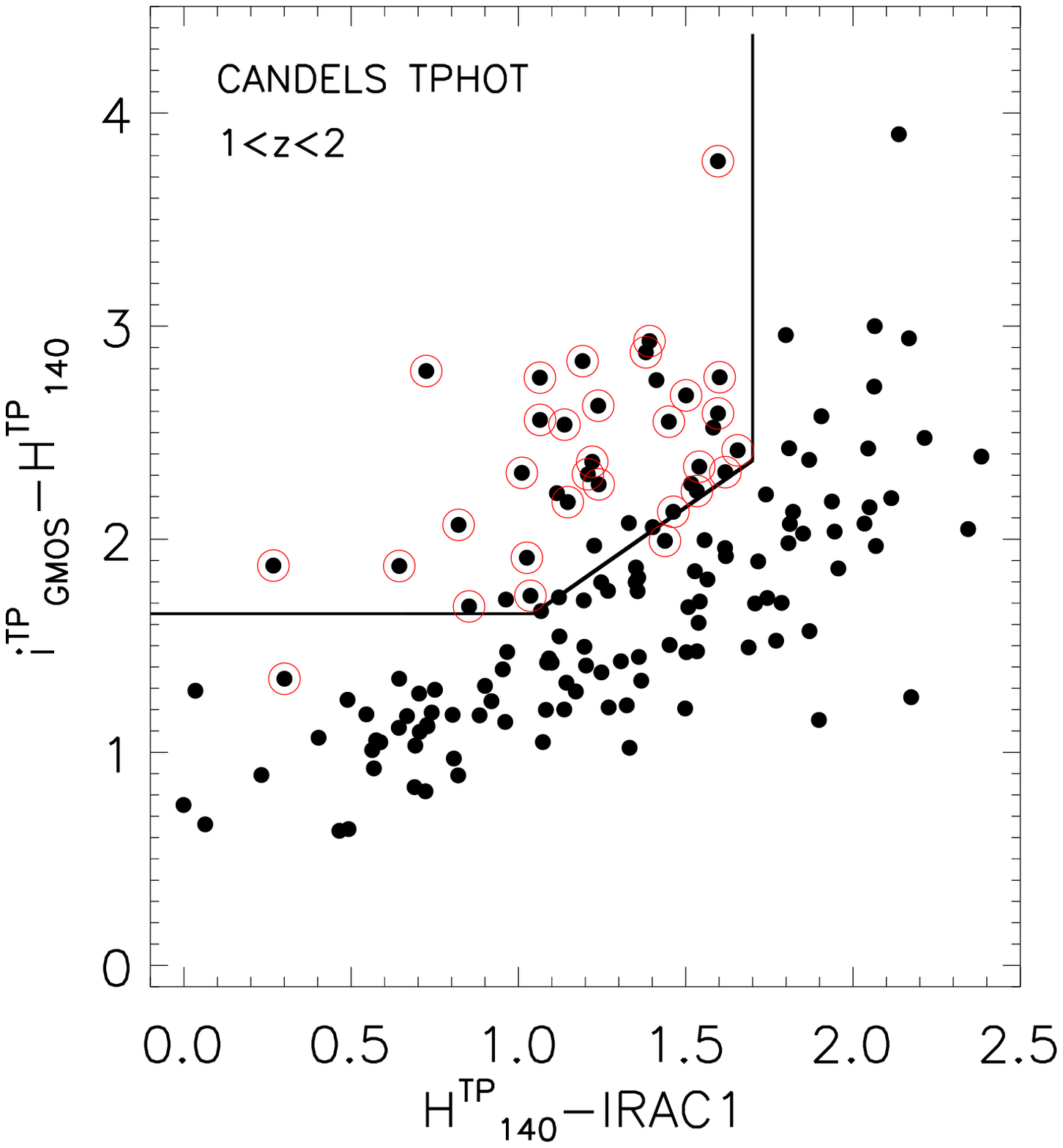}
\caption{Left: Passive galaxy selection using \citet{13guo}'s photometry of the 600 galaxies selected from the combined CANDELS catalogue at $1<z<2$. The empty red circles show galaxies with \citet{15san} ${\rm sSFR} < 10^{-9.5} M_{\odot}$/year.  The lines show the optimized passive galaxy region at $1<z<2$ obtained by using published CANDELS colors from \citet{13guo} and similar to the  Williams et al. (2009)'s {\it UVJ} diagram region.  At our survey redshifts, these apparent magnitudes permit us to select in a very similar way as the Williams et al. (2009)'s {\it UVJ} diagram region. Middle: Passive galaxy selection at $1<z<2$, using the TPHOT photometry modified to ACAM depth and filter response (left) and modified to GMOS depth and filter response (right). The empty red circles show galaxies with \citet{15san} ${\rm sSFR} < 10^{-9.5} M_{\odot}$/year. \label{sel:seltphot}}
\end{figure*}

Therefore, we use our {\it combined CANDELS catalogue} to derive a linear relation between rest-frame {\it U, V} and {\it J} ~bandpasses and apparent magnitudes of our observed bands, and build the equivalent of Williams et al.'s regions in the CANDELS apparent $i_{775} H_{160}IRAC1$ diagram.  Using this empirical conversion and iterating to optimise for both completeness and purity, we find an optimal passive galaxy region defined by:
\begin{eqnarray*}
(i_{775}-H_{160}) &>& 1.85\\
 (H_{160}-IRAC1) &<&1.6 \\
 (i_{775}-H_{160}) &>& 0.4+1.5\times (H_{160}-IRAC1),  
 \end{eqnarray*}
as shown in (Fig.~\ref{sel:seltphot}). This colour-colour cut results in a passive galaxy sample that is $\sim 85$\% complete and pure for galaxies at $1<z<2$.  These results show that a selection of passive/active galaxies based on apparent CANDELS magnitudes in the redshift range $1<z<2$ is equivalent to a {\it UVJ} selection using Williams et al.'s region and published CANDELS {\it U, V} and {\it J}-band magnitudes. This apparent magnitude selection gives much better passive galaxy selection than the use of rest-frame {\it UVJ} diagrams obtained using our four TPHOT bandpasses. 

We also explored a passive region selection within a simpler two line cut, as implemented in other works \citeg{13ilbert,14vanw}, but find no significant difference so we limit our analysis to the three-line passive region.

The last step is to adapt these colours to take into account the filter transmissions of the ground-based ACAM and GMOS $i-band$ and ISAAC $z-$bands of the CARLA data. Using the {\it TPHOT-CANDELS catalogue}, we derive a linear relation between the CANDELS rest-frame {\it U, V} and {\it J} bandpasses and TPHOT apparent magnitudes, and build the equivalent of Williams et al.'s regions in the apparent $i^{TP}_{ACAM}H^{TP}_{140}IRAC1^{TP}$ diagram. Then we optimize the parameters of the passive galaxy empirical regions by changing the four variables a, b, c, d in the relations: $ (i^{TP}_{ACAM}-H^{TP}_{140}) > a$,  $H^{TP}_{140}-IRAC1^{TP} <$ b and $(i^{TP}_{ACAM}-H^{TP}_{140}) > c+d\times (H^{TP}_{140}-IRAC1^{TP})$, to optimize the completeness and purity of passive galaxies in the selected region. 

The optimal choice of parameters for clusters observed with ACAM is:
\begin{eqnarray*}
 (i^{TP}_{ACAM}-H^{TP}_{140}) &>& 1.75 \\
 (H^{TP}_{140}-IRAC1^{TP}) &<&1.65 \\
(i^{TP}_{ACAM} - H^{TP}_{140}) &>& 0.6+1.15\times (H^{TP}_{140}-IRAC1^{TP}), 
 \end{eqnarray*}
 whilst for clusters observed with GMOS, the following parameters are used:
\begin{eqnarray*}
 (i^{TP}_{GMOS}-H^{TP}_{140}) &>& 1.65 \\
 (H^{TP}_{140}-IRAC1^{TP}) &<&1.7 \\
(i^{TP}_{GMOS} - H^{TP}_{140}) &>& 0.5+1.1\times (H^{TP}_{140}-IRAC1^{TP}). 
 \end{eqnarray*}
 
 These regions can be seen in Fig.~\ref{sel:seltphot}. Both selections result in a purity of $\sim 80$\%. The ACAM selection results in a completeness of 80\%, whereas for the clusters with GMOS data the completeness increases to $\sim 90$\%. The active galaxy sample is $\sim 95$\% complete and pure in both selections.

 At $1.5\lesssim z\lesssim2$, the  {\it i}-band  corresponds to the rest frame NUV--band and the  {\it z}-band  to the rest frame U--band. We studied the difference in the passive galaxy sample completeness and purity when using the  {\it z}-band  instead of the  {\it i}-band.
 
Using the same techniques as above, but limiting the galaxy sample to those in the redshift range $1.5<z<2$, we define a passive region through:
\begin{eqnarray*}
(z_{850}-H_{160}) &>& 1.6\\
 (H_{160}-IRAC1) &<&1.6 \\
 (z_{850}-H_{160}) &>& 0.35+1.1\times (H_{160}-IRAC1). 
 \end{eqnarray*}
 
\begin{figure*}
\includegraphics[width=0.45\textwidth]{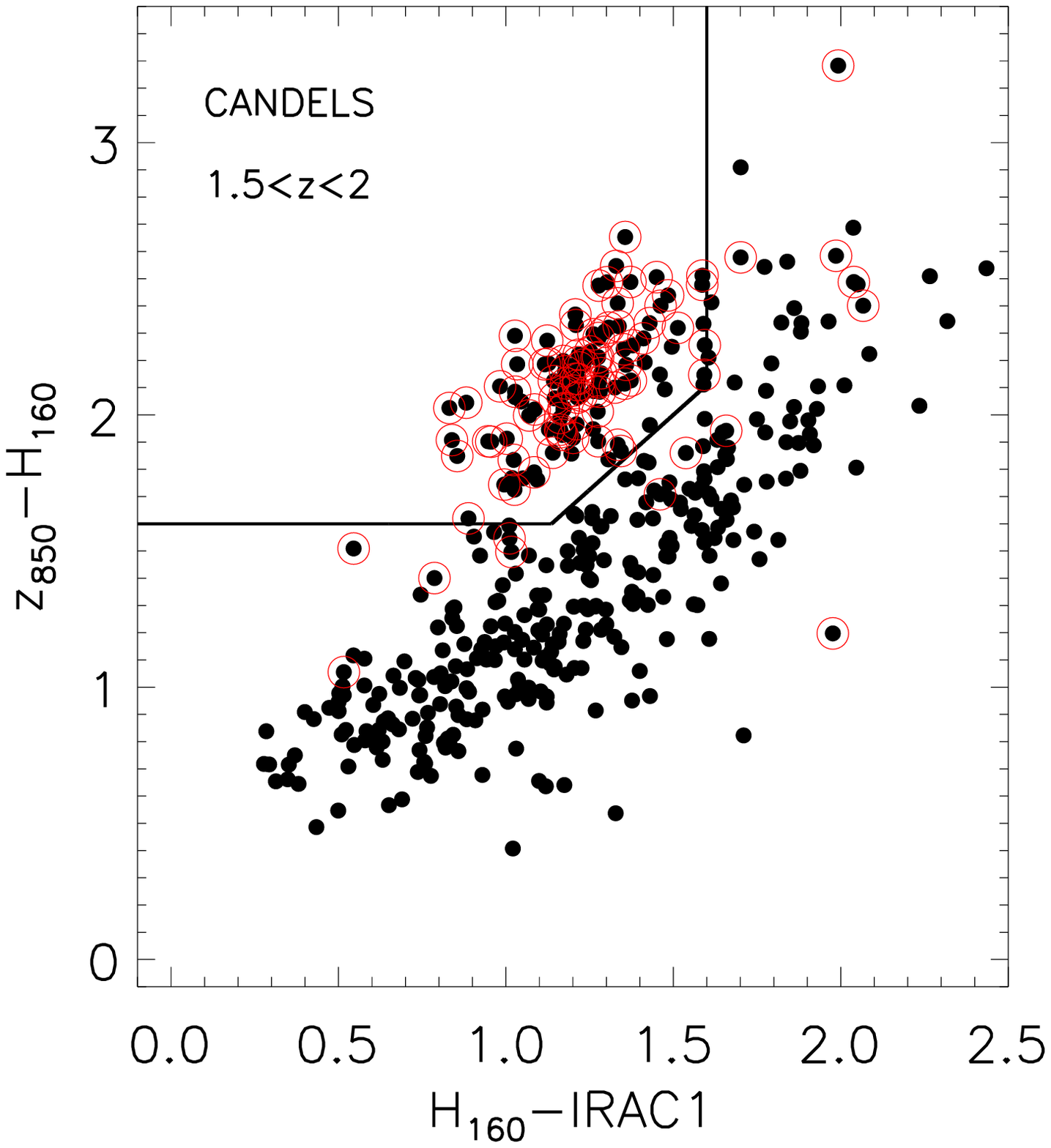}
\includegraphics[width=0.45\textwidth]{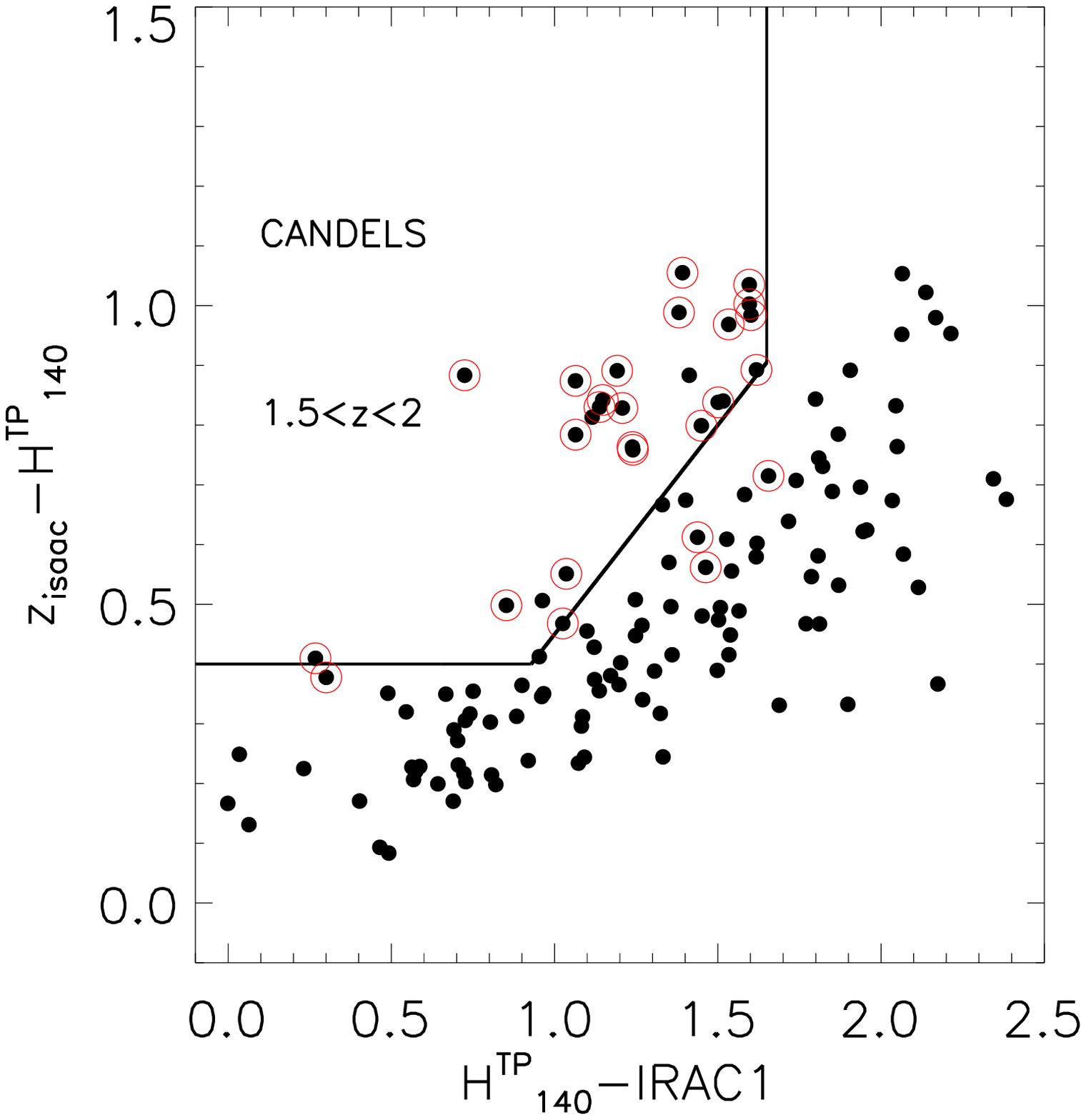}
\caption{Left: Passive galaxy selection using CANDELS photometry. Right: Passive galaxy selection using our TPHOT photometry. The same as Fig.~\ref{sel:seltphot} when using the ISAAC  {\it z}-band  instead of the  {\it i}-band .   \label{sel:selection-zband}}
\end{figure*}

This selection, seen in Fig.~\ref{sel:selection-zband}, results in a passive sample completeness and purity of $\sim 85\%$. The active galaxy sample is $\sim 95$\% complete and pure.

Again we use the same techniques as above with the TPHOT-derived photometry, to transform these relations to our ISAAC  {\it z}-band observation depth and filter response. Our optimal passive region is defined by:
\begin{eqnarray*}
(z_{ISAAC}^{TP}-H^{TP}_{140}) &>& 0.4\\
 (H^{TP}_{140}-IRAC1^{TP}) &<&1.65 \\
 (z_{ISAAC}^{TP}-H^{TP}_{140}) &>& -0.25+0.7\times (H^{TP}_{140}-IRAC1^{TP}). 
 \end{eqnarray*}

This region is shown in Fig.~\ref{sel:selection-zband}, and it selects a passive sample that has a completeness and purity of $\sim 85\%$. The active galaxy sample is $\sim 95$\% complete and pure.

To confirm the completeness and purity of our cluster passive/active selection obtained above, we used our CARLA SFR and mass measurements to calculate the sSFR for the spectroscopically confirmed members in our clusters from \citet{18noirot}, among which 79 have SFR measurements and were detected in all the bandpasses in which their cluster was observed. We measured sSFR using the SFRs from \citet{18noirot} and galaxy masses estimated in this work (only for galaxies detected in {\it IRAC1}), resulting in sSFR measurements for 54 galaxies.  $H_{\alpha}$ SFR are available for 40 of these galaxies, and we use O[III] SFR measurements when the $H_{\alpha}$ flux was not measured. Only 47 galaxies of the 54 were also observed in the i-band or z-band imaging necessary to separate active from passive galaxies using the color-color diagrams obtained in this appendix. The  passive galaxy selection obtained using this last sample is $\sim 80\%$ complete ($\sim 80\%$ of the color-color galaxies selected as passive have log(sSFR)$<-9.5  \ {\rm [yr^{-1}]}$ at $\sim 2 \sigma$)  and pure ($\sim 20\%$ of the color-color galaxies selected as passive have log(sSFR)$ > -9.5  \ {\rm [yr^{-1}]}$ at $\sim 2 \sigma$). Our sample of galaxies selected as active is also  $\sim 80\%$ complete and pure. With these small numbers, we cannot reach firm statistical conclusions about the completeness and purity of our colour-selected passive sample. However, our CARLA spectroscopically confirmed member passive selection is consistent with the completeness and purity of $\sim 80-85\%$ that we estimated in this appendix using CANDELS catalogs.

In the paper main text, we drop the prefix "TP" for the TPHOT magnitudes in Figure~\ref{colcol-clus}, and after Sec. 2.7.

\clearpage


\section{FRACTIONS AS A FUNCTION OF REDSHIFT AND DENSITY CONTRAST }
\begin{figure*}
\center
\includegraphics[angle=90,width=0.45\textwidth]{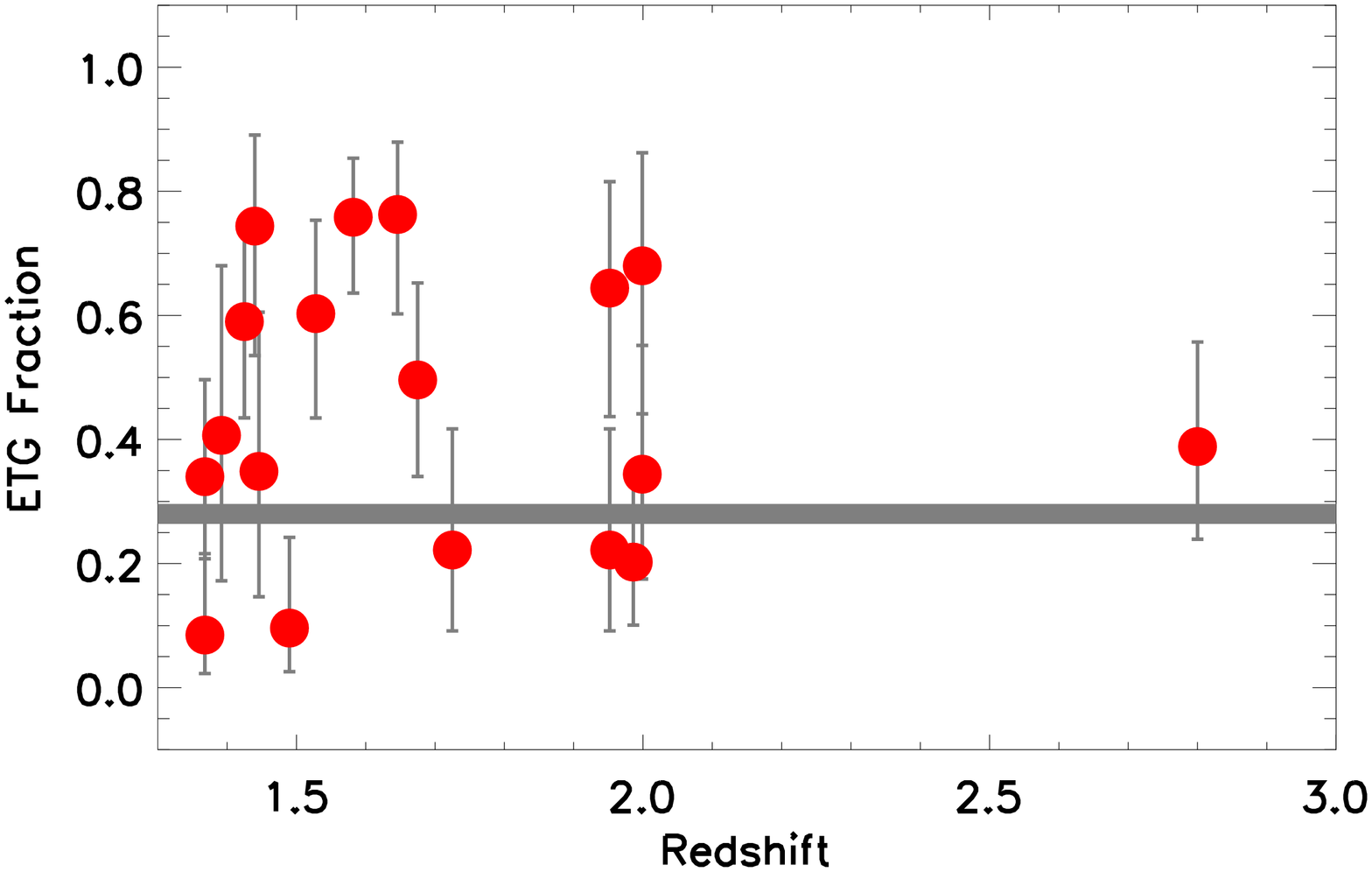}
\includegraphics[angle=90,width=0.45\textwidth]{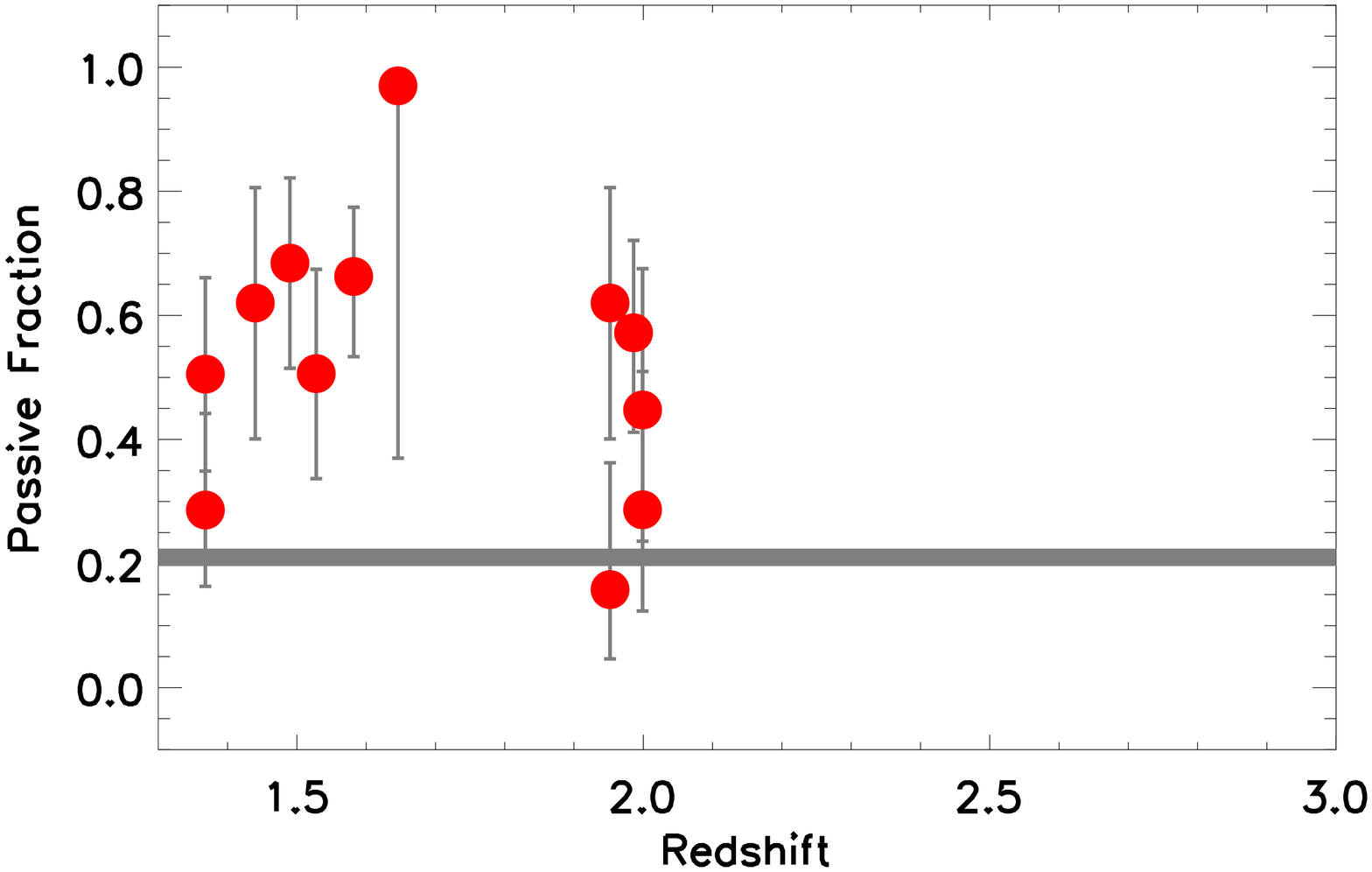}
\includegraphics[angle=90,width=0.45\textwidth]{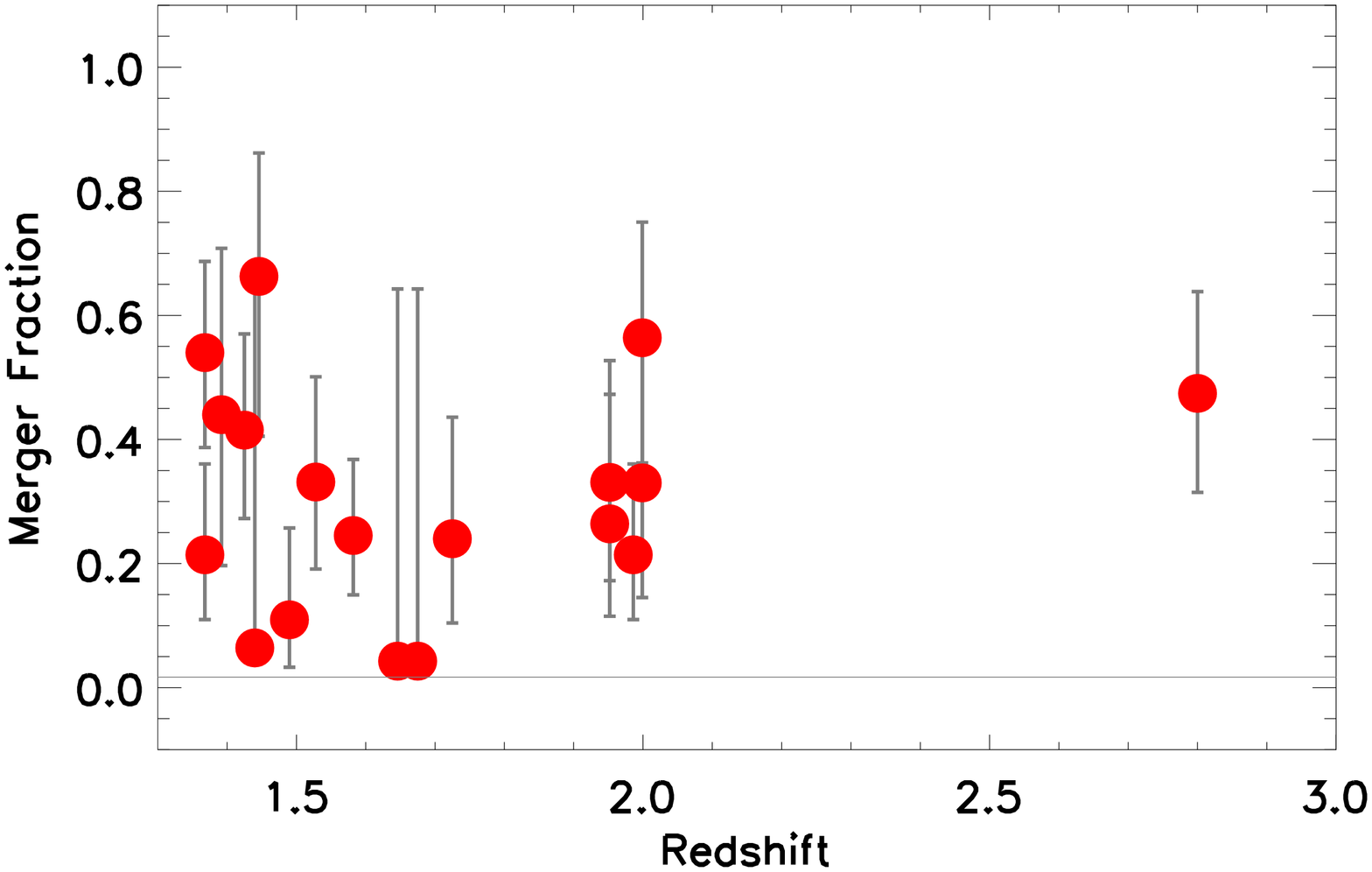}
\caption{ETG, passive, and merger galaxy fractions as a function of redshift. The grey region shows the $\pm1~\sigma$ range of the CANDELS fractions. All the fractions that we considered in this work do not show evolution with redshift. We cannot separate passive and active galaxies in the cluster at $z=2.8$. We don't observe significant correlations.}\label{fig:af0}
\end{figure*}

\begin{figure*}
\center
\includegraphics[angle=90,width=0.45\textwidth]{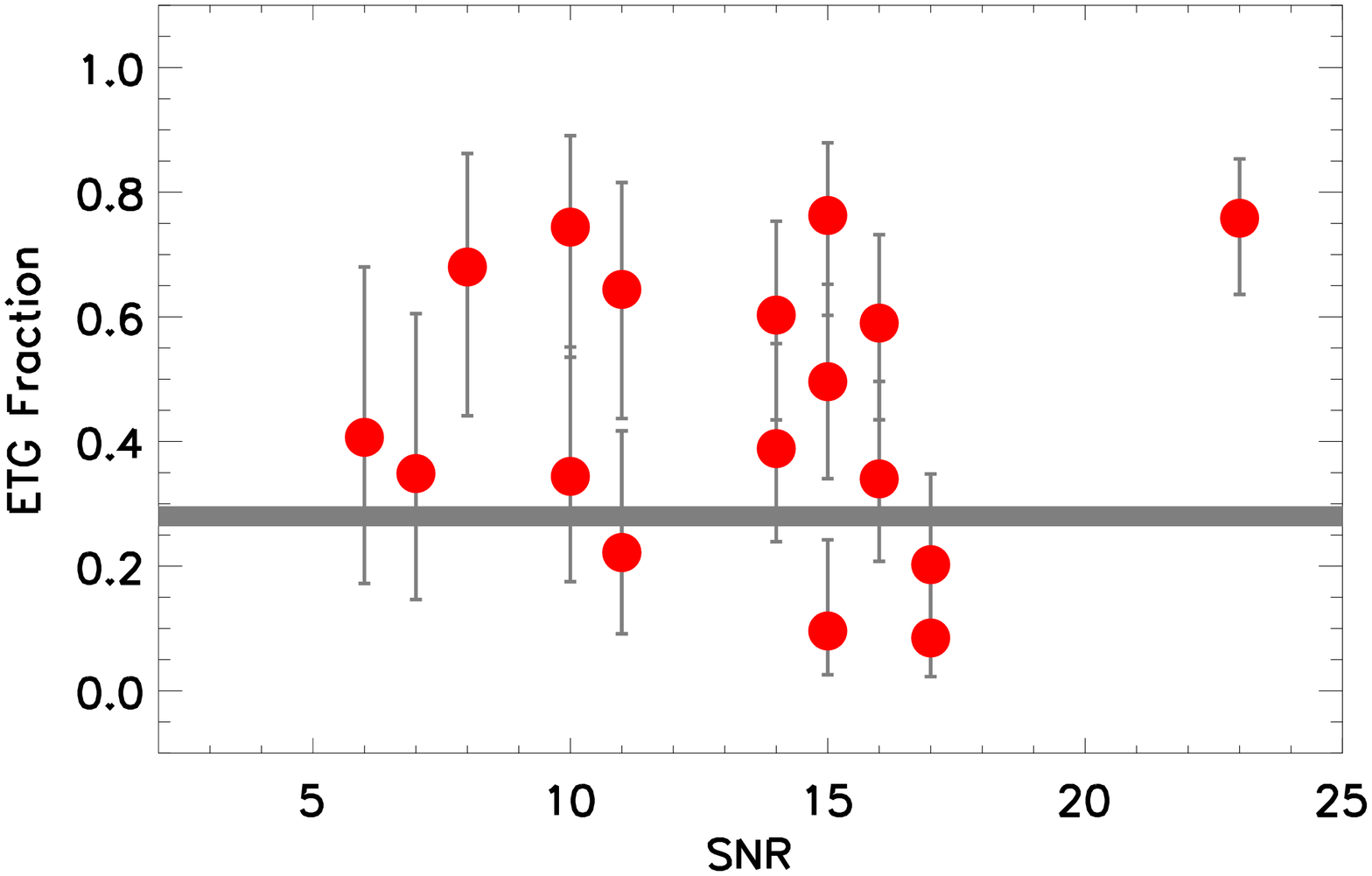}
\includegraphics[angle=90,width=0.45\textwidth]{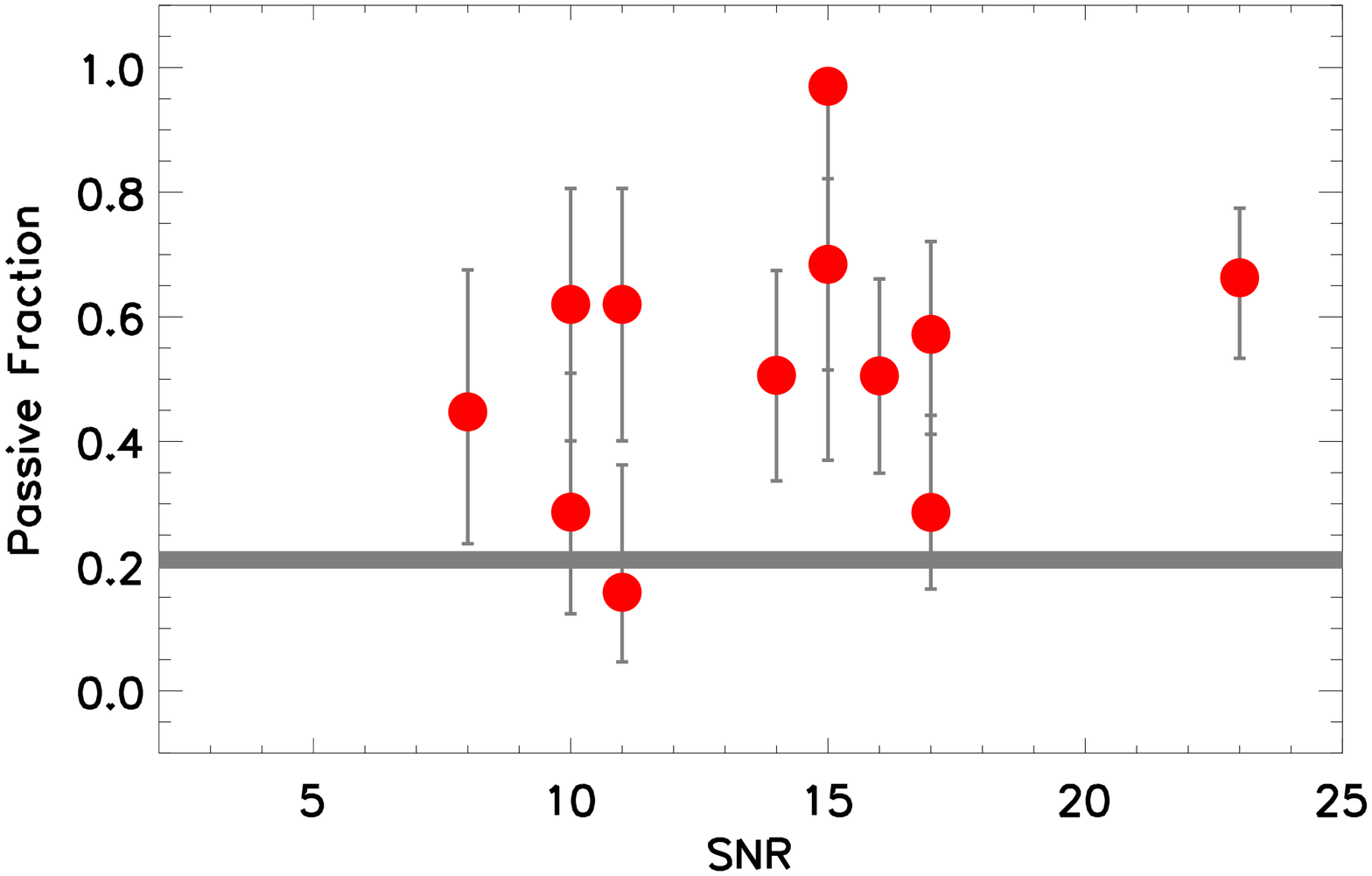}
\includegraphics[angle=90,width=0.45\textwidth]{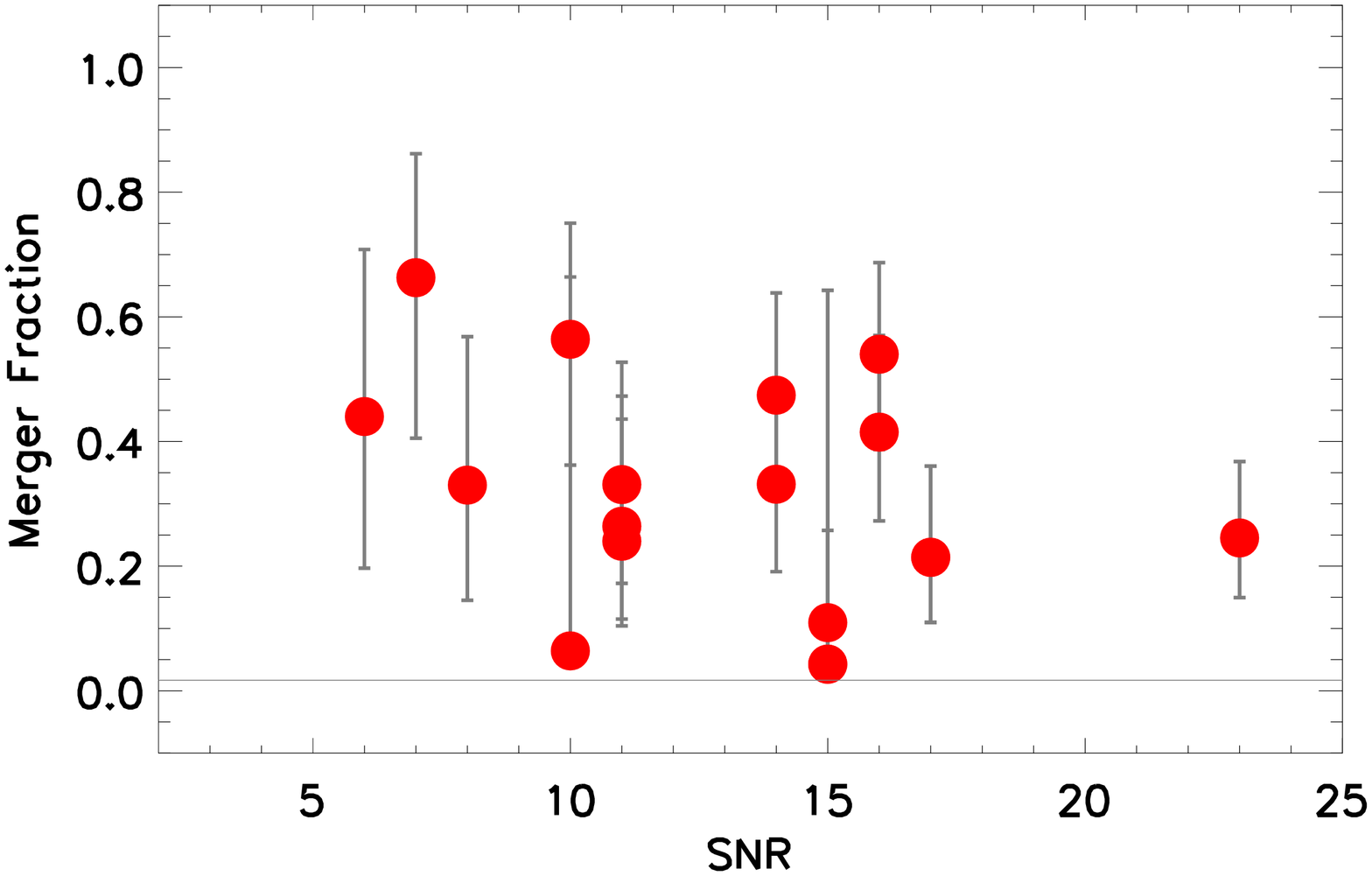}
\caption{ETG, passive,  and merger galaxy fractions as a function of SNR (as a proxy for density constrast; see text). The grey region shows the $\pm1~\sigma$ range of the CANDELS fractions. We don't observe significant correlations.}
\label{fig:af2}
\end{figure*}

\end{appendix}

\clearpage




\begin{thebibliography}{}

\bibitem[Afanasiev et al.(2022)]{Afanasiev2022} Afanasiev, A. V., Mei, S., Fu, H., et al.\ 2022, submitted to \aap
\bibitem[Alberts et al.(2016)]{16alberts} Alberts, S., Pope, A., Brodwin, M., et al.\ 2016, \apj, 825, 72. doi:10.3847/0004-637X/825/1/72
\bibitem[Andreon \& Huertas-Company(2011)]{11andreon} Andreon, S., \& Huertas-Company, M.\ 2011, \aap, 526, A11 
\bibitem[Aoyama et al.(2021)]{21aoyama} Aoyama, K., Kodama, T., Suzuki, T.~L., et al.\ 2021, arXiv:2111.00813
\bibitem[Arnouts et al.(1999)]{99arn} Arnouts, S., Cristiani, S., Moscardini, L., et al.\ 1999, \mnras, 310, 540
\bibitem[Arnouts et al.(2013)]{13arn} Arnouts, S., Le Floc'h, E., Chevallard, J. et al.\ 2013, \aap, 558, 67
\bibitem[Balogh et al.(2017)]{17balogh} Balogh, M.~L., Gilbank, D.~G., Muzzin, A., et al.\ 2017, \mnras, 470, 4168. doi:10.1093/mnras/stx1370
\bibitem[Barden et al.(2012)]{12barden} Barden, M., H{\"a}u{\ss}ler, B., Peng, C.~Y., et al.\ 2012, \mnras, 422, 449. doi:10.1111/j.1365-2966.2012.20619.x
\bibitem[Baronchelli et al.(2016)]{16baro} Baronchelli, I., Scarlata, C., Rodighiero, G., et al.\ 2016, \apjs, 223, 1. doi:10.3847/0067-0049/223/1/1
\bibitem[Barro et al.(2013)]{Barro13} Barro, G., Faber, S.~M., P{\'e}rez-Gonz{\'a}lez, P.~G., et al.\ 2013, \apj, 765, 104. doi:10.1088/0004-637X/765/2/104
\bibitem[Barro et al.(2014)]{Barro14} Barro, G., Faber, S.~M., P{\'e}rez-Gonz{\'a}lez, P.~G., et al.\ 2014, \apj, 791, 52. doi:10.1088/0004-637X/791/1/52
 \bibitem[Bartelmann \& Schneider(2001)]{01bart}Bartelmann, M. \& Schneider, P.  \ 2001, PhR. 340, 291
\bibitem[Behroozi et al.(2019)]{19beh} Behroozi, P., Wechsler, R.~H., Hearin, A.~P. et al.  \ 2019, \mnras, 488, 3143
\bibitem[Bertin \& Arnouts(1996)]{96bertin} Bertin, E., \& Arnouts, S.\ 1996, \aaps, 117, 393 
\bibitem[Bournaud et al.(2011)]{Bournaud11} Bournaud, F., Chapon, D., Teyssier, R., et al.\ 2011, \apj, 730, 4. doi:10.1088/0004-637X/730/1/4
\bibitem[Brammer et al.(2012)]{12bra} Brammer, G.~B., van Dokkum, P.~G., Franx, M., et al.\ 2012, \apjs, 200, 13 
\bibitem[Brodwin et al.(2013)]{13brodwin} Brodwin, M., Stanford, S.~A., Gonzalez, A.~H., et al.\ 2013, \apj, 779, 138. doi:10.1088/0004-637X/779/2/138
\bibitem[Bruzual \& Charlot(2003)]{03bc} Bruzual, G., \& Charlot, S.\ 2003, \mnras, 344, 1000 
\bibitem[Buat et al.(2018)]{18buat} Buat, V., Boquien, M., Ma{\l}ek, K., et al.\ 2018, \aap, 619, A135. doi:10.1051/0004-6361/201833841
\bibitem[Buat et al.(2019)]{19buat} Buat, V., Ciesla, L., Boquien, M., et al.\ 2019, \aap, 632, A79. doi:10.1051/0004-6361/201936643
\bibitem[Calzetti et al.(2000)]{00cal} Calzetti, D., Armus, L., Bohlin, R.~C., et al.\ 2000, \apj, 533, 682. doi:10.1086/308692
\bibitem[Cameron(2011)]{2011PASA...28..128C} Cameron, E.\ 2011, \pasa, 28, 128 
\bibitem[Casey et al.(2015)]{15casey} Casey, C.~M., Cooray, A., Capak, P., et al.\ 2015, \apjl, 808, L33. doi:10.1088/2041-8205/808/2/L33
\bibitem[Casey(2016)]{16casey} Casey, C.~M.\ 2016, \apj, 824, 36. doi:10.3847/0004-637X/824/1/36
\bibitem[Castignani et al.(2014)]{14cast}  Castignani, G., Chiaberge, M.,  Celotti, A. et al.\ 2014, \apj, 792, 114
\bibitem[Cautun et al.(2014)]{2014MNRAS.441.2923C} Cautun, M., van de Weygaert, R., Jones, B.~J.~T., \& Frenk, C.~S.\ 2014, \mnras, 441, 2923 
\bibitem[Chabrier(2003)]{2003PASP..115..763C} Chabrier, G.\ 2003, \pasp, 115, 763 
\bibitem[Chang et al.(2015)]{Chang15} Chang, Y.-Y., van der Wel, A., da Cunha, E., et al.\ 2015, \apjs, 219, 8. doi:10.1088/0067-0049/219/1/8
\bibitem[Charlot \& Fall(2000)]{00charlot} Charlot, S. \& Fall, S.~M.\ 2000, \apj, 539, 718. doi:10.1086/309250
\bibitem[Chiang et al.(2013)]{13chiang} Chiang, Y.-K., Overzier, R., \& Gebhardt, K.\ 2013, \apj, 779, 127 
\bibitem[Chevallard et al.(2013)]{13chevallard} Chevallard, J., Charlot, S., Wandelt, B., et al.\ 2013, \mnras, 432, 2061. doi:10.1093/mnras/stt523
\bibitem[Chiang et al.(2017)]{17chiang} Chiang, Y.-K., Overzier, R.~A., Gebhardt, K., et al.\ 2017, \apjl, 844, L23. doi:10.3847/2041-8213/aa7e7b
\bibitem[Cluver et al.(2017)]{Cluver17} Cluver, M.~E., Jarrett, T.~H., Dale, D.~A., et al.\ 2017, \apj, 850, 68. doi:10.3847/1538-4357/aa92c7
\bibitem[Collet et al.(2015)]{15collet} Collet, C., Nesvadba, N.~P.~H., De Breuck, C., et al.\ 2015, \aap, 579, A89. doi:10.1051/0004-6361/201424544
\bibitem[Contini et al.(2020)]{20contini} Contini, E., Gu, Q., Ge, X., et al.\ 2020, \apj, 889, 156. doi:10.3847/1538-4357/ab6730
\bibitem[Cooke et al.(2015)]{15cooke}  Cooke, E. A.; Hatch, N. A.; Rettura, A. et al.\ 2015, \mnras, 452, 2318
\bibitem[Cooke et al.(2016)]{16cooke} Cooke, E.~A., Hatch, N.~A., Stern, D., et al.\ 2016, \apj, 816, 83. doi:10.3847/0004-637X/816/2/83
\bibitem[Daddi et al.(2017)]{17daddi} Daddi, E., Jin, S., Strazzullo, V. et al.\ 2017, \apjl, 846, 31
\bibitem[Darvish et al.(2015)]{15darvish} Darvish, B., Mobasher, B., Sobral, D., et al.\ 2015, \apj, 805, 121. doi:10.1088/0004-637X/805/2/121
\bibitem[Darvish et al.(2016)]{16darvish} Darvish, B., Mobasher, B., Sobral, D., et al.\ 2016, \apj, 825, 113. doi:10.3847/0004-637X/825/2/113
\bibitem[Darvish et al.(2018)]{18darvish} Darvish, B., Martin, C., Gon{\c{c}}alves, T.~S., et al.\ 2018, \apj, 853, 155. doi:10.3847/1538-4357/aaa5a4
\bibitem[Delaye et al.(2014)]{14dela} Delaye, L., Huertas-Company, M., Mei, S., et al.\ 2014, \mnras, 441, 203. doi:10.1093/mnras/stu496
\bibitem [De Breuck et al.(2001)]{01debreu}De Breuck, C., van Breugel, W., Rottgering, H., et al. \ 2001, \aj, 121, 1241
\bibitem [Dekel et al.(2009)]{09Dekel}Dekel, A. , Birnboim, Y., {Engel}, G, et al. \ 2009, \nat, 457, 451
\bibitem[De Lucia et al.(2012)]{12delucia} De Lucia, G., Weinmann, S., Poggianti, B.~M., et al.\ 2012, \mnras, 423, 1277. doi:10.1111/j.1365-2966.2012.20983.x
\bibitem[Dressler(1980)]{80dressler} Dressler, A.\ 1980, \apj, 258, 351. doi:10.1086/157753
\bibitem[Evrard et al.(2008)]{08ev} Evrard, A.~E., Bialek, J., Busha, M., et al.\ 2008, \apj, 672, 122 
\bibitem[Fang et al.(2018)]{18fang} Fang, J.~J., Faber, S.M., Koo, D.C. et al.\ 2018, \apj, 858, 100
\bibitem[Fassbender et 
al.(2011)]{11fass} Fassbender, R., Nastasi, A., B{\"o}hringer, H., et al.\ 2011, \aap, 527, L10 
\bibitem[Ferreras \& Silk(2000)]{Ferreras2000} Ferreras, I. \& Silk, J.\ 2000, \apjl, 541, L37. doi:10.1086/312898
\bibitem[Galametz et al.(2013)]{13gal} Galametz, A., Grazian, A., Fontana, A., et al.\ 2013, \apjs, 206, 10
\bibitem[Geha et al.(2012)]{12geha} Geha, M., Blanton, M.~R., Yan, R., et al.\ 2012, \apj, 757, 85. doi:10.1088/0004-637X/757/1/85
\bibitem[Gehrels, N. (2006)]{ge06}Gehrels, N. 2006,\apj, 303, 336
\bibitem[George \& Zingade(2015)]{George2015} George, K. \& Zingade, K.\ 2015, \aap, 583, A103. doi:10.1051/0004-6361/201424826
\bibitem[George(2017)]{George2017} George, K.\ 2017, \aap, 598, A45. doi:10.1051/0004-6361/201629667
\bibitem[Gobat et al.(2018)]{18gobat} Gobat, R., Daddi, E., Magdis, G., et al.\ 2018, Nature Astronomy, 2, 239. doi:10.1038/s41550-017-0352-5
\bibitem[Greenslade et al.(2018)]{18green} Greenslade, J., Clements, D.~L., Cheng, T., et al.\ 2018, \mnras, 476, 3336. doi:10.1093/mnras/sty023
\bibitem[Guo et al.(2013)]{13guo} Guo, Y., Ferguson, H.~C., Giavalisco, M., et al.\ 2013, \apjs, 207, 24 
\bibitem[Harshan et al.(2021)]{Harshan2021} Harshan, A., Gupta, A., Tran, K.-V., et al.\ 2021, \apj, 919, 57. doi:10.3847/1538-4357/ac0cf3
\bibitem[Hatch et al.(2014)]{14hatch} Hatch, N.,  Wylezalek, D., Kurk, J. D. et al. \ 2014, \mnras, 445, 280
\bibitem[Hayashi et al.(2011)]{11haya} Hayashi, M., Kodama, 
T., Koyama, Y., Tadaki, K.-I., \& Tanaka, I.\ 2011, \mnras, 415, 2670 
\bibitem[Hayashi et al.(2016)]{16hayashi} Hayashi, M., Kodama, T., Tanaka, I., et al.\ 2016, \apjl, 826, L28. doi:10.3847/2041-8205/826/2/L28
\bibitem[Huertas-Company et al.(2010)]{Huertas2010} Huertas-Company, M., Aguerri, J.~A.~L., Tresse, L., et al.\ 2010, \aap, 515, A3. doi:10.1051/0004-6361/200913188
\bibitem[Huertas-Company et al.(2015)]{15huertas} Huertas-Company, M., Gravet, R., Cabrera-Vives, G. et al. \ 2015, \apjs, 221, 8
\bibitem[Hwang et al.(2012)]{Hwang12} Hwang, H.~S., Geller, M.~J., Kurtz, M.~J., et al.\ 2012, \apj, 758, 25. doi:10.1088/0004-637X/758/1/25
\bibitem[Ilbert et al.(2006)]{06ilbert} Ilbert, O., Arnouts, S., McCracken, H.~J., et al.\ 2006, \aap, 457, 841
\bibitem[Ilbert et al.(2013)]{13ilbert} Ilbert, O., McCracken, H.~J., Le F{\`e}vre, O., et al.\ 2013, \aap, 556, A55
\bibitem[Kawinwanichakij et al.(2017)]{Kawi2017} Kawinwanichakij, L., Papovich, C., Quadri, R.~F., et al.\ 2017, \apj, 847, 134. doi:10.3847/1538-4357/aa8b75

\bibitem[Jaff{\'e} et al.(2011)]{Jaffe2011} Jaff{\'e}, Y.~L., Arag{\'o}n-Salamanca, A., De Lucia, G., et al.\ 2011, \mnras, 410, 280. doi:10.1111/j.1365-2966.2010.17445.x
\bibitem[Kaviraj et al.(2011)]{Kaviraj2011} Kaviraj, S., Schawinski, K., Silk, J., et al.\ 2011, \mnras, 415, 3798. doi:10.1111/j.1365-2966.2011.19002.x
\bibitem[Kaviraj et al.(2013)]{Kaviraj2013} Kaviraj, S., Cohen, S., Ellis, R.~S., et al.\ 2013, \mnras, 428, 925. doi:10.1093/mnras/sts031
\bibitem[Kartaltepe et al.(2015)]{15karta} Kartaltepe, J.~S., Mozena, M., Kocevski, D., et al.\ 2015, \apjs, 221, 11 
\bibitem[Koyama et al.(2021)]{21koyama} Koyama, Y., Polletta, M. del C., Tanaka, I., et al.\ 2021, \mnras, 503, L1. doi:10.1093/mnrasl/slab013
\bibitem[Kubo et al.(2017)]{17kubo} Kubo, M., Yamada, T., Ichikawa, T., et al.\ 2017, \mnras, 469, 2235. doi:10.1093/mnras/stx920
\bibitem[Izquierdo-Villalba et al.(2018)]{18iz} Izquierdo-Villalba, D., Orsi, {\'A}. A., Bonoli, S., et al.\ 2018, \mnras, 480, 1340. doi:10.1093/mnras/sty1941
\bibitem[Laidler et al.(2007)]{07lai} Laidler, V.~G., Papovich, C., Grogin, N.~A., et al.\ 2007, \pasp, 119, 1325 
\bibitem[Laigle et al.(2018)]{18laigle} Laigle, C., Pichon, C., Arnouts, S., et al.\ 2018, \mnras, 474, 5437. doi:10.1093/mnras/stx3055
\bibitem[Lee et al.(2006)]{Lee2006} Lee, J.~H., Lee, M.~G., \& Hwang, H.~S.\ 2006, \apj, 650, 148. doi:10.1086/507121
\bibitem[Lee et al.(2009)]{09lee} Lee, S.-K., Idzi, R., Ferguson, H.~C., et al.\ 2009, \apjs, 184, 100 
\bibitem[Leja et al.(2019)]{19leja} Leja, J., Tacchella, S.,  Conroy, C. et al.\ 2019, \apj, 880, 9
\bibitem[Lemaux et al.(2019)]{19lemaux} Lemaux, B.~C., Tomczak, A.~R., Lubin, L.~M., et al.\ 2019, \mnras, 490, 1231. doi:10.1093/mnras/stz2661
\bibitem[{\L}okas \& Mamon(2001)]{01lokas} {\L}okas, E.~L. \& Mamon, G.~A.\ 2001, \mnras, 321, 155. doi:10.1046/j.1365-8711.2001.04007.x
\bibitem[Lovell et al.(2018)]{18lovell} Lovell, C.~C., Thomas, P.~A., \& Wilkins, S.~M.\ 2018, \mnras, 474, 4612. doi:10.1093/mnras/stx3090
\bibitem[Lubin et al.(2009)]{09lubin} Lubin, L.~M., Gal, R.~R., Lemaux, B.~C., et al.\ 2009, \aj, 137, 4867. doi:10.1088/0004-6256/137/6/4867
\bibitem[Mancini et al.(2019)]{19manci} Mancini, C., Daddi, E., Juneau, S., et al.\ 2019, \mnras, 489, 1265
\bibitem[Mansheim et al.(2017)]{Mansheim2017} Mansheim, A.~S., Lemaux, B.~C., Dawson, W.~A., et al.\ 2017, \apj, 834, 205. doi:10.3847/1538-4357/834/2/205
\bibitem[Maraston et al.(2010)]{10mar} Maraston, C., Pforr, J., Renzini, A., et al.\ 2010, \mnras, 407, 830 
\bibitem[Marinello et al.(2020)]{20mari} Marinello, M., Overzier, R.~A., R{\"o}ttgering, H.~J.~A., et al.\ 2020, \mnras, 492, 1991. doi:10.1093/mnras/stz3333=
\bibitem[Markwardt(2009)]{09mpfit} Markwardt, C.~B.\ 2009,Astronomical Data Analysis Software and Systems XVIII, 411, 251
\bibitem[Markov et al.(2020)]{20markov} Markov, V., Mei, S., Salom{\'e}, P., et al.\ 2020, \aap, 641, A22
\bibitem[Makovoz, D., \& Khan 2005]{05mar} Makovoz, D., \& Khan, I. 2005, in ASP Conf. Ser. 347, Astronomical Data
Analysis Software and Systems XIV, ed. P. Shopbell, M. Britton, \& R. Ebert (San Francisco, CA: ASP), 81
\bibitem[Martig et al.(2009)]{Martig2009} Martig, M., Bournaud, F., Teyssier, R., et al.\ 2009, \apj, 707, 250. doi:10.1088/0004-637X/707/1/250
\bibitem[Martinache et al.(2018)]{18marti} Martinache, C., Rettura, A., Dole, H., et al.\ 2018, \aap, 620, A198. doi:10.1051/0004-6361/201833198
\bibitem[Mei et al.(2006)]{06meia} Mei, S., Blakeslee, J.~P., Stanford, S.~A., et al.\ 2006a, \apj, 639, 81 
\bibitem[Mei et al.(2006)]{06meib} Mei, S., Holden, B.~P.,  Blakeslee, J.~P., et al.\ 2006b, \apj, 644, 759 
\bibitem[Mei et al.(2009)]{09mei} Mei, S., Holden, B.~P., Blakeslee, J.~P., et al.\ 2009, \apj, 690, 42 
\bibitem[Mei et al.(2012)]{12mei} Mei, S., Stanford, S.~A., Holden, B.~P., et al.\ 2012, \apj, 754, 141 
\bibitem[Mei et al.(2015)]{15mei} Mei, S., Scarlata, C., Pentericci, L., et al.\ 2015, \apj, 804, 117. doi:10.1088/0004-637X/804/2/117
\bibitem[Merlin et al.(2015)]{15merlin} Merlin, E., Fontana, A., Ferguson, H.~C., et al.\ 2015, \aap, 582, A15 
\bibitem[Merlin et al.(2016a)]{16merlin_a} Merlin, E., Amor{\'{\i}}n, R., Castellano, M., et al.\ 2016, \aap, 590, A30 
\bibitem[Merlin et al.(2016b)]{16merlin_b} Merlin, E., Bourne, N., Castellano, M., et al.\ 2016, \aap, 595, A97 
\bibitem[Merten et al.(2015)]{15merten} Merten, J., Meneghetti, M., Postman, M., et al.\ 2015, \apj, 806, 4. doi:10.1088/0004-637X/806/1/4
\bibitem[Miley \& De Breuck(2008)]{08miley}  Miley, G. \& De Breuck, C., 2008, A\&ARv, 15, 67
\bibitem[Mo, van den Bosch \& White(2010)]{10mo} Mo, H., van den Bosch, F. \& White S., 2010, Galaxy Formation \& Evolution;  ISBN-10: 0521857937 
\bibitem[Muldrew, Hatch \& Cooke(2018)]{18mul} Muldrew S.~I., Hatch N.~A., Cooke E.~A., 2018, \mnras, 473, 2335
\bibitem[Muzzin et al.(2013)]{13muzzin} Muzzin, A., Marchesini, D., Stefanon, M., et al.\ 2013, \apj, 777, 18. doi:10.1088/0004-637X/777/1/18
\bibitem[Navarro et al.(1997)]{97nfw} Navarro, J.~F., Frenk, C.~S., \& White, S.~D.~M.\ 1997, \apj, 490, 493. doi:10.1086/304888
\bibitem[Newman et al.(2013)]{13newman} Newman, A.~B., Ellis, 
\bibitem[Noirot et al.(2016)]{16noirot} Noirot, G., Vernet, J., De Breuck, C., et al.\ 2016, \apj, 830, 90 
\bibitem[Noirot et al.(2018)]{18noirot} Noirot, G.,  al.\ 2018, \apj, 859, 38
\bibitem[Oke \& Gunn(1983)]{1983ApJ...266..713O} Oke, J.~B., \& Gunn, J.~E.\ 1983, \apj, 266, 713 
\bibitem[Orsi et al.(2016)]{16orsi} Orsi, {\'A}. A., Fanidakis, N., Lacey, C.~G., et al.\ 2016, \mnras, 456, 3827. doi:10.1093/mnras/stv2919
\bibitem[Pannella et al.(2015)]{15pan} Pannella, M., Elbaz, D., Daddi, E., et al.\ 2015, \apj, 807, 141. doi:10.1088/0004-637X/807/2/141
\bibitem[Paterno-Mahler et al.(2017)]{17pat} Paterno-Mahler, R., Blanton, E.~L., Brodwin, M., et al.\ 2017, \apj, 844, 78. doi:10.3847/1538-4357/aa7b89
\bibitem[Peng et al.(2002)]{02peng} Peng, C.~Y., Ho, L.~C., Impey, C.~D., \& Rix, H.\ 2002, \aj, 124, 266 
\bibitem[Peng et al.(2010)]{10peng} Peng, Y.-. jie ., Lilly, S.~J., Kova{\v{c}}, K., et al.\ 2010, \apj, 721, 193. doi:10.1088/0004-637X/721/1/193
\bibitem[Peng et al.(2012)]{12peng} Peng, Y.-. jie ., Lilly, S.~J., Renzini, A., et al.\ 2012, \apj, 757, 4. doi:10.1088/0004-637X/757/1/4
\bibitem[Pforr et al.(2012)]{12pforr} Pforr, J., Maraston, C., \& Tonini, C.\ 2012, \mnras, 422, 3285 
\bibitem[Polletta et al.(2021)]{21polletta} Polletta, M., Soucail, G., Dole, H., et al.\ 2021, arXiv:2109.04396
\bibitem[Postman et al.(2005)]{05postman} Postman, M., Franx, M., Cross, N.~J.~G., et al.\ 2005, \apj, 623, 721 
\bibitem[Raichoor et al.(2011)]{11raic} Raichoor, A., Mei, S., Nakata, F., et al.\ 2011, \apj, 732, 12 
\bibitem[Reddy et al.(2015)]{15reddy} Reddy, N.~A., Kriek, M., Shapley, A.~E., et al.\ 2015, \apj, 806, 259. doi:10.1088/0004-637X/806/2/259
\bibitem[Reddy et al.(2018)]{18reddy} Reddy, N.~A., Oesch, P.~A., Bouwens, R.~J., et al.\ 2018, \apj, 853, 56. doi:10.3847/1538-4357/aaa3e7
\bibitem[Rettura, et al.(2014)]{14rettura} Rettura A., et al., 2014, \apj, 797, 109
\bibitem[Santini et al.(2015)]{15san} Santini, P., Ferguson, H.~C., Fontana, A., et al.\ 2015, \apj, 801, 97 
\bibitem[Sarron \& Conselice(2021)]{21sarron} Sarron, F. \& Conselice, C.~J.\ 2021, \mnras, 506, 2136. doi:10.1093/mnras/stab1844
\bibitem[Sazonova et al.(2020)]{Sazonova2020} Sazonova, E., Alatalo, K., Lotz, J., et al.\ 2020, \apj, 899, 85. doi:10.3847/1538-4357/aba42f
\bibitem[Scoville et al.(2007)]{07scoville} Scoville, N., Aussel, H., Brusa, M., et al.\ 2007, \apjs, 172, 1. doi:10.1086/516585
\bibitem[Sersic(1968)]{68ser} Sersic, J. L. \ 1968, Atlas de Galaxias Australes
\bibitem[Skelton et al.(2014)]{14ske} Skelton, R.~E., Whitaker, K.~E., Momcheva, I.~G., et al.\ 2014, \apjs, 214, 24 
\bibitem[Sheen et al.(2016)]{Sheen2016} Sheen, Y.-K., Yi, S.~K., Ree, C.~H., et al.\ 2016, \apj, 827, 32. doi:10.3847/0004-637X/827/1/32
\bibitem[Shi et al.(2021)]{2021ApJ...911...46S} Shi, K., Toshikawa, J., Lee, K.-S., et al.\ 2021, \apj, 911, 46. doi:10.3847/1538-4357/abe62e
\bibitem[Shi et al.(2021)]{21shi} Shi, K., Toshikawa, J., Lee, K.-S., et al.\ 2021, \apj, 911, 46. doi:10.3847/1538-4357/abe62e
\bibitem[Shimakawa et al.(2018)]{18shima} Shimakawa, R., Koyama, Y., R{\"o}ttgering, H.~J.~A., et al.\ 2018, \mnras, 481, 5630. doi:10.1093/mnras/sty2618
\bibitem[Smith et al.(2005)]{05smith} Smith, G.~P., Treu, T., Ellis, R.~S., et al.\ 2005, \apj, 620, 78. doi:10.1086/426930
\bibitem[Sirianni et al. (2005)]{05si}Sirianni, M. et al. 2005, PASP, 117, 1049
\bibitem[Sorba, \& Sawicki(2018)]{18sor} Sorba, R., \& Sawicki, M.\ 2018, \mnras, 476, 1532
\bibitem[Stanford et al.(2012)]{12stan} Stanford, S.~A., Brodwin, M., Gonzalez, A.~H., et al.\ 2012, \apj, 753, 164. doi:10.1088/0004-637X/753/2/164
\bibitem[Strazzullo et al.(2019)]{19stra} Strazzullo, V., Pannella, M., Mohr, J.~J., et al.\ 2019, \aap, 622, A117. doi:10.1051/0004-6361/201833944
\bibitem[Strazzullo et al.(2018)]{18stra} Strazzullo, V., Coogan, R.~T., Daddi, E., et al.\ 2018, \apj, 862, 64. doi:10.3847/1538-4357/aacd10
\bibitem[Strazzullo et al.(2016)]{16stra} Strazzullo, V., Daddi, E., Gobat, R., et al.\ 2016, \apjl, 833, L20. doi:10.3847/2041-8213/833/2/L20
\bibitem[Strazzullo et al.(2015)]{15stra} Strazzullo, V., Daddi, E., Gobat, R., et al.\ 2015, \aap, 576, L6. doi:10.1051/0004-6361/201425038
\bibitem[Tadaki et al.(2012)]{12tada} Tadaki, K., Kodama, 
T., Ota, K., et al.\ 2012, \mnras, 423, 2617 
\bibitem[Tomczak et al.(2014)]{14tom} Tomczak, A.~R., Quadri, R.~F., Tran, K.-V.~H., et al.\ 2014, \apj, 783, 85. doi:10.1088/0004-637X/783/2/85
\bibitem[Tomczak et al.(2017)]{17tom} Tomczak, A.~R., Lemaux, B.~C., Lubin, L.~M., et al.\ 2017, \mnras, 472, 3512. doi:10.1093/mnras/stx2245
\bibitem[Tomczak et al.(2019)]{19tom} Tomczak, A.~R., Lemaux, B.~C., Lubin, L.~M., et al.\ 2019, \mnras, 484, 4695. doi:10.1093/mnras/stz342
\bibitem[Tran et al.(2010)]{10tran} Tran, K.-V.~H., Papovich, 
C., Saintonge, A., et al.\ 2010, \apjl, 719, L126 
\bibitem[Trayford et al.(2020)]{20trayford} Trayford, J.~W., Lagos, C. del P., Robotham, A.~S.~G., et al.\ 2020, \mnras, 491, 3937. doi:10.1093/mnras/stz3234
\bibitem[van der Burg et al.(2020)]{20burg} van der Burg, R.~F.~J., Rudnick, G., Balogh, M.~L., et al.\ 2020, \aap, 638, A112. doi:10.1051/0004-6361/202037754
\bibitem[van Dokkum et al.(2013)]{13vand} van Dokkum, P., Brammer, G., Momcheva, I., et al.\ 2013, arXiv:1305.2140 
\bibitem[van der Wel et al.(2006)]{06vanw} van der Wel, A., Franx, M., Wuyts, S., et al.\ 2006, \apj, 652, 97 
\bibitem[van der Wel et al.(2012)]{12vanw} van der Wel, A., Bell, E.~F., H{\"a}ussler, B., et al.\ 2012, \apjs, 203, 24
\bibitem[van der Wel et al.(2014)]{14vanw} van der Wel, A., Franx, M., van Dokkum, P.~G., et al.\ 2014, \apj, 788, 28
\bibitem[Wang et al.(2016)]{16wang} Wang, T., Elbaz, D., Daddi, E., et al.\ 2016, \apj, 828, 56. doi:10.3847/0004-637X/828/1/56
\bibitem[Wen Han (2013)]{13wen} Wen, Z.~L. \& Han J.~L.\ 2013, \mnras, 436, 275
\bibitem[West (1988)]{88west} West, M.~J., Oemler, A. \& Dekel, A.\ 1988, \apj,  327,1 
\bibitem[Wetzel et al.(2013)]{13wetzel} Wetzel, A.~R., Tinker, J.~L., Conroy, C., et al.\ 2013, \mnras, 432, 336. doi:10.1093/mnras/stt469
\bibitem[Whitaker et al.(2014)]{14whita} Whitaker, K.~E., Franx, M., Leja, J., et al.\ 2014, \apj, 795, 104
\bibitem[Williams et al.(2009)]{09will} Williams, R.~J.; Quadri, R.~F.; Franx, M. et al.\ 2009,  \apj, 691, 1879
\bibitem[Wright et al.(2010)]{Wright10} Wright, E.~L., Eisenhardt, P.~R.~M., Mainzer, A.~K., et al.\ 2010, \aj, 140, 1868. doi:10.1088/0004-6256/140/6/1868
\bibitem[Wylezalek et al.(2013)]{13wyle} Wylezalek, D., Galametz, A., Stern, D., et al.\ 2013, \apj, 769, 79 
\bibitem[Wylezalek et al.(2014)]{14wyle} Wylezalek, D., Vernet, J., De Breuck, C., et al.\ 2014, \apj, 786, 17 
\bibitem[Zavala et al.(2019)]{Zavala2019} Zavala, J.~A., Casey, C.~M., Scoville, N., et al.\ 2019, \apj, 887, 183. doi:10.3847/1538-4357/ab5302
\bibitem[Zeimann et al.(2012)]{12zei} Zeimann, G.~R., 
Stanford, S.~A., Brodwin, M., et al.\ 2012, \apj, 756, 115 
\bibitem[Zheng et al.(2021)]{21zheng} Zheng, X.~Z., Cai, Z., An, F.~X., et al.\ 2021, \mnras, 500, 4354. doi:10.1093/mnras/staa2882


\end{thebibliography}



\end{document}